\documentclass[prd,twocolumn,amsmath,amssymb,floatfix,superscriptaddress,showpacs]{revtex4}
\usepackage{mathrsfs}
\usepackage{graphicx}
\usepackage{color}
\usepackage{dcolumn}
\usepackage{bm}
\usepackage{float}
\usepackage[caption=false]{subfig} 
\usepackage{multirow}
\usepackage{url}
\usepackage{comment}
\usepackage{enumitem}
\usepackage{calc}
\usepackage{booktabs}
\usepackage{threeparttable}

\allowdisplaybreaks

\begin{document}
\newcommand{\tabincell}[3][1]{\renewcommand\arraystretch{#1}\begin{tabular}{@{}#2@{}}#3\end{tabular}}

\title{Accuracy of source localization for eccentric inspiraling binary mergers using a ground-based detector network}

\author{Hsing-Po Pan}
\email{hppan@phys.ncku.edu.tw}
\affiliation{Department of Physics, National Cheng-Kung University,
Tainan 701, Taiwan}
\author{Chun-Yu Lin}
\affiliation{National Center for High-Performance Computing,
Hsinchu 300, Taiwan}
\author{Zhoujian Cao}
\affiliation{Department of Astronomy, Beijing Normal University, Beijing 100875, China}
\author{Hwei-Jang Yo}
\affiliation{Department of Physics, National Cheng-Kung University,
Tainan 701, Taiwan}

\date{\today}

\begin{abstract}
The problem of gravitational wave parameter estimation and source localization
is crucial in gravitational wave astronomy.
Gravitational waves emitted by compact binary coalescences in the sensitivity
band of second-generation ground-based detectors could have non-negligible
eccentricities.
Thus it is an interesting topic to study how the eccentricity of a binary
source affects and improves the accuracy of its localization
(and the signal-to-noise ratio). 
In this work we continue to investigate this effect with the enhanced
postcircular waveform model.
Using the Fisher information matrix method,
we determine the accuracy of source localization with three ground-based
detector networks.
As expected, the accuracy of source localization is improved considerably
with more detectors in a network.
We find that the accuracy also increases significantly by increasing the
eccentricity for the large total mass ($M\ge 40M_\odot$) binaries
with all three networks.
For the small total mass ($M<40 M_\odot$) binaries, this effect is negligible.
For the smaller total mass ($M<5 M_\odot$) binaries, the accuracy could be even
worse at some orientations with increasing eccentricity.
This phenomenon comes mainly from how well the frequency of the higher
harmonic modes induced by increasing eccentricity coincides with the 
sensitive bandwidth of the detectors.
For the case of the $100 M_\odot$ black hole binary,
the improvement factor is about $2$ in general when the eccentricity grows
from $0.0$ to $0.4$.
For the cases of the $22 M_\odot$ black hole binary and the $2.74 M_\odot$
neutron star binary,
the improvement factor is less than $1.1$,
and it may be less than 1 at some orientations.
\end{abstract}

\maketitle

\section{\label{intro}introduction}
Since 2015, advanced LIGO \cite{LIGO}, with Virgo \cite{VIRGO} joining later, has identified 11 gravitational wave events \cite{PhysRevX.9.031040} of compact binary coalescence during the first and the second observation runs (O1 and O2). As their sensitivities are getting improved and detectors such as KAGRA \cite{0264-9381-29-12-124007} and LIGO-India \cite{LIGO-M1100296-v2} will also join the detection network \cite{Abbott_2016_LRR_19_1}, one would expect even more frequent events that will definitely bring a promising future for gravitational wave astronomy \cite{10.1093/nsr/nwx029,1904.03187}. Furthermore, third-generation detectors like the Einstein Telescope \cite{Freise_ET} and Cosmic Explorer \cite{Abbott_Cosmic_Explorer}, space-based detectors, such as LISA \cite{1702.00786}, Taiji \cite{Gong_2011_Taiji}, or Tianqin \cite{Luo_2016_Tianqin}, and detectors based on novel designs beyond interferometry will make the gravitational wave community even more dynamic and bring us unexpected puzzles and resolutions about nature as well.

Compact binary coalescence is a major type of gravitational wave source. The binary black hole’s inspiral, coalescence, and ringdown waveforms as the simplest two-body problem in the General Relativity (GR) have also been extensively studied analytically and numerically. Despite that all observations so far are consistent with GR \cite{Ghosh_2017_CQG_35_014002}, the fact that circular binary waveforms are primarily used as the matched filtering template in most of the current detection and parameter estimation tasks \cite{PhysRevD.46.5236,PhysRevD.47.2198,PhysRevD.48.4738,PhysRevD.49.2658,PhysRevD.89.042004} is still a fly in the ointment. A more systematic and precise analysis is therefore always sought in the attempt to more closely approach the underlying reality. Using a circular binary waveform template is computationally cheaper and physically sound, since the trajectory as a binary gets closer will be circularized efficiently \cite{PhysRev.131.435,PhysRev.136.B1224}, due to the high gravitational radiation before it’s gravitational wave has entered the sensitivity band of a ground-based detector. However, there are astrophysical scenarios in which a binary has a non-negligible eccentricity, and it may contribute observational features in the sensitivity band. 
Two closely bound black holes and a third one orbiting the mass center of the
first two can compose a hierarchical triple system \cite{Kozai_1962_AJ_67_591K},
which is common in globular clusters.
The Kozai-Lidov mechanism in these systems may also lead to non-negligible
eccentricity. 
Wen in Ref.~\cite{Wen_2003_AJ_598_419} argued that approximately 30\% of the inner
binaries in globular cluster could merge with eccentricities larger than 0.1
as they enter the advanced LIGO’s frequency band.
However, the later work in Ref.~\cite{Silsbee_2017, Antonini_2017} found that a lower
percentage of these inner binaries in isolated triple systems have nonzero eccentricities.
And Liu \textit{et al.} \cite{1905.00427} found that about 7\% of binary black
holes and 18\% of neutron star - black hole binaries in hierarchical triple
systems with spin-orbit misalignment could merge with eccentricities larger
than 0.1 at 10 Hz.
In addition, resonant binary–single interaction in globular clusters
\cite{Samsing_2014, PhysRevD.97.103014, Samsing_2018},
gravitational wave capture events in globular clusters
\cite{PhysRevLett.120.151101, PhysRevD.98.123005} and in galactic nuclei
\cite{10.1111/j.1365-2966.2009.14653.x, Gond_n_2018_APJ_860_5, Rasskazov_2019},
binary-binary encounters in globular clusters \cite{Zevin_2019},
nonhierarchical triples in globular clusters and in nuclear star clusters
\cite{1805.06458}, and bound quadruples \cite{10.1093_mnras_stz1175}
can result in eccentric binary neutron stars, neutron star–black hole binaries,
and binary black holes that will reside in the advanced LIGO’s frequency band.

An important step for gravitational data analysis is parameter estimation, by which source parameters like masses, spins, orientation, polarization, location, the phase at the merger, and their uncertainty are approximated. In the Bayesian framework of parameter estimation \cite{PhysRevD.93.024013,thrane_talbot_2019_PASA_36_e010}, those are quantified by underlying, usually intractable, posterior distributions that could be approximated by a Markov chain Monte Carlo (MCMC) simulation
\cite{Nissanke_2011,PhysRevD.89.102005,Cokelaer_2008_CQG_25_184007, PhysRevD.89.042004}
and similar algorithms. 
This is, however, numerically prohibitively expensive for the systems with a large survey of parameters. For example, the eccentric spinning binaries depend on 17 parameters \cite{PhysRevD.70.042001}.
In contrast, the Fisher information matrix allows for approximating the uncertainty \cite{PhysRevD.49.2658, PhysRevD.46.5236, PhysRevD.47.2198} without knowing the central values. The inverse of the Fisher information matrix approximates well the standard deviation of and correlations among the parameters for the case of a large signal-to-noise ratio and Gaussian noise \cite{PhysRevD.79.084032}. Simply speaking, given a set of templates, the more sensitively the waveform depends on the parameters, the larger the corresponding components of the Fisher matrix are, and therefore the smaller the standard deviation and uncertainty are. Eccentric waveforms have been used beyond the circular one to improve the parameter estimation accuracy; Kyutoku and Seto studied \cite{10.1093/mnras/stu698} the premerger localization accuracy of eccentric binary neutron stars. 
Gond{\'{a}}n \textit{et al.} investigated high-eccentricity binary black holes \cite{Gond_n_2018_APJ_855_34} and extended to binary neutron stars, neutron star–black hole binaries, and binary black holes with the black hole masses up to $110 M_\odot$ \cite{Gond_n_2019_APJ_871_178}. 
Mik\'oczi \textit{et al.} found that template waveforms with higher eccentricity could increase the accuracy of source localization for LISA \cite{PhysRevD.86.104027}. For a single detector on the ground, Sun \textit{et al.} also found \cite{PhysRevD.92.044034} that the source parameters can be estimated generally with improved uncertainty if a more eccentric waveform were used for the Fisher information matrix.

Our primary objective is to systematically assess the ability of source localization under the detector network, which is essential for multimessenger observations. A gravitational wave interferometric detector is essentially omnidirectional. A ground-based detector network is essential for the triangle localization of gravitational wave sources \cite{Fairhurst_2011_CQG_28_105021, PhysRevD.74.082004,PhysRevD.81.082001}. Note that the situation is different from that of the space-based detectors of which the position changes a lot within a period of the gravitational wave strain, and thereby the strain itself contains information about source location. In our previous work \cite{PhysRevD.96.084046}, we have found a more precise localization for an eccentric binary under a three-detector network with eccentric gravitational waveforms. In this work, we will, therefore, extend the study systematically to the four- and five-detector network with an eccentric waveform template. The enhanced postcircular (EPC) frequency domain model \cite{PhysRevD.90.084016} proposed by Huerta \textit{et al.} was used to calculate the Fisher matrix, which is based on the Yunes \textit{et al.} postcircular (PC) model \cite{PhysRevD.80.084001}.

As summarized in Ref. \cite{PhysRevD.90.084016},
the EPC model has some appealing features of the two waveform families, i.e.,
the $x$ model \cite{PhysRevD.82.024033} at 2 post-Newtonian (PN) order and the TaylorF2 model
\cite{PhysRevD.87.082004,PhysRevD.80.084043} at 3.5 PN order,
taken as reference points.
This indicates that the EPC waveform can be treated with two aspects:
one corresponds to the quasicircular part, which is up to 3.5 PN order from
the TaylorF2 model,
and the other corresponds to the eccentric part which is up to 2 PN order
from the $x$ model.
Since the effect of the pericenter precession appears at 1 PN order,
the pericenter precession has been accounted for already
in the EPC model.

From this point of view, the EPC model has two limitations:
One of the limitations is that the EPC model is only a phenomenological extension of the
PC model which makes the EPC model lack of physical explanation to eccentric
binary systems.
The other limitation is that it only describes the inspiral stage of an
eccentric binary system without the consideration of its merger and ringdown
stages yet.
For the aim of gravitational wave (GW) source parameter estimation, the first limitation is not a major concern
from the viewpoint of observation.
However, the second limitation might make the result from the EPC model on the
accuracy of parameter estimation weaker than the one from the real situation.
We will describe this model in detail in Sec.~\ref{sec:EPC}.

This paper is organized as follows. In Sec.~\ref{sec:EPC}, we introduce the EPC waveform model and define the variables. Then we describe related information about the network of advanced detectors. The noise models and the location information of the detectors that are used in this work are also presented there. In Sec.~\ref{sec:FIM}, we describe the source localization accuracy estimation method. In Sec.~\ref{sec:results}, we present our result of the source localization accuracy for eccentric binaries. The eccentric waveform model can improve the source localization accuracy quite well for the large mass binary systems, but not for the small mass systems. Finally, we summarize our conclusions in Sec.~\ref{sec:conclusion}.

We use the geometric units $G=c=1$ in this paper.
$M_\odot$ is used to denote the solar mass.
We denote the mass of the two objects of the binary as $m_1$ and $m_2$,
the total mass as $M=m_1+m_2$,
the symmetric mass ratio as $\eta =\displaystyle\frac{m_1 m_2}{M^2}$ and
the chirp mass as
$\mathcal{M}=\displaystyle\frac{(m_1 m_2)^{3/5}}{M^{1/5}}=\eta^{3/5}M$.

\section{Eccentric Model with Detector Network}\label{sec:EPC}
\subsection{\label{subsec:EPC_in_sigle}Enhanced postcircular waveform model}
The PC waveform model by Yunes \textit{et al.}~\cite{PhysRevD.80.084001}
is a waveform model for an eccentric binary coalescence in the frequency domain.
In the PC model, the conservative and dissipative orbital dynamics are treated
with the post-Newtonian approximation.
The effect of the small eccentricity is treated through a high-order spectral
decomposition.
Then the waveform is computed via the stationary-phase-approximation method
\cite{PhysRevD.49.2658}.
Later, the above result was generalized to a higher-order PN
approximation \cite{PhysRevD.82.124064,Tessmer_2010,doi:10.1002/andp.201100007}.
Huerta \textit{et al.}~extended the PC model into the EPC model
\cite{PhysRevD.90.084016}.
The EPC model is constructed with the two following requirements:
\begin{enumerate}[itemsep=1pt,partopsep=1pt,parsep=\parskip ,topsep=1pt]
\item In the zero-eccentricity limit, the model reduces to the TaylorF2
model at 3.5 PN order.
\item In the zeroth PN order, the model recovers the PC expansion,
including eccentricity corrections up to order $O(e^8)$.
\end{enumerate}

The waveform of the EPC model can be written as
\begin{equation}
\tilde{h}(f)=C\frac{\mathcal{M}^{5/6}}{D_L}f^{-7/6}\sum_{\ell = 1}^{10}
\xi_\ell\left(\frac{\ell}{2}\right)^{2/3}e^{-i\Psi_\ell}, \label{Eq_EPC}
\end{equation}
where $C=-\displaystyle\frac{1}{8\pi^{2/3}}\sqrt{\frac{5}{6}}$,
$f$ is the frequency of the gravitational wave and $\ell$ is the harmonic.
The phase $\Psi_\ell$ is defined as 
\begin{equation}
\Psi_\ell=2\pi f t_c-\ell\phi_c+\left(\frac{\ell}{2}\right)^{8/3}
\frac{3}{128\eta\nu_{\rm ecc}^5}\sum_{i = 0}^7 a_i\nu_{\rm ecc}^i.
\label{Eq_Psi_l}
\end{equation}
The EPC waveform involves 11 parameters ($e_0$, $D_L$, $\mathcal{M}$, $\eta$,
$t_c$, $\phi_c$, $\theta$, $\phi$, $\psi$, $\iota$, $\beta$),
where 
\begin{itemize}[itemsep=1pt,partopsep=1pt,parsep=\parskip ,topsep=1pt]
\item[$e_0$:] Initial eccentricity which corresponds to the initial frequency
$f_0$ of the gravitational wave.
\item[$D_L$:] Luminosity distance between the detector and the
gravitational-wave source. 
\item[$\mathcal{M}$:] Chirp mass of the binaries. 
\item[$\eta$:] Symmetric mass ratio of of the binaries. 
\item[$t_c$:] Arrival time of the coalescence signal.
\item[$\phi_c$:] Initial orbital phase of the coalescence.
\item[$\theta$:] Polar angle of the source base on the detector coordinate.
\item[$\phi$:] Azimuthal angle of the source base on the detector coordinate.
\item[$\psi$:] Polarization angle with respect to the detector.
\item[$\iota$:] Inclination angle, the polar angle between the orbital angular
momentum and the line joining the source to the detector. 
\item[$\beta$:] Azimuthal angle on the orbital plane around the line joining
the source to the detector.
\end{itemize}
The angles $\theta$ and $\phi$ describe the orientation of the source.
The explicit expression of the amplitude $\xi_\ell$ in Eq.~\eqref{Eq_EPC} is
listed in Eq.~(4.31) of Ref.~\cite{PhysRevD.80.084001},
in which it depends on $e_0$ and the angle parameters ($\theta$, $\phi$, $\psi$, $\iota$,
$\beta$).
In Eq.~\eqref{Eq_Psi_l}, $\nu_{\rm ecc}$ is the orbital velocity of the
binary object, and it depends on $e_0$ and $f$.
The explicit expression for $\nu_{\rm ecc}$ can be found in Eq.~(13) of
Ref.~\cite{PhysRevD.90.084016}.
The coefficients $a_i$ are listed in Eq.~(3.18) of Ref.~\cite{PhysRevD.80.084043}.
The three parameters $\theta$, $\phi$, and $\psi$ appear in the antenna
pattern function, which is defined as
\begin{align}
F_{+}=&-\frac{1+\cos^2\theta}{2}\cos 2\phi\cos 2\psi-
\cos\theta\sin 2\phi\sin 2\psi,\\
F_\times=&+\frac{1+\cos^2\theta}{2}\cos 2\phi\sin 2\psi-
\cos\theta\sin 2\phi\cos 2\psi.
\end{align}
Regarding to the reliability of this model, Fig.~3 of
Ref.~\cite{PhysRevD.90.084016} gives a good illustration.
As mentioned in the Introduction, the $x$ model is reliable for
eccentric binaries up to 2 PN order.
The TaylorF2/TaylorT4 models are reliable for quasicircular binaries up
to 3.5 PN order.
In this plot, the overlap for the results from the EPC model and from the
TaylorF2/TaylorT4 models decays when the initial eccentricity increases.
This indicates that the TaylorF2/TaylorT4 models break down in predicting
eccentric binaries.
In the same plot, the overlap for the results from the EPC model and from
the $x$ model almost keeps constant in the eccentricity range [$0$, $0.4$],
which tells us that the EPC model is as reliable at $e_0=0.4$ as at $e_0=0$.
As stated in Ref.~\cite{PhysRevD.90.084016}, the phase prescription of the EPC
model is reliable for $e_0\le 0.6$ for a $6 M_\odot+6 M_\odot$ system and
for $e_0\le 0.4$ for a $1.4 M_\odot+ 1.4 M_\odot$ system.
Therefore, the EPC waveform model is reliable for the initial eccentricity
up to $0.4$.

\subsection{Waveform model in the detector network}
\label{subsec:EPC_in_network}
Most of the waveform expressions for the EPC model shown in the literature are for
a single detector.
Sun {\it et al.} have used the EPC model to study the parameters estimation for
an eccentric binary in Ref.~\cite{PhysRevD.92.044034}.
We have extended the study of this model for three detectors 
\cite{PhysRevD.96.084046}.
In this work, we would like to further extend the study to a global network,
including KAGRA \cite{0264-9381-29-12-124007} in Japan and LIGO-India
\cite{LIGO-M1100296-v2}, which will be built in India.

To make our discussion self-contained, we have written out the EPC
waveform model for the detector networks by setting up an Earth coordinate system
\cite{PhysRevD.96.084046} in our previous work.
In the Earth coordinates, the EPC model involves 11 parameters
($e_0$, $D_{Le}$, $\mathcal{M}$, $\eta$, $t_{ce}$, $\phi_c$, $\theta_e$,
$\phi_e$, $\psi_e$, $\iota_e$, $\beta_e$), where 
\begin{itemize}[itemsep=1pt,partopsep=1pt,parsep=\parskip ,topsep=1pt]
\item[$D_{Le}$:] Luminosity distance between the center of the Earth and the
gravitational wave source.
\item[$t_{ce}$:] Arrival time of the coalescence signal with respect to
the center of the Earth.
\item[$\theta_e$:] Polar angle of the source based on the Earth coordinates.
\item[$\phi_e$:] Azimuthal angle of the source based in the Earth coordinates.
\item[$\psi_e$:] Polarization angle with respect to the Earth coordinates.
\item[$\iota_e$:] Inclination angle, the polar angle between the orbital
angular momentum and the line joining the source to the center of the Earth. 
\item[$\beta_e$:] Azimuthal angle in the orbital plane around the line
joining the source to the center of the Earth.
\end{itemize}
\begin{table}
    \centering
    \caption{The location of detectors and the orientation of their arms \cite{PhysRevD.64.042004, 0264-9381-28-12-125023, 0264-9381-30-15-155004, PhysRevD.90.024053}. The azimuth of their arm's rotation is from the north direction to the west direction.}
    \linespread{1.3}\selectfont
    \begin{tabular}{c c c c c} 
     \bottomrule[0.5pt]
        Detector & Latitude &Longitude & $x$ arm & $y$ arm \\ \midrule[0.5pt]
        
        LIGO-Hanford & $46^{\circ}27'19"$ & $-119^{\circ}24'28"$ & $36^{\circ}$ & $126^{\circ}$ \\ \hline

        LIGO-Livingston & $30^{\circ}33'46"$ & $-90^{\circ}46'27"$ & $108^{\circ}$ & $198^{\circ}$ \\ \hline

        VIRGO & $43^{\circ}37'53"$ & $10^{\circ}30'16"$ & $341^{\circ}$ & $71^{\circ}$ \\ \hline
        
        KAGRA & $36^{\circ}24'46"$ & $137^{\circ}18'13"$ & $298.3^{\circ}$ & $28.3^{\circ}$ \\ \hline
        
        LIGO-India & $19^{\circ}5'47"$ & $74^{\circ}2'59"$ & $45^{\circ}$ & $135^{\circ}$ \\ 
     \bottomrule[0.5pt]
     \end{tabular}     
\label{det}
\end{table}
Here, we assume that $\theta_i$ and $\phi_i$ are the location of the $i$th
detector in the network,
and the $x$ arm rotates from the north direction to the west direction with $\psi_i$.
The angles ($\theta_i$, $\phi_i$) in terms of the longitude $\lambda$
(a negative value indicates the Southern Hemisphere) and latitude $\alpha$
(a negative value indicates west of the Prime Meridian) on the Earth are
\begin{align}
\theta_i & = \frac{\pi}{2} - \lambda,\\
\phi_i & = \left\{
\begin{array}{ll}
\alpha, & \mbox{if }\alpha \ge 0 \\
2\pi+\alpha, & \mbox{if }\alpha < 0.
\end{array} \right.
\end{align}
The relations between the detector-based quantities mentioned in the last
paragraph and the Earth-based quantities can be written as follows:
\begin{widetext}
\begin{align}
D_{L}& = D_{Le} + \Delta D_L, \, t_c = t_{ce} + \Delta t_c, \, \iota \approx \iota_e, \, \beta \approx \beta_e, \, \cos\theta \approx \sin\theta_{e}\sin\theta_{i}\cos(\phi_{e}-\phi_{i}) + \cos\theta_{e}\cos\theta_{i},\label{Eq_DEr_Da}\\
\tan\phi &= \frac{\sin\theta_{e}\sin\psi_i\cos\theta_{i}\cos(\phi_{e}-\phi_{i})-\cos\theta_{e}\sin\theta_{i}
\sin\psi_i-\sin\theta_{e}\cos\psi_i\sin(\phi_{e}-\phi_{i})}
{-\sin\theta_{e}\cos\psi_i\cos\theta_{i}\cos(\phi_{e}-\phi_{i})+\cos\theta_{e}\sin\theta_{i}\cos\psi_i +\sin\theta_{e}\sin\psi_i\sin(\phi_{i}-\phi_{e})}, \\
\cos\psi &\approx \sin\theta_{e}\sin\psi_{e}\sin\theta_{i}\sin(\phi+\psi_i)-\cos(\phi+\psi_i)\cos\psi_{e}
\cos(\phi_{e}-\phi_{i})-\cos(\phi+\psi_i)\sin\psi_{e}\cos\theta_{e}\sin(\phi_{e}-\phi_{i}) \notag \\
&-\sin(\phi+\psi_i)\cos\psi_{e}\cos\theta_{i}\sin(\phi_{e}-\phi_{i})+\cos\theta_{e}\sin\psi_{e}
\cos\theta_{i}\sin(\phi+\psi_i)\cos(\phi_{e}-\phi_{i}).\label{Eq_DEr_cospsia} \\
\sin \psi & \approx \cos\psi_e \cos\theta \cos\theta_i \cos(\phi+\psi_i) \sin(\phi_e-\phi_i) - \cos\psi_e \cos\theta \sin(\phi+\psi_i) \cos(\phi_e-\phi_i) + \cos\psi_e \sin\theta \sin\theta_i \sin(\phi_e-\phi_i) \notag \\
& - \sin\psi_e \cos\theta_e \cos\theta \cos\theta_i \cos(\phi+\psi_i) \cos(\phi_e-\phi_i) - \sin\psi_e \cos\theta_e \cos\theta \sin(\phi+\psi_i) \sin(\phi_e-\phi_i) \notag \\
& - \sin\psi_e \cos\theta_e \sin\theta \sin\theta_i \cos(\phi_e-\phi_i) - \sin\psi_e \sin\theta_e \cos\theta \sin\theta_i \cos(\phi+\psi_i) + \sin\psi_e \sin\theta_e \sin\theta \cos\theta_i, \label{Eq_DEr_sinpsia}
\end{align}
\end{widetext}
where
\begin{align}
\Delta D_L=\Delta t_c\approx\frac{R^2_E}{2 D_{Le}}-R_E[&
\sin\theta_e\sin\theta_i\cos(\phi_e-\phi_i)\nonumber\\
+&\cos\theta_e\cos\theta_i],
\end{align}
$R_E$ is the radius of the Earth.
These relations result from the Appendix in Ref.~\cite{PhysRevD.96.084046}.

\begin{figure}[htbp]
\begin{tabular}{c}
\includegraphics[width=0.49\textwidth]{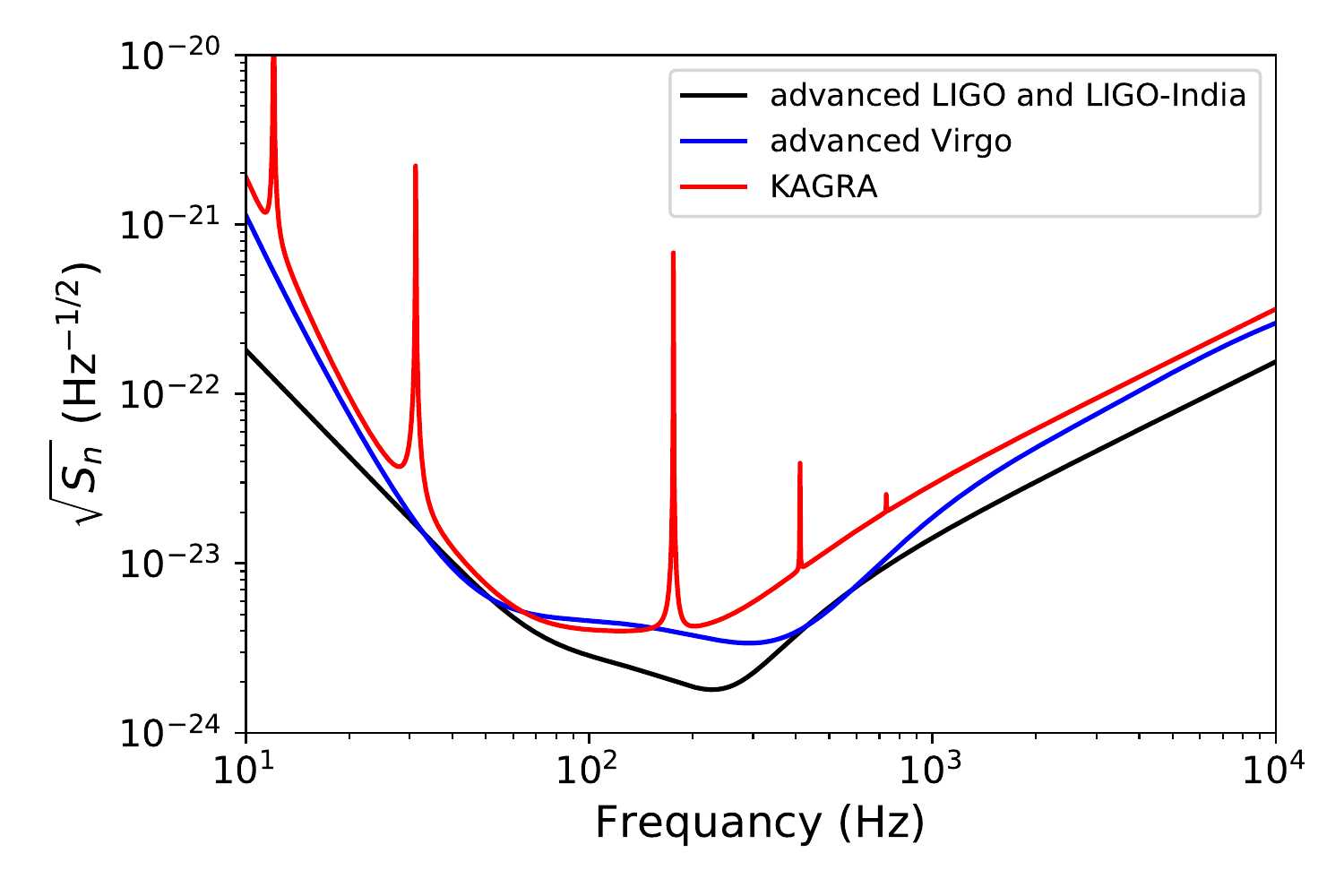}
\end{tabular}
\caption{The sensitivity curves for the advanced LIGO, LIGO-India, advanced Virgo, and KAGRA.} \label{fig:ASD}
\end{figure}

In the current paper, we consider three network configurations with
I) three, II) four, and III) five advanced detectors.
In case I, we consider the LIGO-Livingston (L), LIGO-Hanford (H) \cite{LIGO}, and advanced Virgo detector (V) \cite{VIRGO} in Cascina, denoted by LHV.
In case II, KAGRA (K) \cite{0264-9381-29-12-124007} in Gifu Prefecture is
joined to the network in addition to LHV, denoted by LHVK.
And we add the LIGO-India (I) \cite{LIGO-M1100296-v2} in the
network for case III, denoted by LHVKI.
The information for these detectors is listed in Table \ref{det}.
We use the one-sided noise power spectral density (PSD) for the advanced LIGO
and LIGO-India as follows \cite{PhysRevD.71.084008}:
when $f\ge20$ Hz,
\begin{equation}
S_n(f)=S_0\left(x^{-4.14}-5 x^{-2} +111\frac{2-2 x^2+x^4}{2+x^2}\right),
\label{Eq_SnLIGO}
\end{equation}
where $x=f/f_0$, $f_0=215$ Hz, and $S_0=10^{-49}/{\rm Hz}$.
When $f<20$ Hz, $S_n(f)=\infty$.
For the advanced Virgo, we use Eq.~(6) in Ref.~\cite{1202.4031}:
when $f\ge10$ Hz,
\begin{align}
S_{n}(f)=S_0[&0.07\exp(-0.142-1.437x+0.407x^{2}) \notag \\
+&3.10\exp(-0.466-1.043x-0.548x^{2}) \notag \\
+&0.40\exp(-0.304+2.896x-0.293x^{2}) \notag \\
+&0.09\exp(1.466+3.722x-0.984x^{2})]^2,  \label{Eq_SnVIRGO}
\end{align}
where $x=\ln(f/f_0)$, $f_0=300$ Hz, and $S_0=1.585081\times 10^{-48}/{\rm Hz}$.
When $f<10$ Hz, $S_n(f)=\infty$.
For KAGRA, we use the KAGRA design curve in Ref.~\cite{LALSuiteKAGRA}.
Figure \ref{fig:ASD} shows the PSD for the advanced LIGO/LIGO-India,
the advanced Virgo and KAGRA.

\begin{figure*}[htbp]
\begin{tabular}{cc}
\subfloat[]{\includegraphics[width=0.5\textwidth]{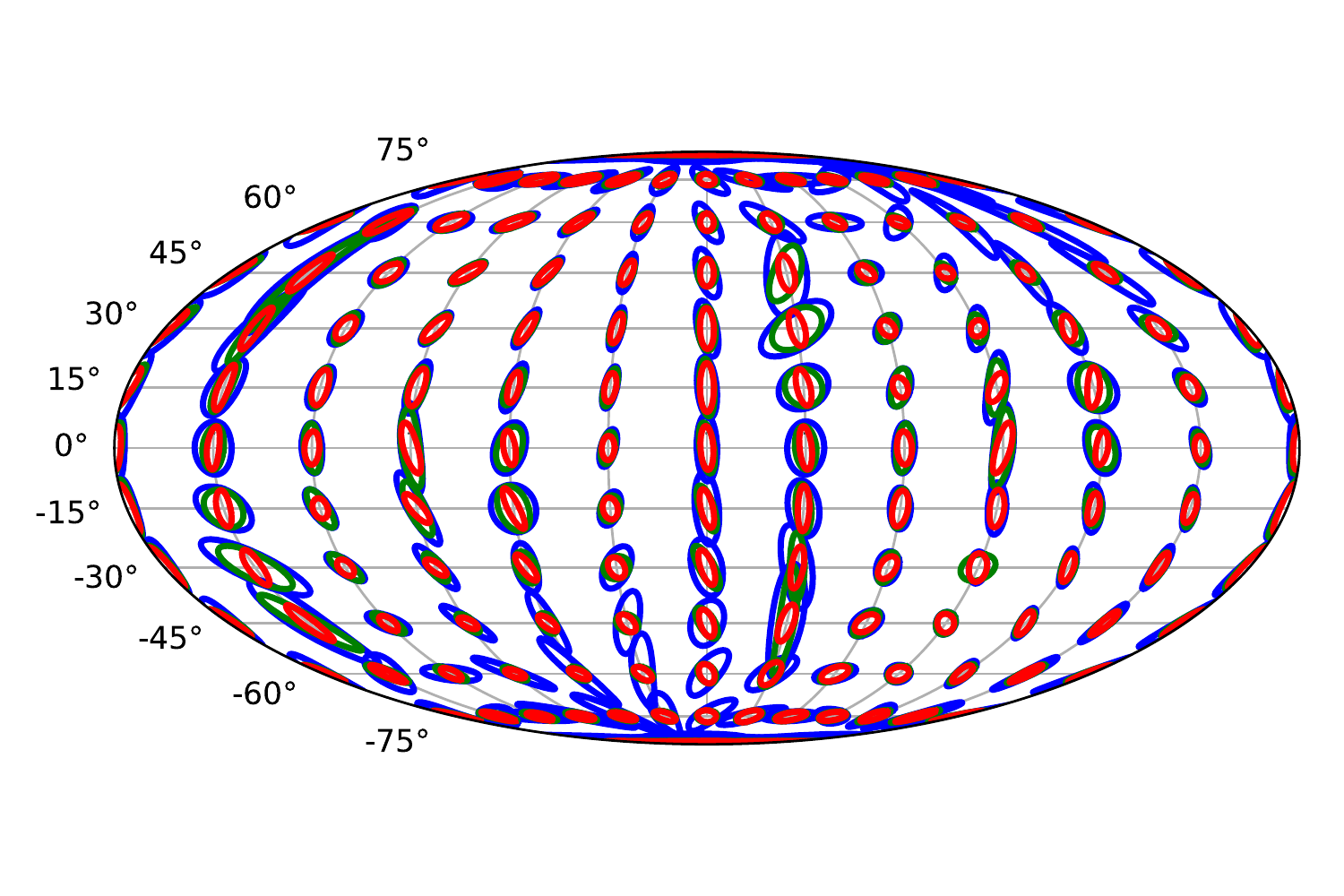}
\label{fig:100-ellipse-0}}& 
\subfloat[]{\includegraphics[width=0.5\textwidth]{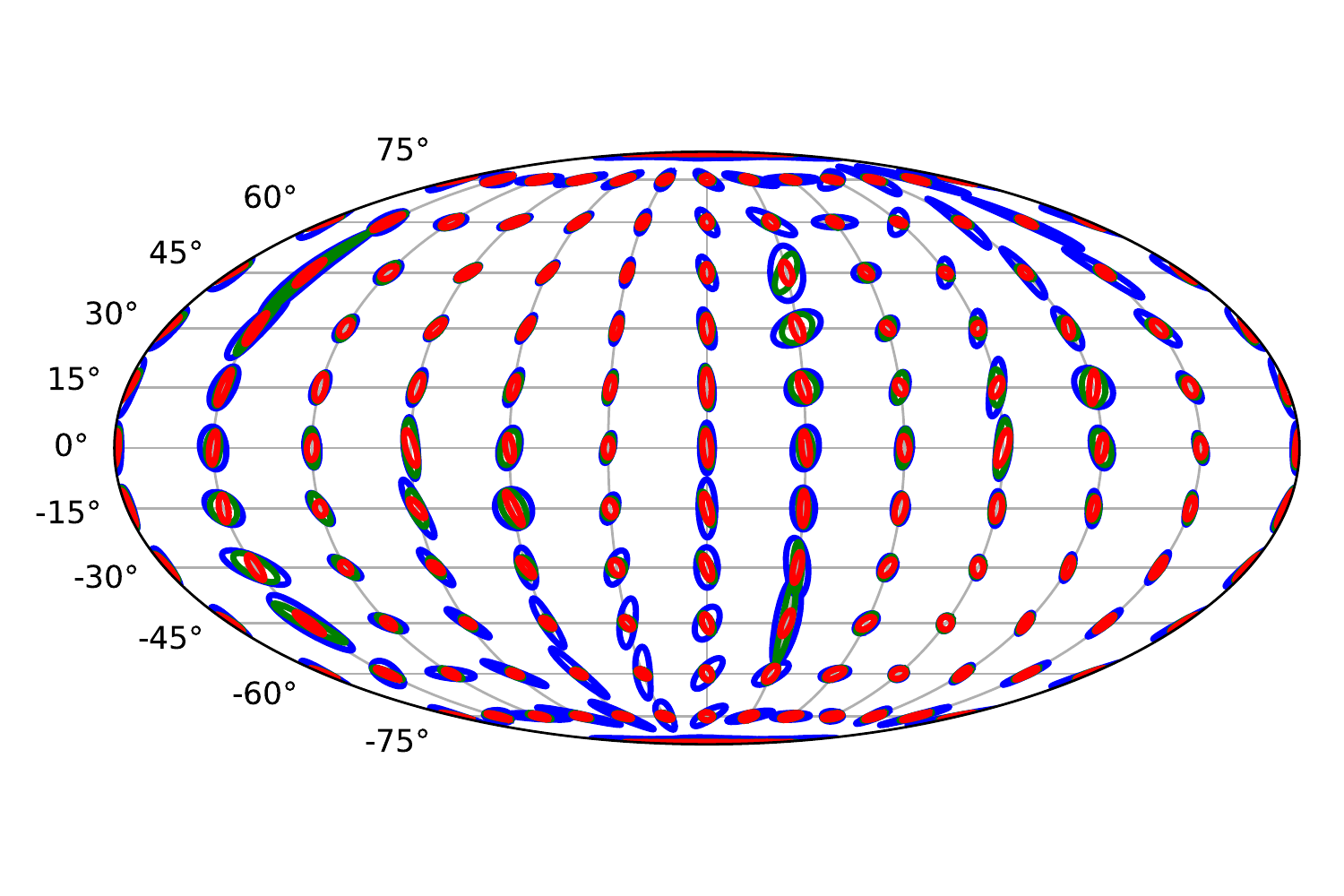}
\label{fig:100-ellipse-0.4}}
\end{tabular}
\caption{Error ellipses of the source localization in the big BBH case for
(a) $e_{0.0}$ and (b) $e_{0.4}$.
The blue, green, and red ellipses correspond to the LHV, LHVK, and LHVKI cases,
respectively.
This shows that the network with more detectors gives a more accurate
localization and thus smaller error ellipses.
It also indicates that a binary system with higher initial eccentricity gives a 
more accurate localization, and thus smaller error ellipses, than the ones with
smaller initial eccentricity.}
\label{fig:100-ellipse}
\end{figure*}

\section{Source localization accuracy estimate method}\label{sec:FIM}
In this paper, we use the Fisher information matrix method \cite{PhysRevD.49.2658, PhysRevD.46.5236, PhysRevD.47.2198} to estimate the source localization accuracy. We first define the matched filter signal-to-noise ratio (SNR) of a network for $N$ detectors as
\begin{equation}
\rho^2\equiv\sum_{k=1}^N(h_k|h_k)_k=\sum_{k=1}^N 4{\rm Re}
\int_{f_{\rm min}}^{f_{\rm max}}
\frac{\tilde{h}_k^{*}(f)\tilde{h}_k(f)}{S_{kn}(f)}{\rm d}f, \label{Eq_SNR}
\end{equation}
where $(|)_k$ denotes the inner product for the $k$th detector,
$\tilde{h}_k$ is the waveform for the $k$th detector in frequency
domain, $S_{kn}(f)$ is the one-sided power spectral density of the noise
for the $k$th detector, and ${}^*$ denotes complex conjugation.
The limits of integration, $f_{\rm min}$ and $f_{\rm max}$, correspond to
the frequency bound of the detectors and the nature of the signal.
Considering the frequency bound of the detectors we are using in this work,
we set the lower limit of integration as 20 Hz.
For the upper limit, since the EPC model is an inspiral waveform,
which is valid until the last stable orbit frequency
$F_{\rm LSO}\approx\displaystyle \frac{1}{6^{3/2} (2 \pi M)}$, we can set
$F_{\rm LSO}$ as the upper orbital frequency bound of the integration.
Corresponding to the orbital frequency $F$,
the $\ell$th harmonic component results in a gravitational wave with
frequency $\ell F$.
Hence, we assume that the upper cutoff frequency of the $\ell$th harmonic
is $\ell F_{\rm LSO}$.
Since the EPC model has ten harmonics, we set the upper limit of integration
to be 10$F_{\rm LSO}$.

Let $\Delta p^a$ denote the errors in the estimation of the parameters $p^a$.
If the SNR is high enough, $\Delta p^a$ obeys the Gaussian probability
distribution, which can be written as
\begin{equation}
g(\Delta p^a)=G\exp\left(-\frac{1}{2}\Gamma_{bc}\Delta p^b\Delta p^c\right),
\label{Eq_Gaussian}
\end{equation}
where $G$ is a normalization constant.
The quantity $\Gamma_{ab}$ is the Fisher information matrix.
For a network with $N$ detectors, the Fisher information matrix is defined as
\begin{equation}
\Gamma_{ab}\equiv\sum_{k=1}^N(\partial_a h_k|\partial_b h_k)_k, \label{Eq_FIM}
\end{equation}
where $\partial_a$ means $\partial/\partial p^a$.
In this work, $p^a$ denotes any of the parameters among 
$e_0$, $D_{Le}$, $\mathcal{M}$, $\eta$, $t_{ce}$, $\phi_c$, $\theta_e$,
$\phi_e$, $\psi_e$, $\iota_e$, and $\beta_e$.
So $\Gamma_{ab}$ is an 11 by 11 matrix.
The covariance matrix is defined as
\begin{equation}
\Sigma^{ab}\equiv\langle\Delta p^a\Delta p^b\rangle=(\Gamma^{-1})^{ab},
\label{Eq_InvFIM}
\end{equation}
where $\langle\cdot\rangle$ denotes the average with respect to
the probability distribution function in Eq. \eqref{Eq_Gaussian}.
We can estimate the root-mean-square error, which is given by
\begin{equation}
\sigma_a=\sqrt{\langle (\Delta p^a)^2\rangle}=\sqrt{\Sigma^{aa}}.
\label{Eq_error}
\end{equation}
Here $\Sigma^{aa}$ is the diagonal element of the covariance matrix
with respect to the parameter $p^a$.
In this work, we focus on the source localization accuracy,
defined as the measurement error of the sky position solid angle,
which is given by
\begin{equation}
\Delta\Omega=2\pi\sqrt{(\sigma_{\phi_e}\sigma_{\cos\theta_e})^2-
(\Sigma^{\phi_e\cos\theta_e})^2}, \label{Eq_Omega}
\end{equation}
where $\Sigma^{\phi_e\cos\theta_e}$ is the nondiagonal element of
the covariance matrix with respect to the parameters $\cos\theta_e$ and $\phi_e$.
If $\Delta\Omega$ is smaller, the source localization is more accurate.

\section{results}\label{sec:results}
In this section, we use the EPC waveform model, described in Sec.~\ref{sec:EPC},
to show the influence of the initial eccentricity, as well as the effect of
different gravitational wave detector networks, on the accuracy of the source
localization.
Here we investigate two binary black hole (BBH) systems, one with a total mass
$100 M_\odot$ and the other with $22 M_\odot$, and a binary neutron star (BNS)
system with a total mass $2.74 M_\odot$ and thus a chirp mass
$\mathcal{M}=1.188 M_\odot$.
We call the one with $100 M_\odot$ the big BBH, the one with $22 M_\odot$ the
GW151226-like BBH, and the one with $2.74 M_\odot$ the GW170817-like BNS.
We define $e_x$ as $e_0 = x$ for convenience.

\subsection{\label{big_BBH}Big BBH case}
First, we consider the big BBH with a total mass $100 M_\odot$.
We fix the parameters $D_{Le}=410$Mpc, $\eta=1/4$,
$\mathcal{M}=M\eta^{3/5}=43.53 M_\odot$,
and $t_{ce}=\phi_{c}=\iota_{e}=\beta_{e}=\psi_{e}=0$, while varying $\theta_{e}$,
$\phi_{e}$, and $e_0$ to investigate the resulting accuracy of the source
localization.
In Fig.~\ref{fig:100-ellipse}, we show the results of the error ellipses for
$e_{0.0}$ (Fig.~\ref{fig:100-ellipse-0}) and $e_{0.4}$
(Fig.~\ref{fig:100-ellipse-0.4}).
We compare the result for the cases of LHV, LHVK and LHVKI in each subgraph.
Every ellipse in Fig.~\ref{fig:100-ellipse} represents the 5$\sigma$ error
region in the $\theta_e$-$\phi_e$ sphere.
We can see from Fig.~\ref{fig:100-ellipse} that the accuracy of the source
localization is improved when we use more detectors.
It is about two times better when we use the LHVK network instead of the LHV
network and three times better when we use the LHVKI network, by comparing the area of the ellipses.
It is also improved by about 2.5 times better when the initial eccentricity
changes from 0.0 to 0.4.
In general, this shows that the network with more detectors gives a more accurate
localization and thus smaller error ellipses, as people expect.
It also indicates that a binary system with higher initial eccentricity gives
more accurate localization, and thus smaller error ellipses, than the one with
smaller initial eccentricity.

\begin{figure*}[htbp]
\begin{tabular}{ccc}
\includegraphics[width=0.33\textwidth]{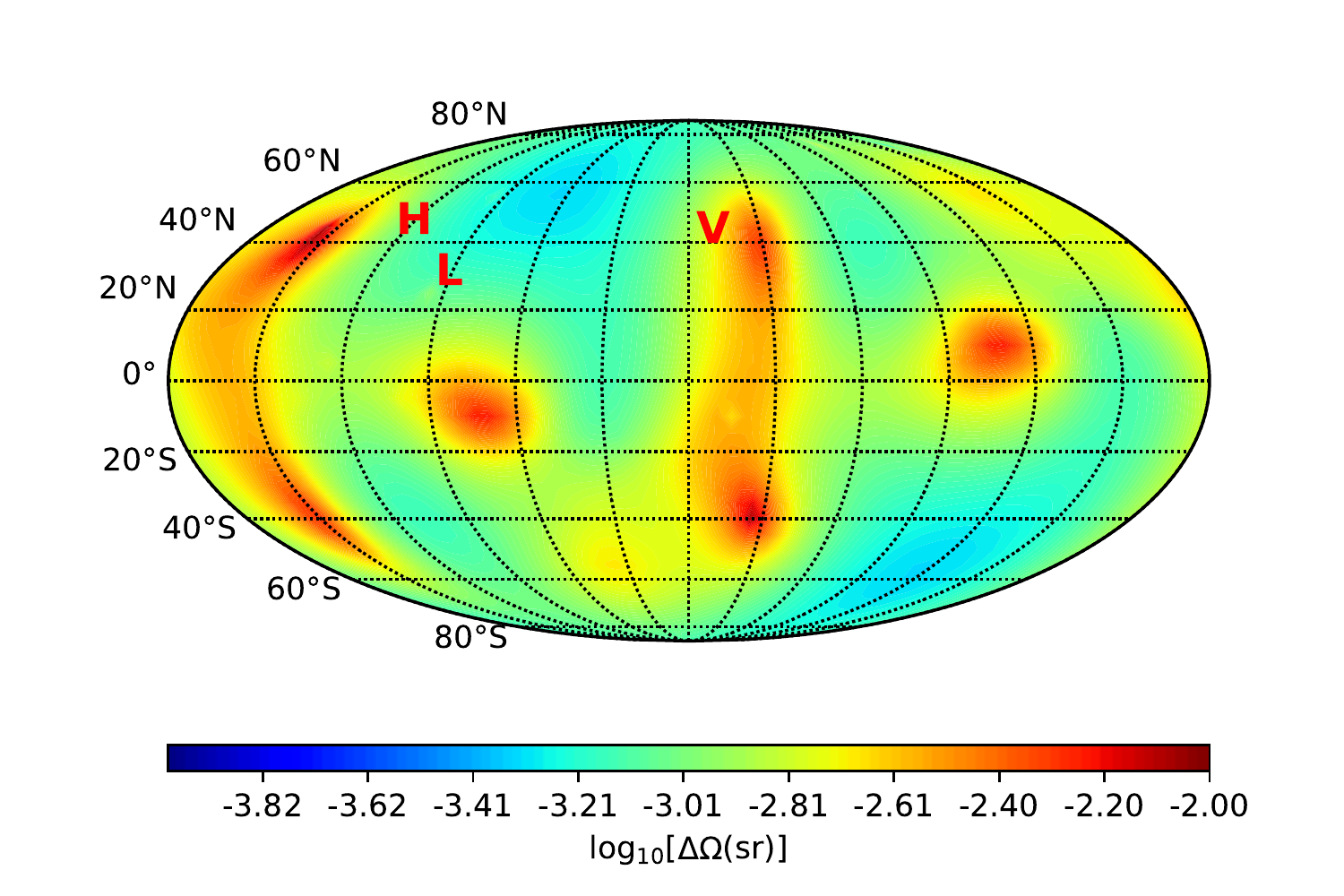}
\label{fig:100-space-0-LHV} &
\includegraphics[width=0.33\textwidth]{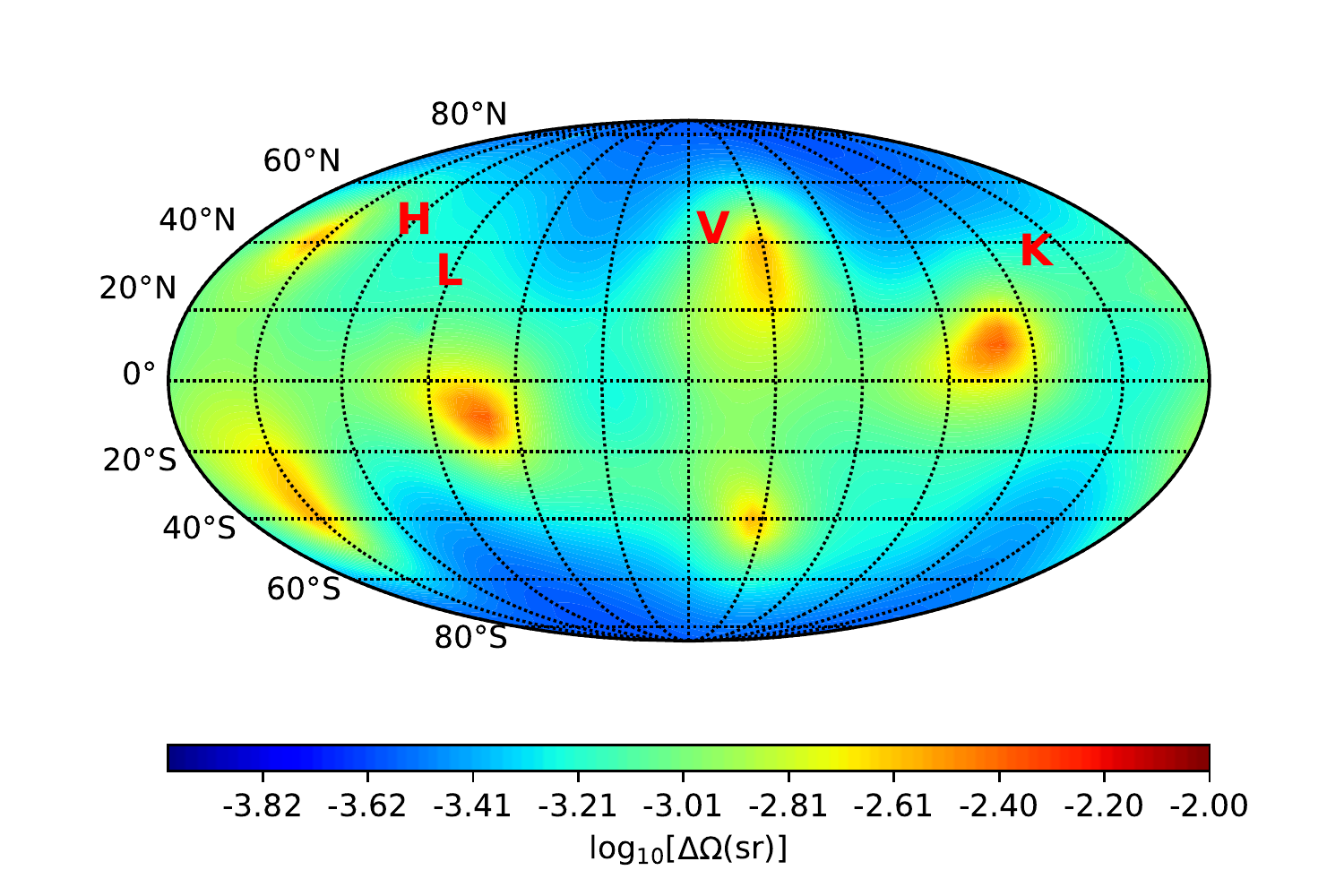}
\label{fig:100-space-0-LHVK} &
\includegraphics[width=0.33\textwidth]{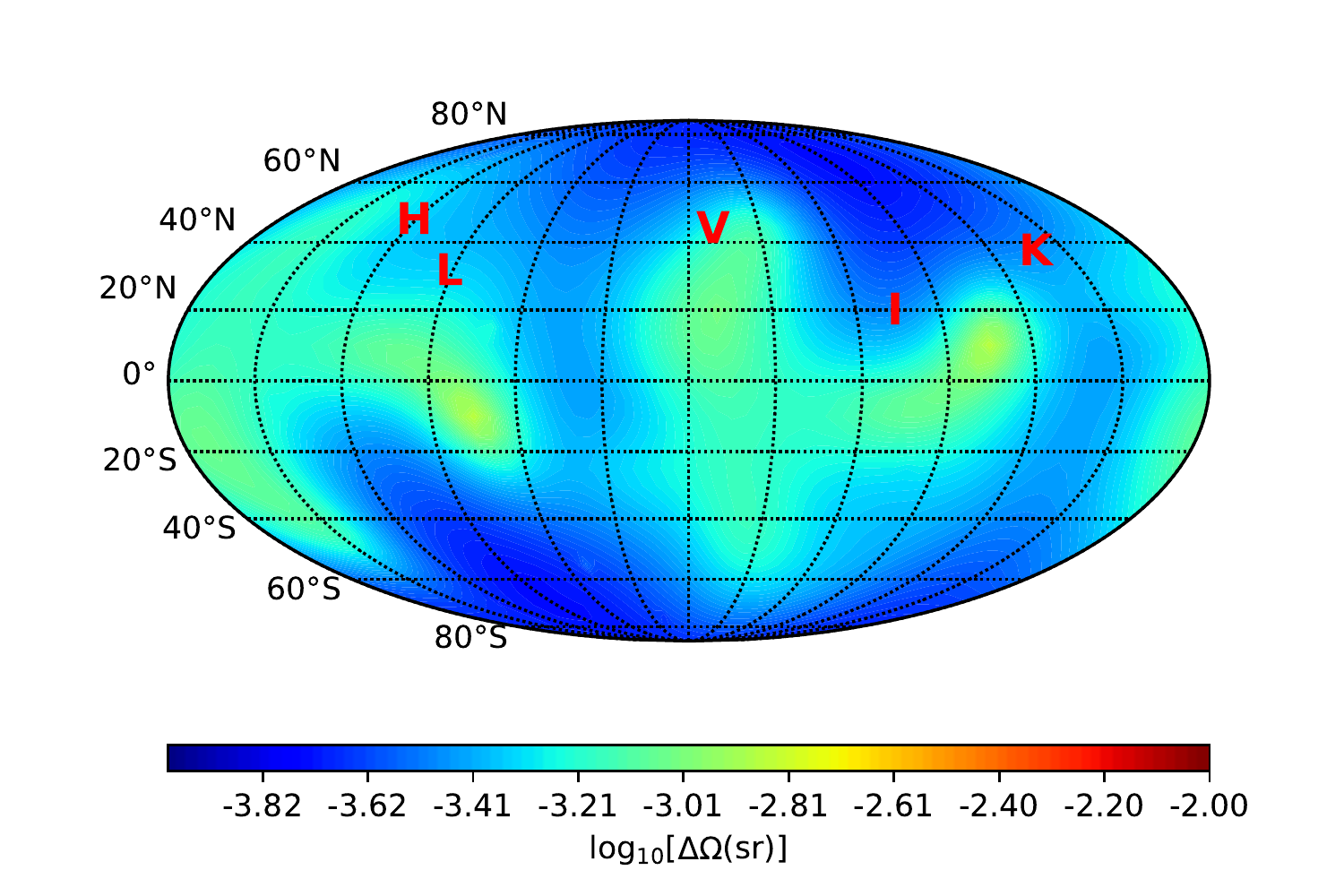}
\label{fig:100-space-0-LHVKI}\\
\includegraphics[width=0.33\textwidth]{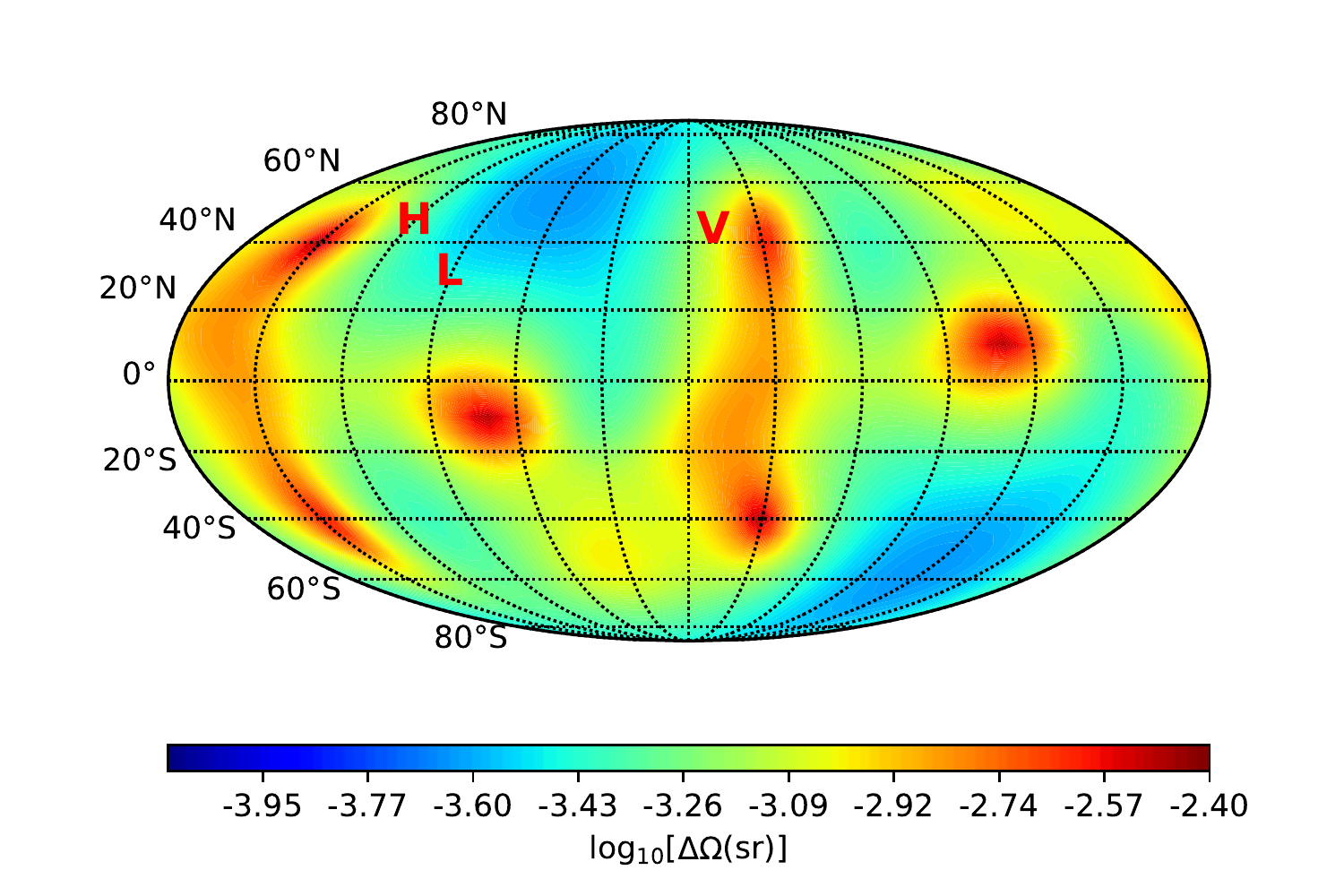}
\label{fig:100-space-0.4-LHV} &
\includegraphics[width=0.33\textwidth]{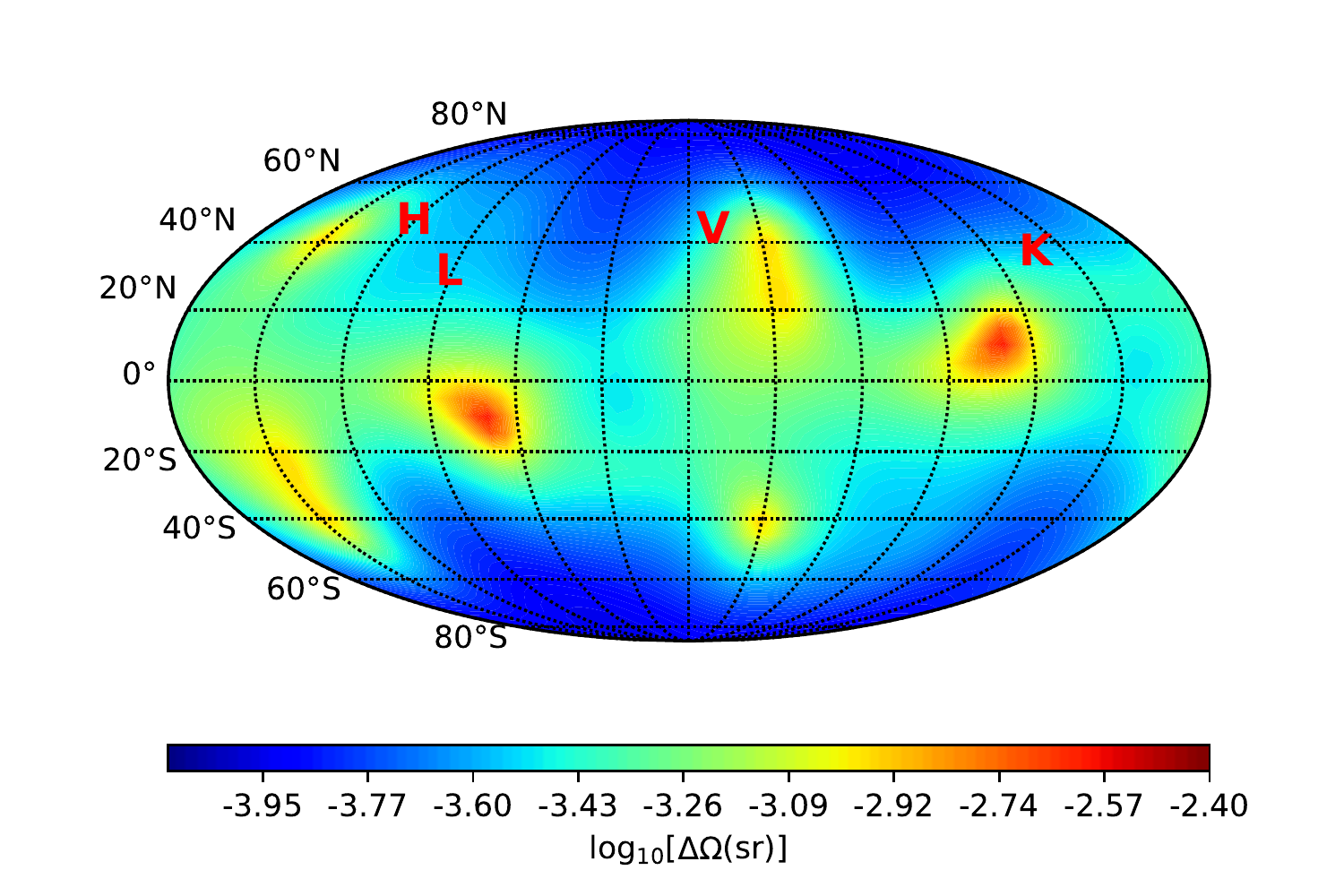}
\label{fig:100-space-0.4-LHVK} &
\includegraphics[width=0.33\textwidth]{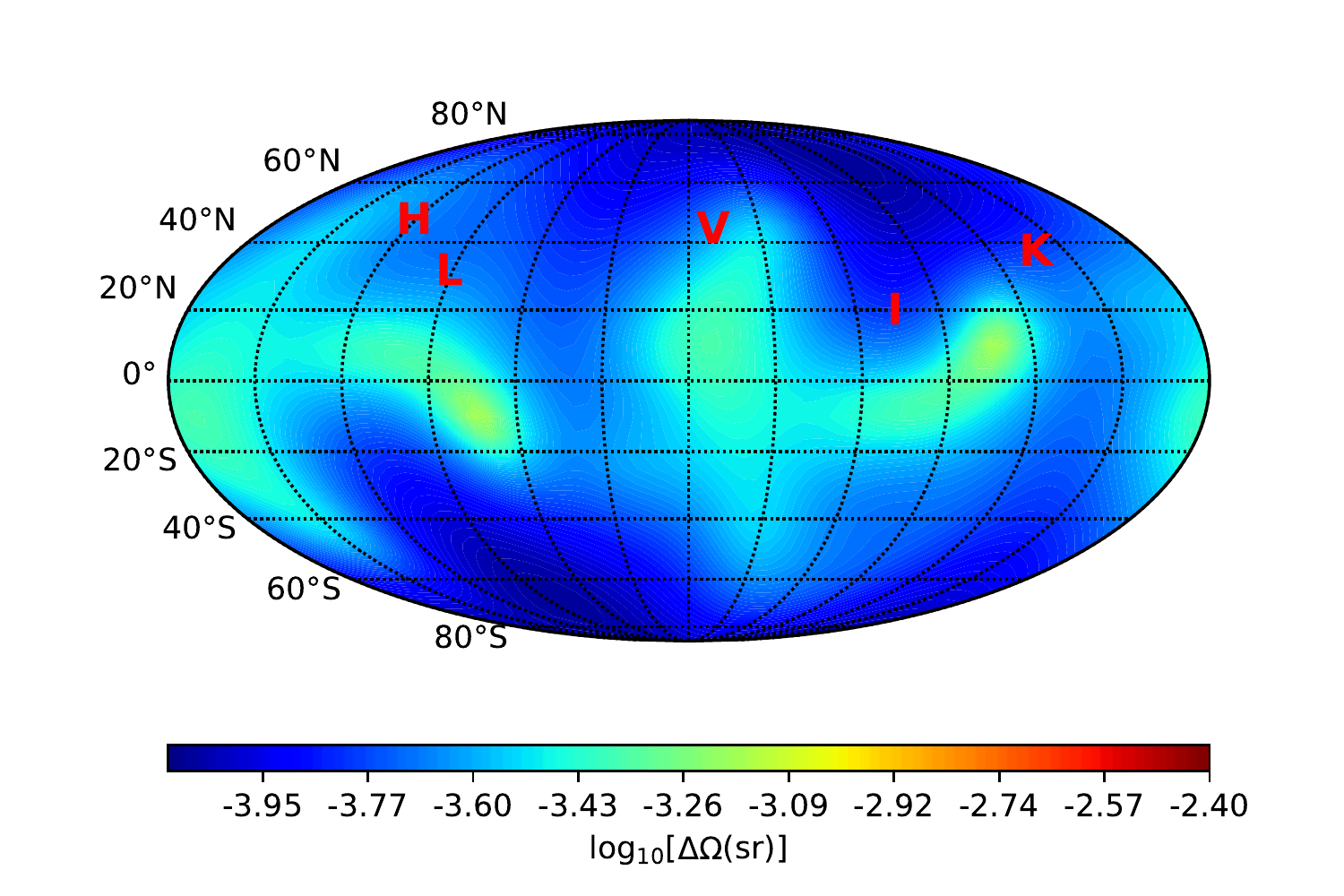}
\label{fig:100-space-0.4-LHVKI}
\end{tabular}
\caption{Estimated error $\Delta\Omega$ of the source localization for the big
BBH case.
The panels in the upper and the lower rows correspond to the eccentricities
$e_{0.0}$ and $e_{0.4}$, respectively.
We show the $\Delta\Omega$'s for the LHV, LHVK, and LHVKI cases in the left,
middle, and right columns, respectively.}
\label{fig:100-space}
\end{figure*}

\begin{figure*}
\begin{tabular}{ccc}
\includegraphics[width=0.33\textwidth]{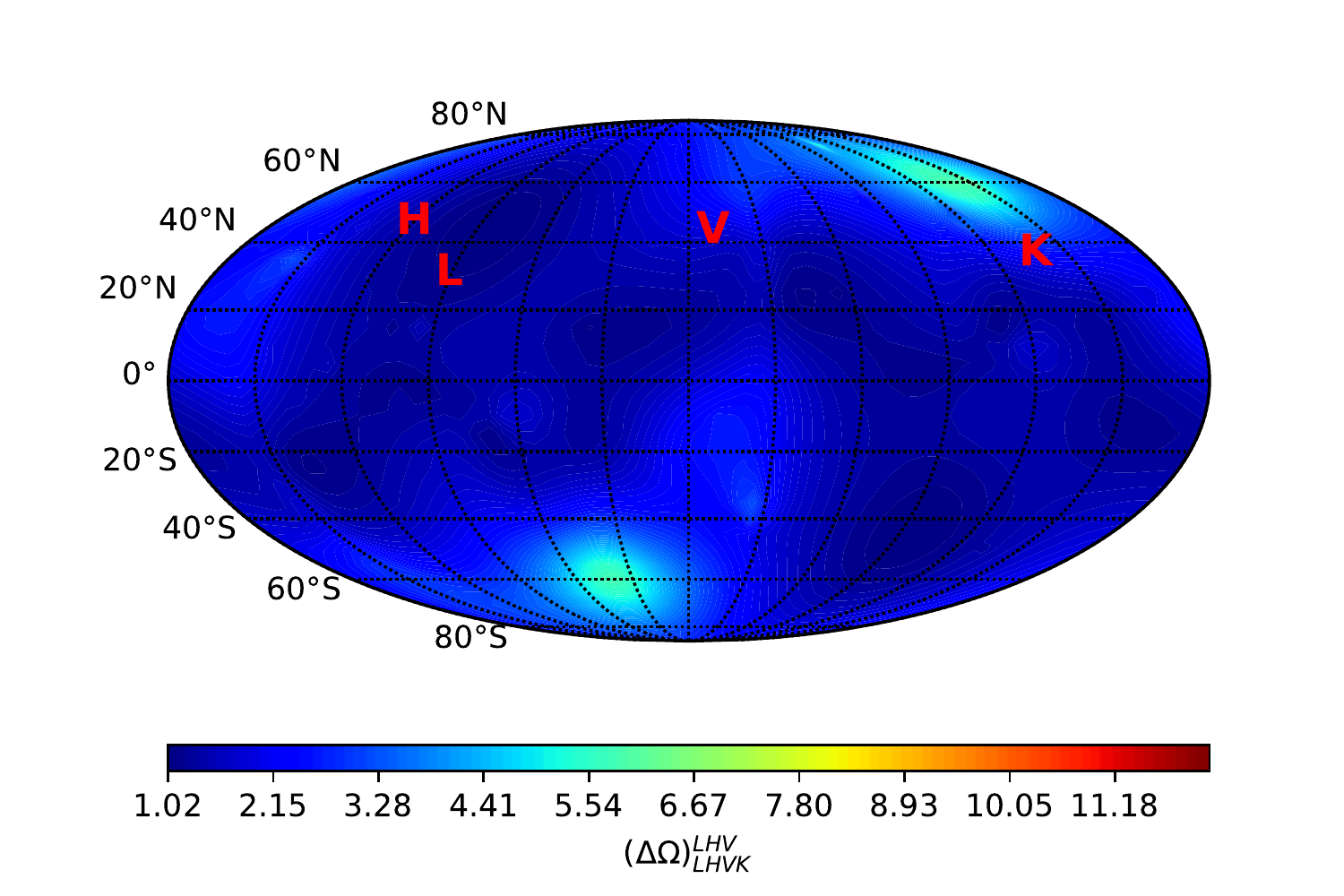}
\label{fig:100-space-0-3-4} &
\includegraphics[width=0.33\textwidth]{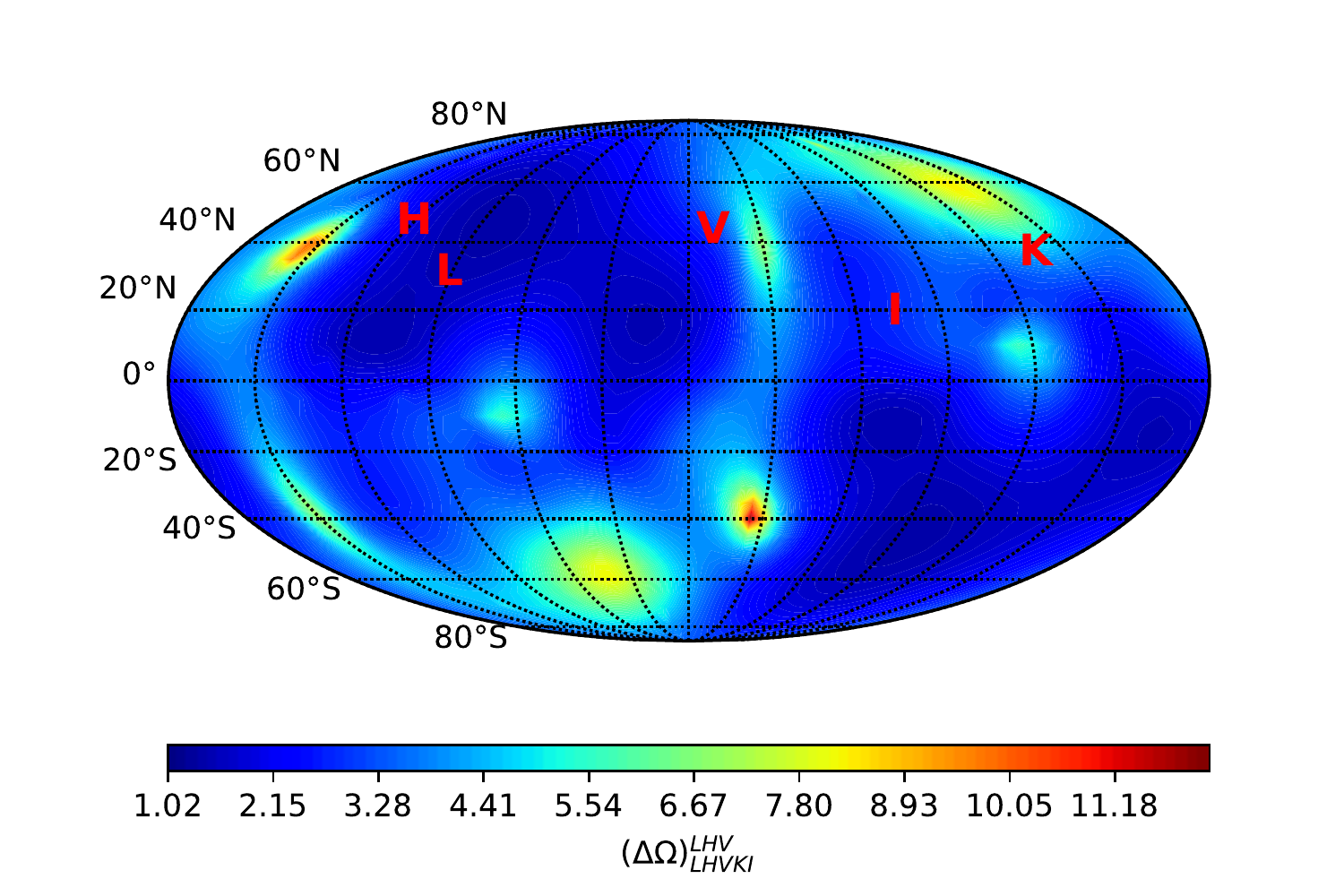}
\label{fig:100-space-0-3-5} &
\includegraphics[width=0.33\textwidth]{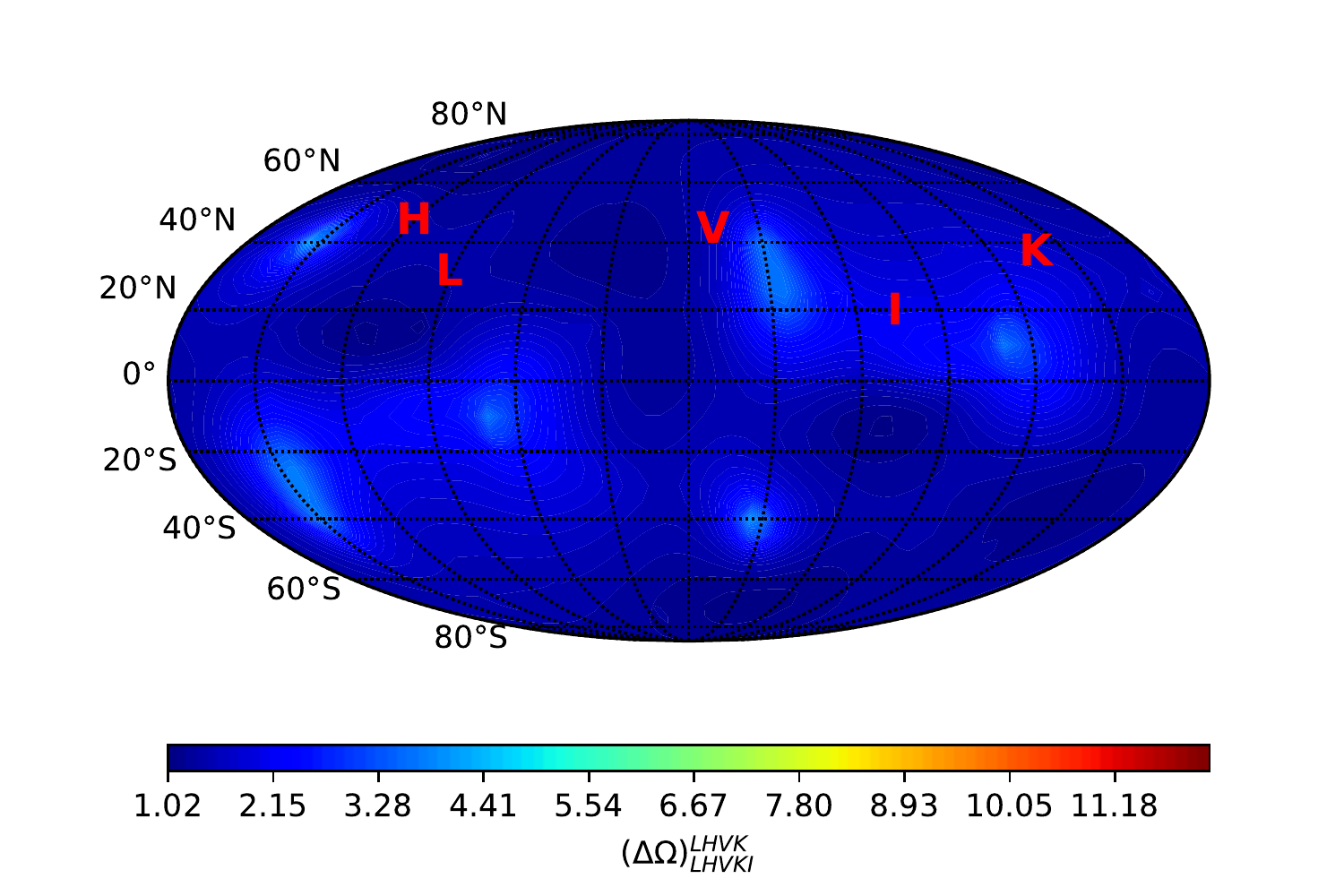}
\label{fig:100-space-0-4-5} \\
\includegraphics[width=0.33\textwidth]{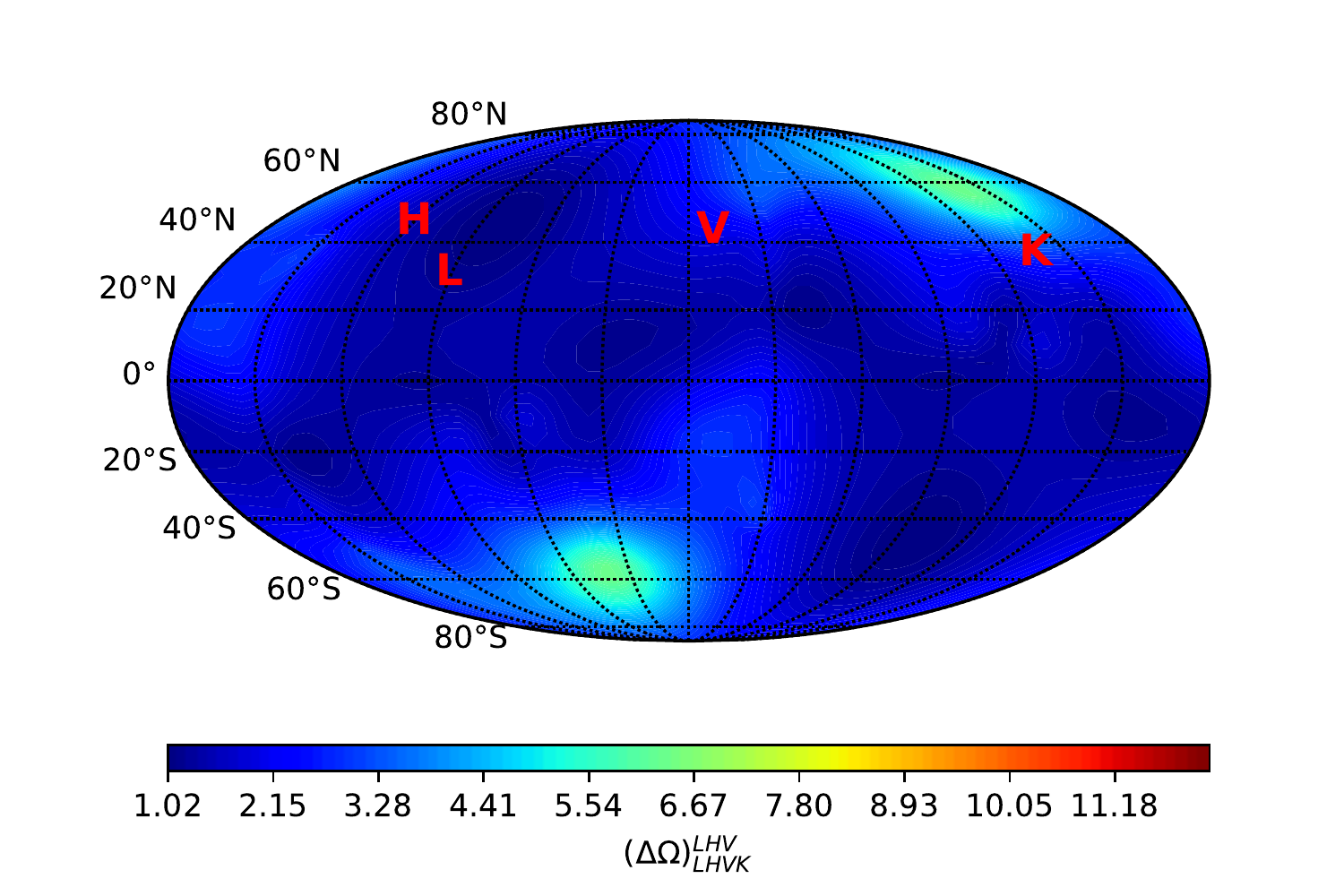}
\label{fig:100-space-0.4-3-4} &
\includegraphics[width=0.33\textwidth]{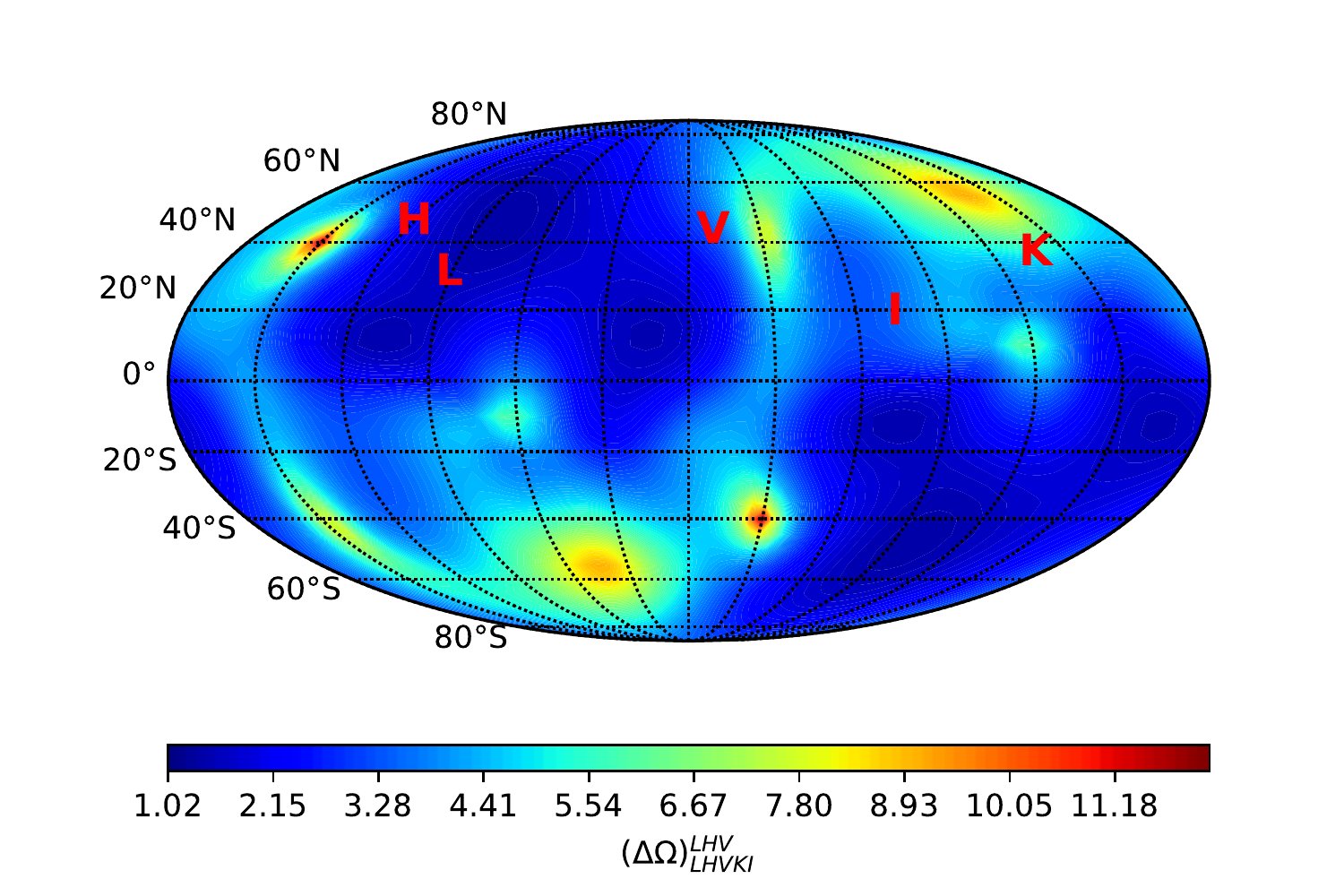}
\label{fig:100-space-0.4-3-5} &
\includegraphics[width=0.33\textwidth]{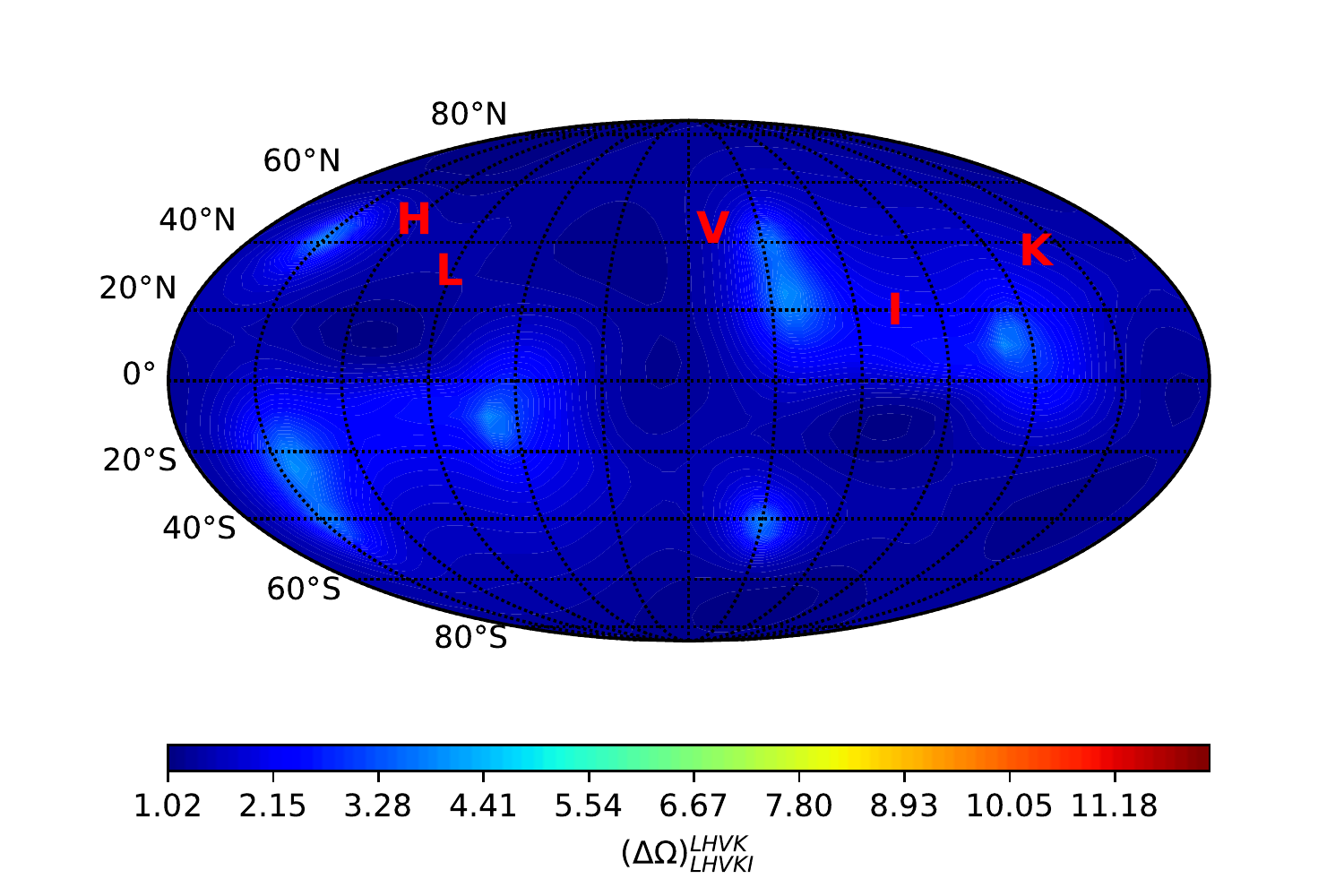}
\label{fig:100-space-0.4-4-5}
\end{tabular}
\caption{Ratios of $\Delta\Omega$ among different networks, defined in
Eq.~(\ref{omegaratio}), for the big BBH case.
The panels in the upper and the lower rows correspond to the eccentricities
$e_{0.0}$ and $e_{0.4}$, respectively.
$\Delta\Omega^{\text{LHV}}_{\text{LHVK}}$,
$\Delta\Omega^{\text{LHV}}_{\text{LHVKI}}$, and
$\Delta\Omega^{\text{LHVK}}_{\text{LHVKI}}$ are shown in the left, middle,
and right columns, respectively.}
\label{fig:100-space-ratio-net}
\end{figure*}

\begin{figure}
\begin{tabular}{c}
\includegraphics[width=0.5\textwidth]{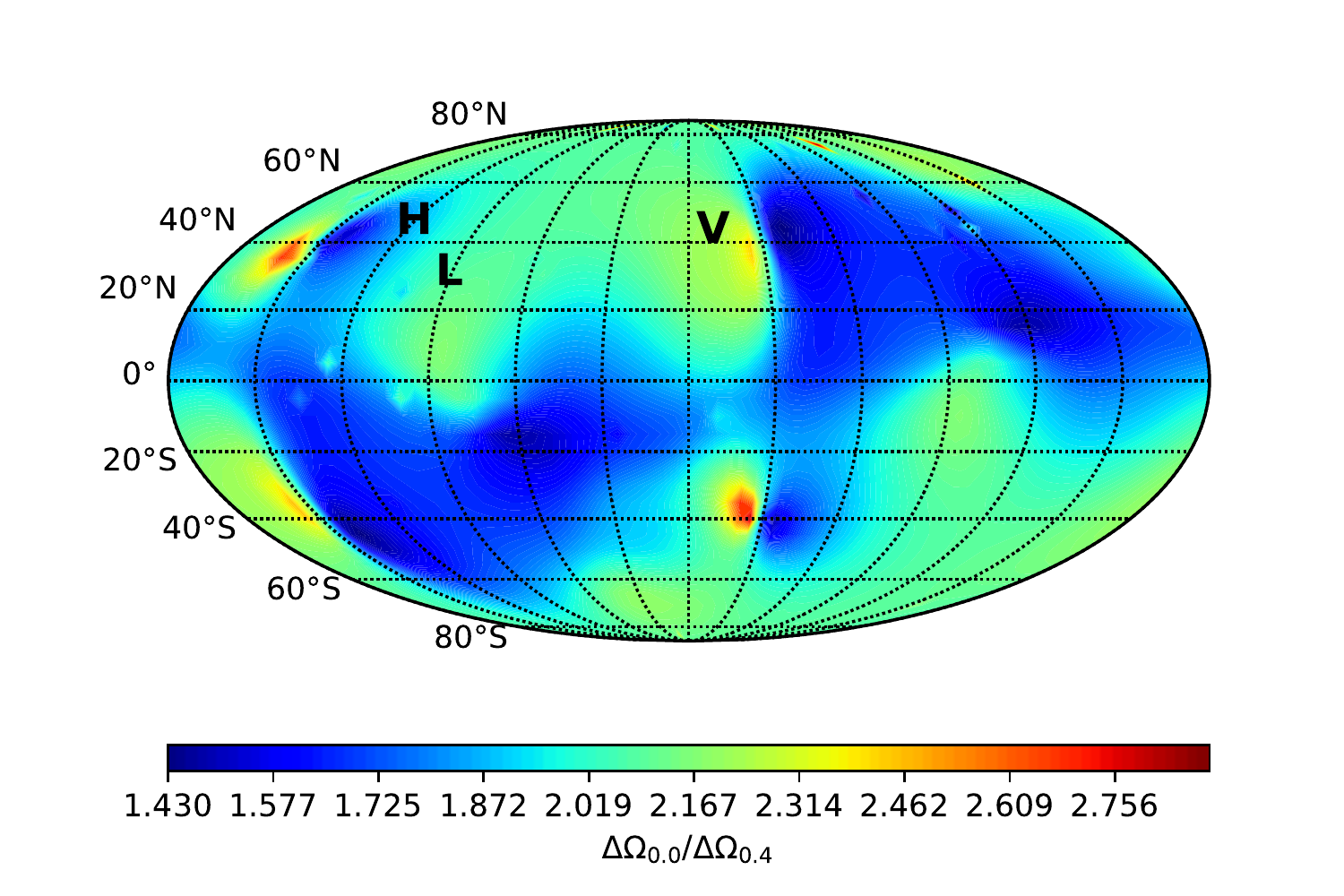}
\label{fig:100-space-ratio-LHV} \\
\includegraphics[width=0.5\textwidth]{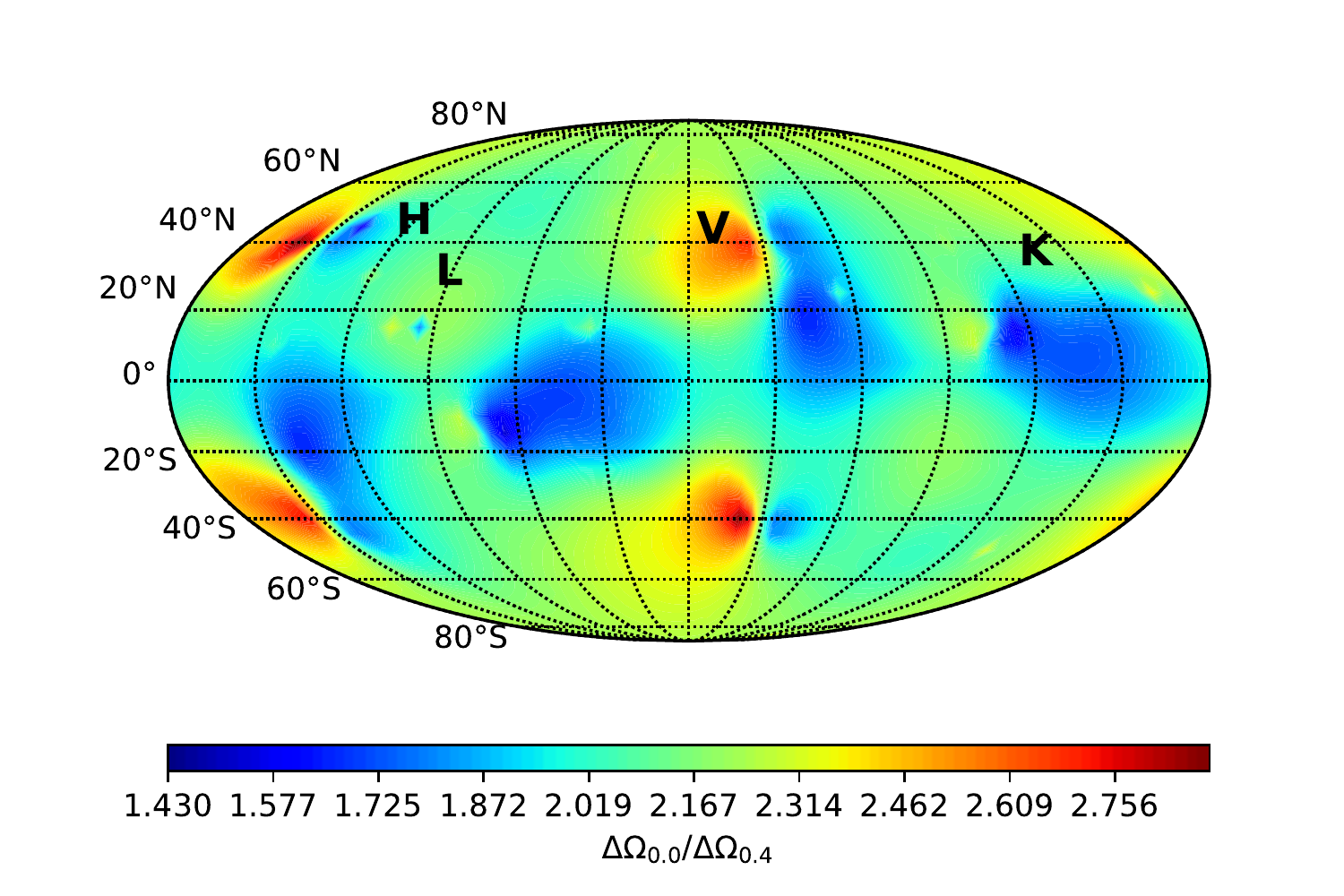}
\label{fig:100-space-ratio-LHVK} \\
\includegraphics[width=0.5\textwidth]{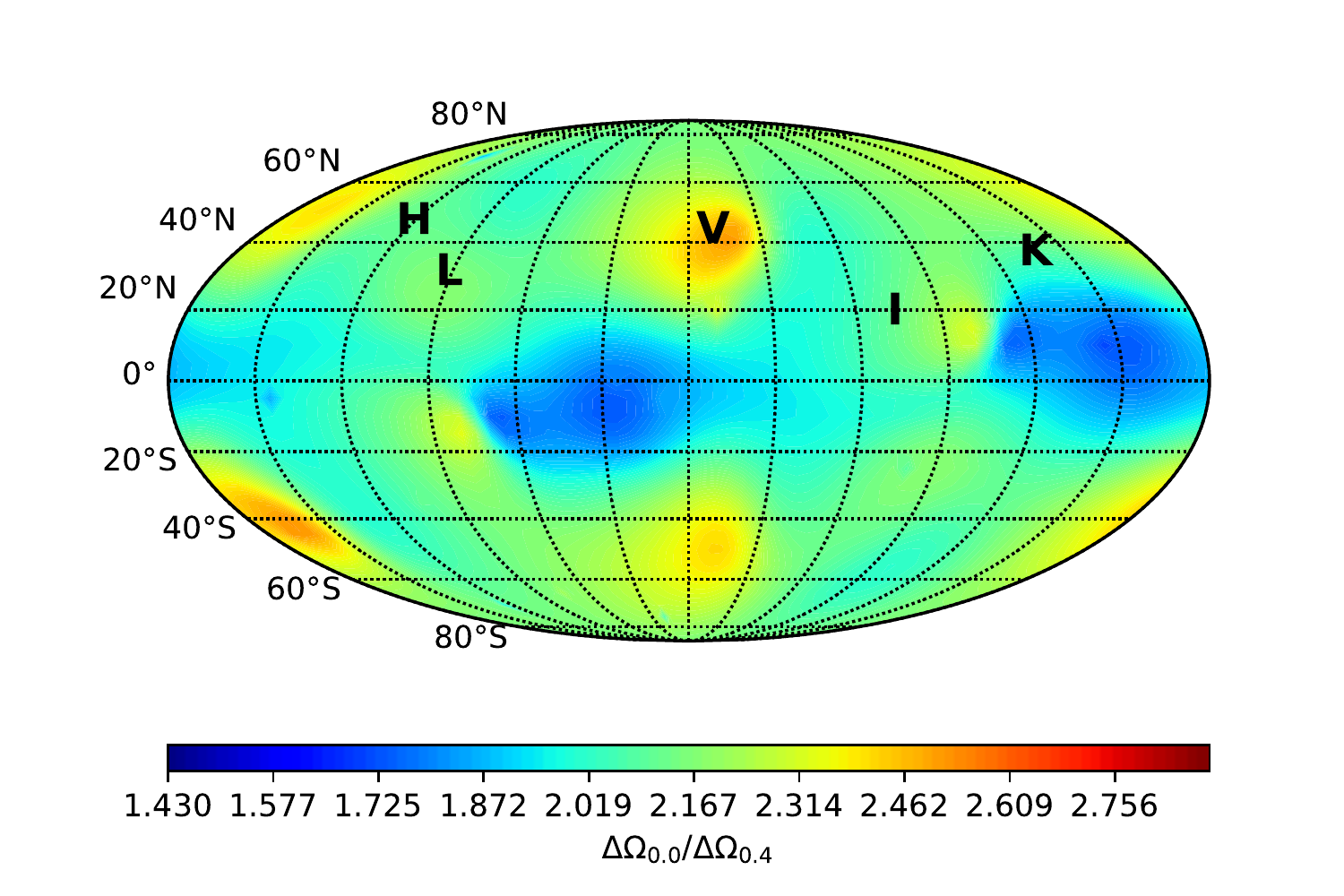}
\label{fig:100-space-ratio-LHVKI}
\end{tabular}
\caption{$\Delta\Omega^{0.0}_{0.4}$ for the big BBH case.
The plots in the upper, middle, and lower panels correspond to the LHV, LHVK,
and LHVKI cases, respectively.}
\label{fig:100-space-ratio-e0}
\end{figure}

\begin{figure}
\begin{tabular}{c}
    \includegraphics[width=0.5\textwidth]{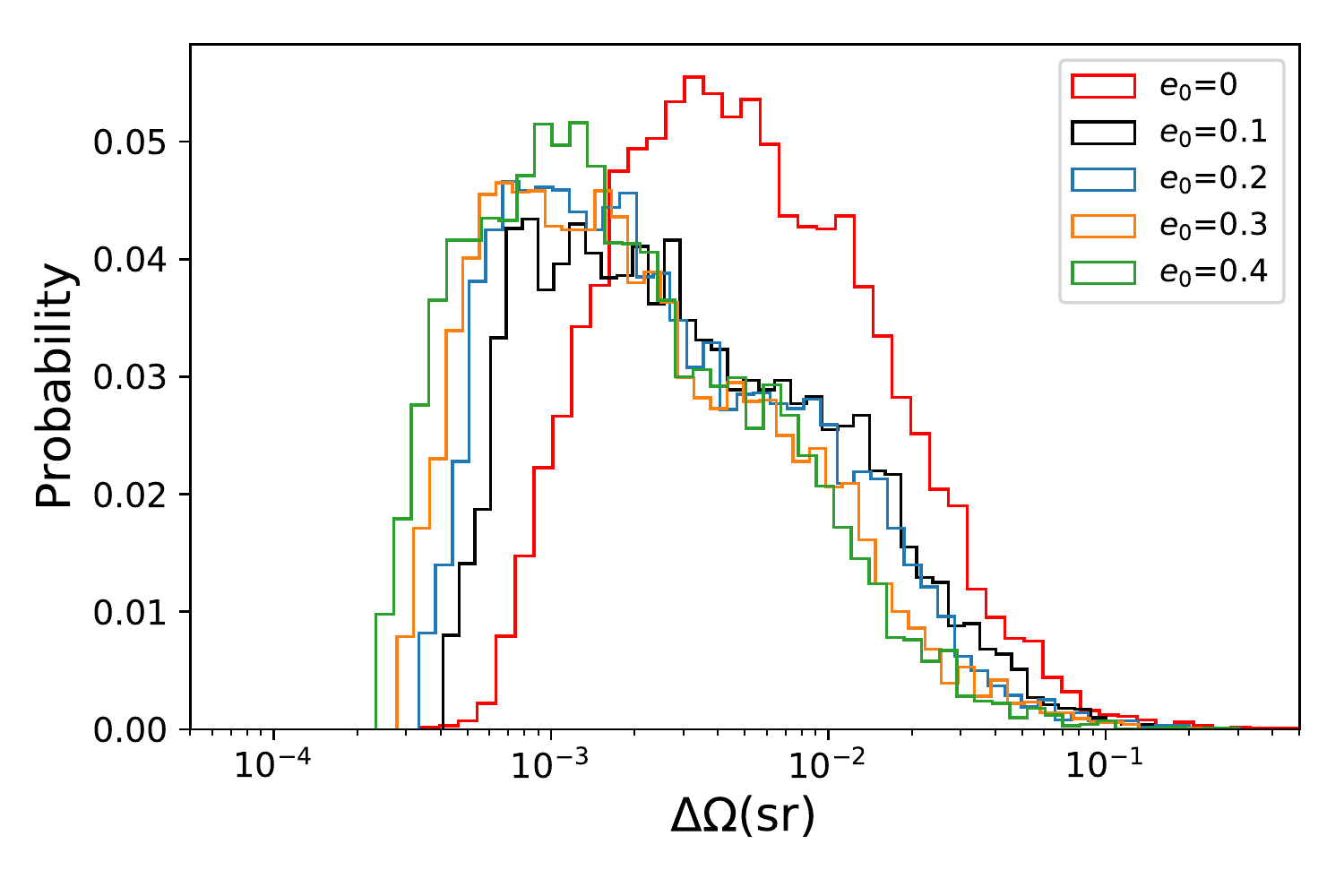} \\
    \includegraphics[width=0.5\textwidth]{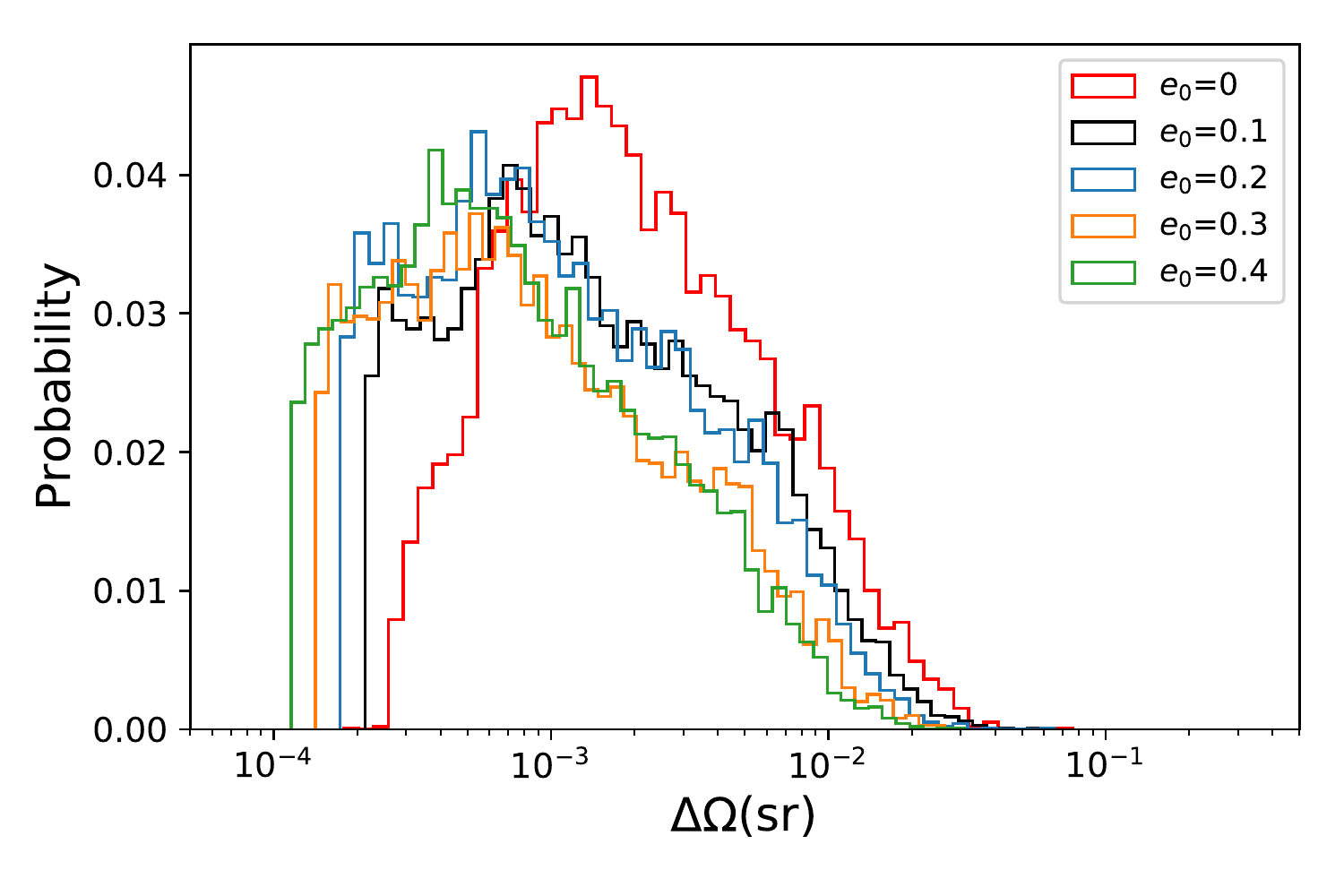} \\
    \includegraphics[width=0.5\textwidth]{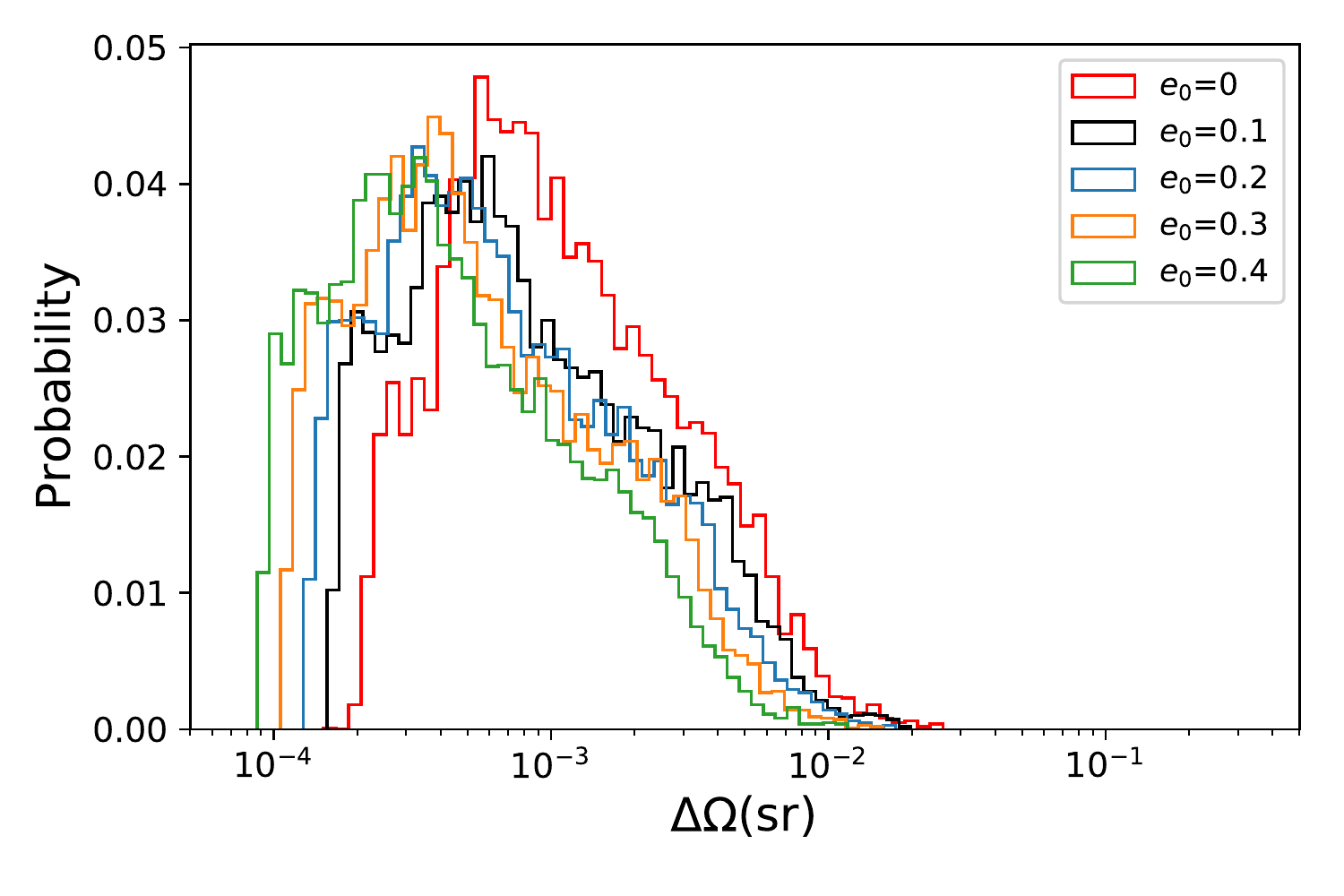}
\end{tabular}
\caption{Histograms of $\Delta\Omega$ for varying angle parameters $\iota_e$,
$\beta_e$, $\psi_e$, $\theta_e$ and $\phi_e$ with $10^4$ Monte Carlo samples,
for the big BBH case.
The plots in the upper, middle, and lower panels correspond to the LHV, LHVK,
and LHVKI case, respectively.}
\label{fig:100-histograms}
\end{figure}

\begin{table}
\caption{The best/worst accuracy of source localization and the corresponding
sky location for the big BBH case.}
\resizebox{0.49\textwidth}{\height}{
\linespread{1.3}\selectfont
\begin{tabular}{c|c|c|c} 
\toprule[0.5pt]
Network & $e_0$ &($\theta_e$, $\phi_e$)& $\Delta\Omega$  \\ \hline
\multirow{2}{*}{LHV}&0.0&(2.53, 2.01)/(2.27, 0.44)&
$4.84\times 10^{-4}$/$8.34\times 10^{-3}$ \\
&0.4&(2.53, 2.09)/(0.87, 3.67)&$2.31\times 10^{-4}$/$3.70\times 10^{-3}$
\\ \hline
\multirow{2}{*}{LHVK}&0.0&(2.97, 4.80)/(1.40, 1.92)&
$2.58\times 10^{-4}$/$4.22\times 10^{-3}$\\ 
&0.4&(2.97, 4.89)/(1.75, 5.06)&$1.15\times 10^{-4}$/$2.53\times 10^{-3}$
\\ \hline
\multirow{2}{*}{LHVKI}&0.0&(2.71, 4.71)/(1.75, 4.97)&
$1.87\times 10^{-4}$/$1.42\times 10^{-3}$\\ 
&0.4&(2.79, 4.80)/(1.75, 4.97)&$8.66\times 10^{-5}$/$6.91\times 10^{-4}$\\
\bottomrule[0.5pt]
\end{tabular}}     
\label{tab:BigBBH_Omega}
\end{table}

\begin{table}
\caption{The best/worst improvement of source localization accuracy among
the networks.}
\linespread{1.3}\selectfont
\begin{tabular}{c|c|c|c|c}
\toprule[0.5pt]
Case&$e_0$&$\Delta\Omega^{\rm LHV}_{\rm LHVK}$
&$\Delta\Omega^{\rm LHV}_{\rm LHVKI}$&$\Delta\Omega^{\rm LHVK}_{\rm LHVKI}$\\
\hline
\multirow{2}{*}{Big}&0.0&6.58/1.02&12.2/1.41&4.18/1.02\\
                    &0.4&6.40/1.02&12.3/1.40&4.03/1.06\\ \hline
\multirow{2}{*}{GW151226-like}&0.0&7.85/1.01&19.8/1.28&7.06/1.02\\
                              &0.4&7.96/1.01&21.1/1.26&7.44/1.02\\ \hline
\multirow{2}{*}{GW170817-like}&0.0&8.58/1.00&29.2/1.21&7.00/1.01\\
                              &0.4&8.70/1.00&30.0/1.21&6.99/1.01\\
\bottomrule[0.5pt]
\end{tabular}
\label{tab:improve_det}
\end{table}

\begin{figure*}[htbp]
\begin{tabular}{cc}
\subfloat[]{\includegraphics[width=0.5\textwidth]{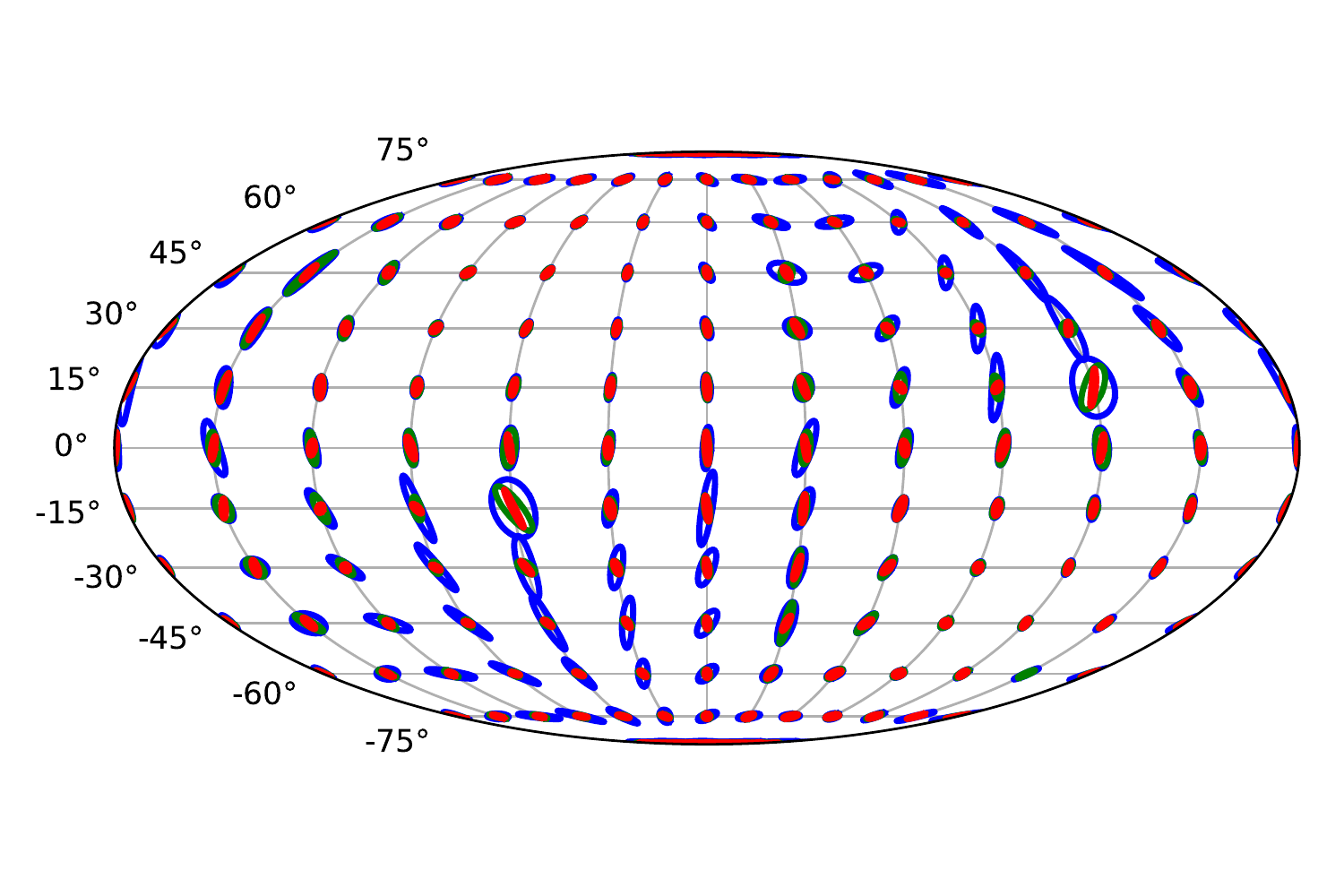}
\label{fig:22-ellipse-0}}&
\subfloat[]{\includegraphics[width=0.5\textwidth]{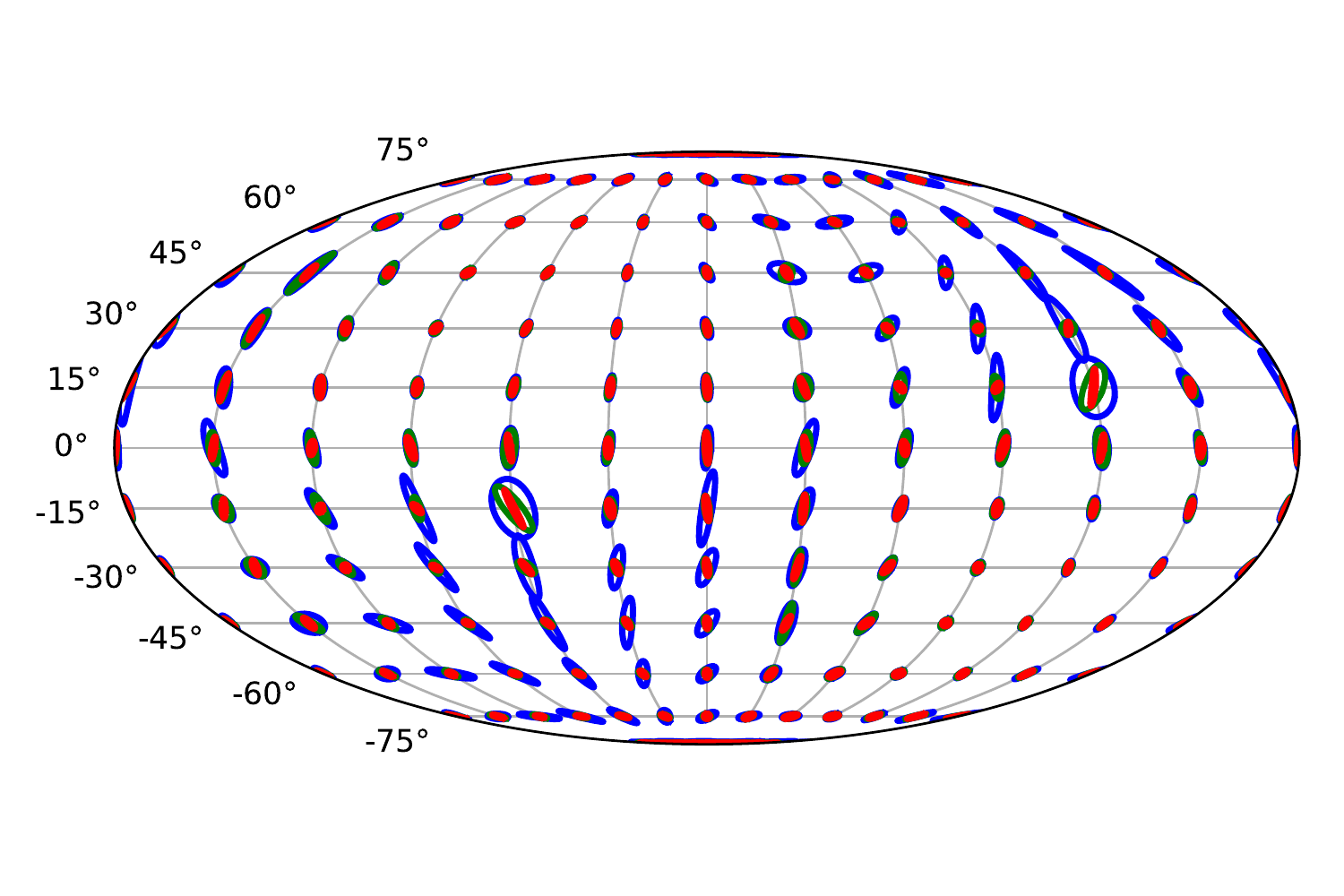}
\label{fig:22-ellipse-0.4}}
\end{tabular}
\caption{Error ellipses of the source localization in the GW151226-like BBH case
for (a) $e_{0.0}$ and (b) $e_{0.4}$.
The blue, green, and red ellipses correspond to the LHV, LHVK, and LHVKI cases,
respectively.}
\label{fig:22-ellipse}
\end{figure*}

We compared $\Delta\Omega$ for the LHV, LHVK, and LHVKI cases with $e_{0.0}$ and
$e_{0.4}$ for all ($\theta_e$, $\phi_e$) in Fig.~\ref{fig:100-space}.
Here, sr means ``square radian,'' and $1\mbox{sr}=(180/\pi)^2\approx 3282.81$
$\text{deg}^2$.
In the upper row, we show the result of $\Delta\Omega$ for $e_{0.0}$.
The left, middle, and right columns correspond to the networks LHV, LHVK,
and LHVKI, respectively.
One can see that $\Delta\Omega$ is smaller for a network with more detectors
in general.
In the lower row of Fig.~\ref{fig:100-space}, we show the result of
$\Delta\Omega$ for $e_{0.4}$.
One can see that the distribution behavior is similar to that of the $e_{0.0}$ case.
When more detectors are used, a better accuracy of source localization is obtained.
The best and worst $\Delta\Omega$'s and the corresponding sky locations are
listed in Table \ref{tab:BigBBH_Omega}.
One can see that the accuracy of source localization in the $e_{0.4}$ case is
better than that in the $e_{0.0}$ case.

In Fig.~\ref{fig:100-space-ratio-net}, we compare $\Delta\Omega$ among the
three networks by plotting the ratio of $\Delta\Omega$ among them, as
\begin{align}
\Delta\Omega^{\rm LHV}_{\rm LHVK}&=\frac{\Delta\Omega_{\rm LHV}}
{\Delta\Omega_{\rm LHVK}},\quad
\Delta\Omega^{\rm LHV}_{\rm LHVKI}=\frac{\Delta\Omega_{\rm LHV}}
{\Delta\Omega_{\rm LHVKI}},\notag \\
\Delta\Omega^{\rm LHVK}_{\rm LHVKI}&=\frac{\Delta\Omega_{\rm LHVK}}
{\Delta\Omega_{\rm LHVKI}},\label{omegaratio}
\end{align}
for each $(\theta_e,\phi_e)$.
We show the results of these ratios for $e_{0.0}$ and that for $e_{0.4}$ in
the upper and lower row, respectively.
The distribution behavior in the $e_{0.0}$ case and in the $e_{0.4}$ case is
similar.
The best and worst improvement factors for the big BBH case are listed in the 
second row of Table \ref{tab:improve_det}.
One can see that the improvement factors between $e_{0.0}$ and $e_{0.4}$ are
close,
and the accuracy is improved significantly by having more detectors in the
network.
The results in Fig.~\ref{fig:100-space} and Fig.~\ref{fig:100-space-ratio-net}
say that the network with more detectors gives a smaller $\Delta\Omega$ and 
thus more accurate localization in this scenario.

In Fig.~\ref{fig:100-space-ratio-e0}, we plot the improvement factor
$\displaystyle\Delta\Omega_{0.4}^{0.0}\equiv
\frac{\Delta\Omega_{0.0}}{\Delta\Omega_{0.4}}$ for each $(\theta_e,\phi_e)$,
where the subscripts 0.0 and 0.4 mean $e_{0.0}$ and $e_{0.4}$, respectively.
We show the best and worst improvement factors between the two eccentricities
for the three networks in the second row of Table.~\ref{tab:improve_e},
for this case.
The results in Fig.~\ref{fig:100-space-ratio-e0} and
Table \ref{tab:improve_e} emphasize that the cases with higher initial
eccentricity give a smaller $\Delta\Omega$
and thus a more accurate localization than the ones with smaller initial
eccentricity, in this scenario.

\begin{table}
\caption{The best/worst improvement of source localization accuracy between
the eccentricities.}
\linespread{1.3}\selectfont
\begin{tabular}{c|c|c} 
\toprule[0.5pt]
Case&Network&$\Delta\Omega^{0.0}_{0.4}$\\ \hline
\multirow{3}{*}{Big}&LHV&2.84/1.43\\ 
                    &LHVK&2.89/1.53\\ 
                    &LHVKI&2.52/1.70\\ \hline
\multirow{3}{*}{GW151226-like}&LHV&1.10/0.99\\ 
                              &LHVK&1.07/1.00\\ 
                              &LHVKI&1.08/1.04\\ \hline
\multirow{3}{*}{GW170817-like}&LHV&1.03/0.98\\ 
                              &LHVK&1.01/0.99\\ 
                              &LHVKI&1.01/0.99\\
\bottomrule[0.5pt]
\end{tabular}
\label{tab:improve_e}
\end{table}

\begin{figure*}
\begin{tabular}{ccc}
\includegraphics[width=0.33\textwidth]{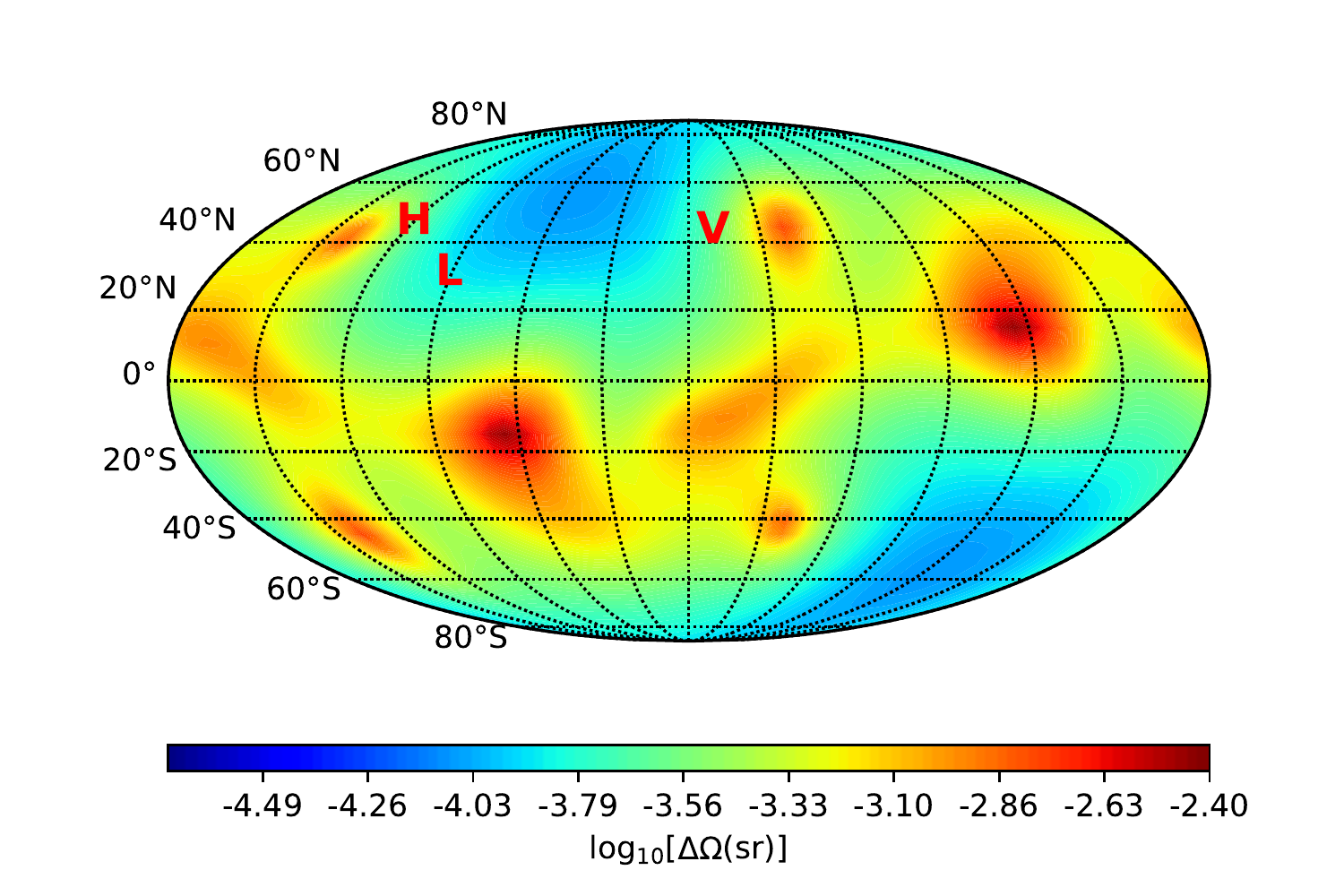}
\label{fig:22-space-0-LHV} &
\includegraphics[width=0.33\textwidth]{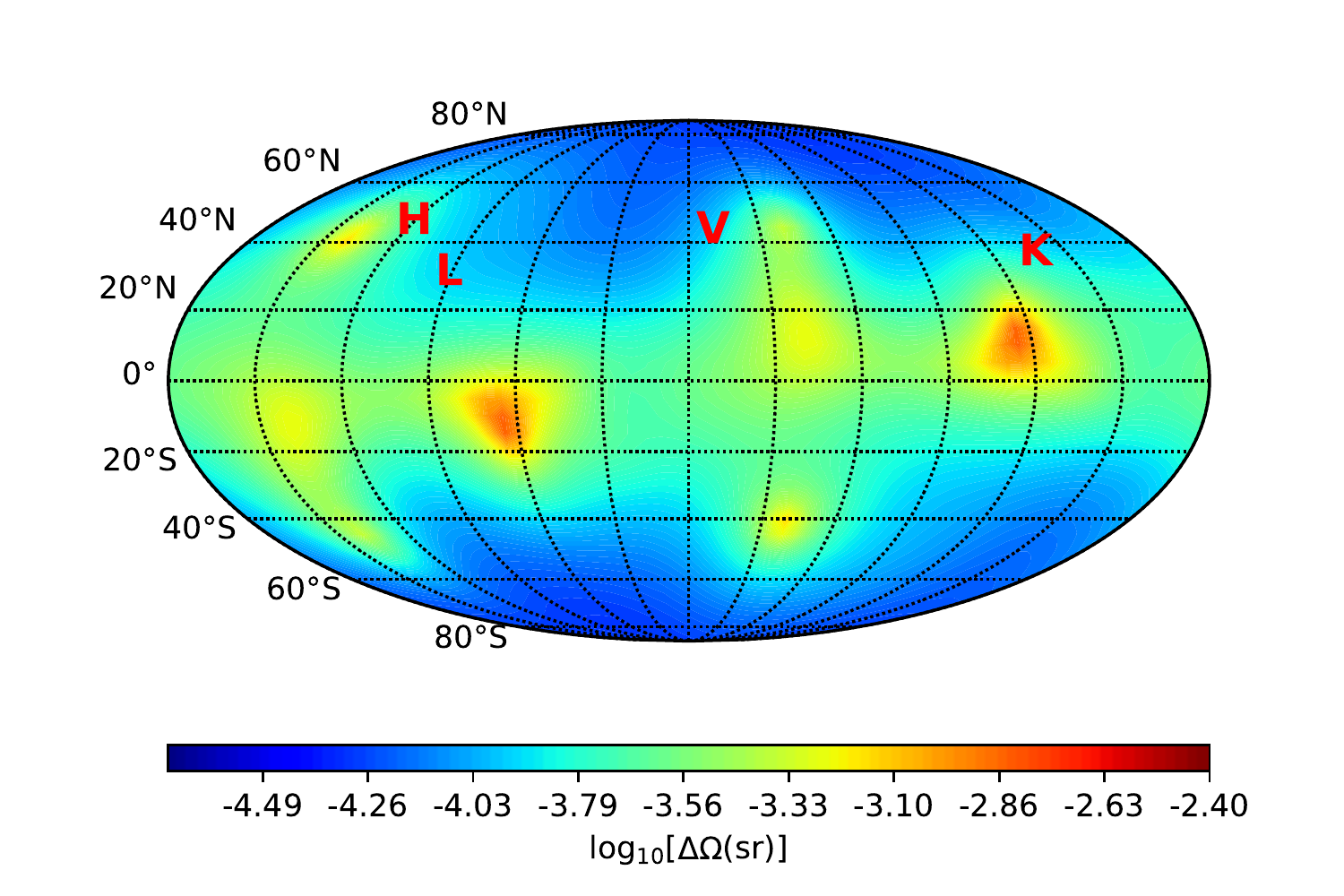}
\label{fig:22-space-0-LHVK} &
\includegraphics[width=0.33\textwidth]{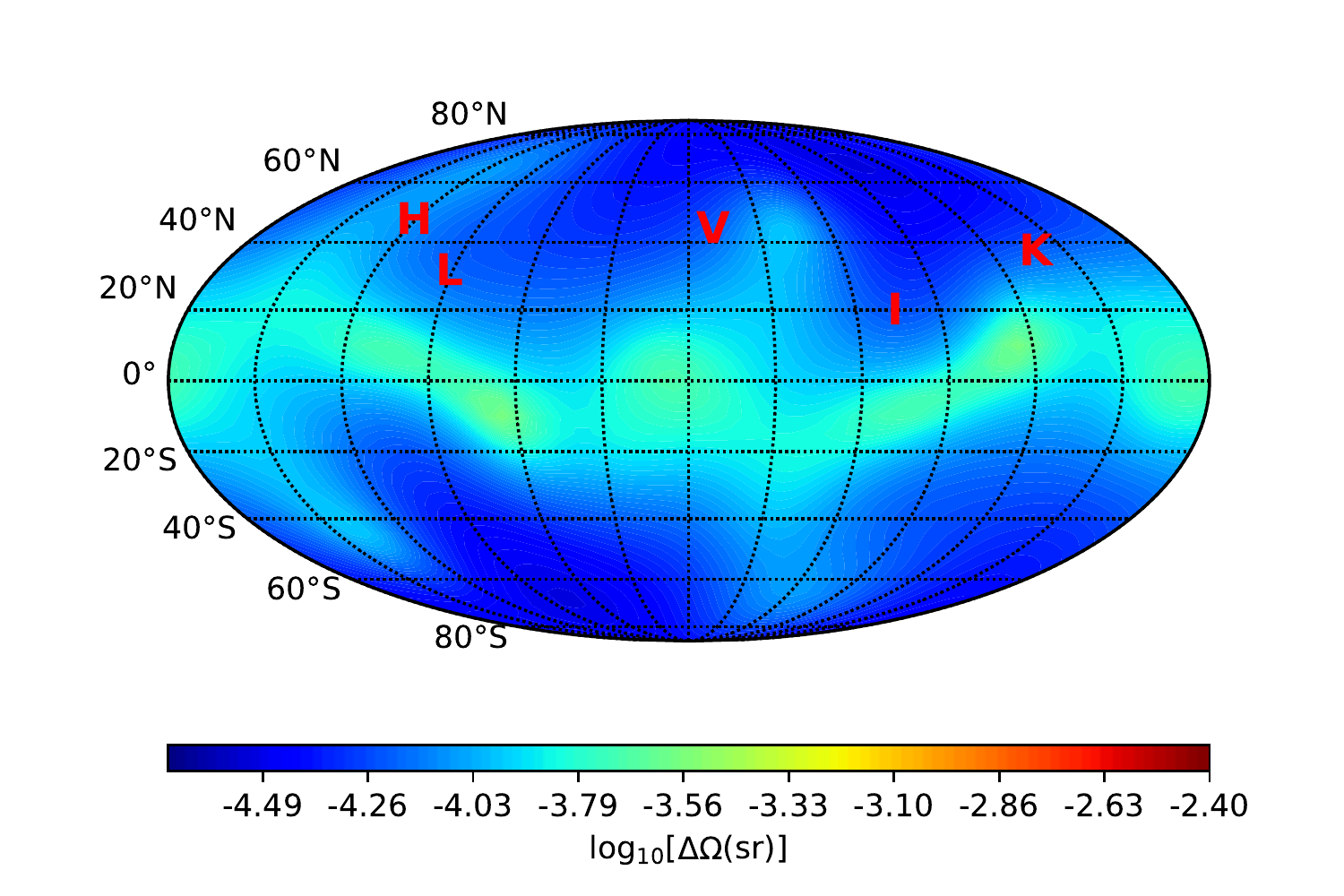}
\label{fig:22-space-0-LHVKI} \\
\includegraphics[width=0.33\textwidth]{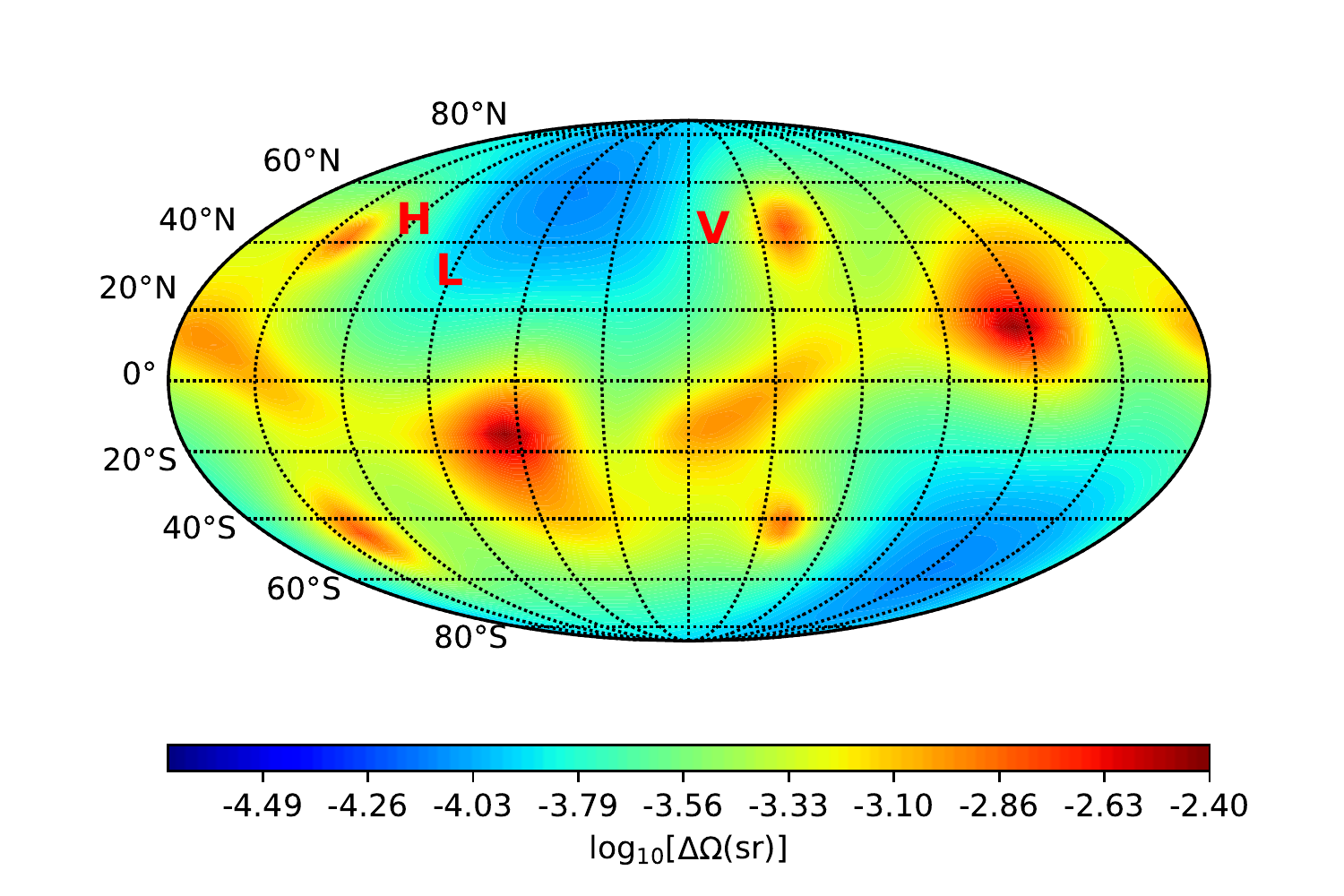}
\label{fig:22-space-0.4-LHV} &
\includegraphics[width=0.33\textwidth]{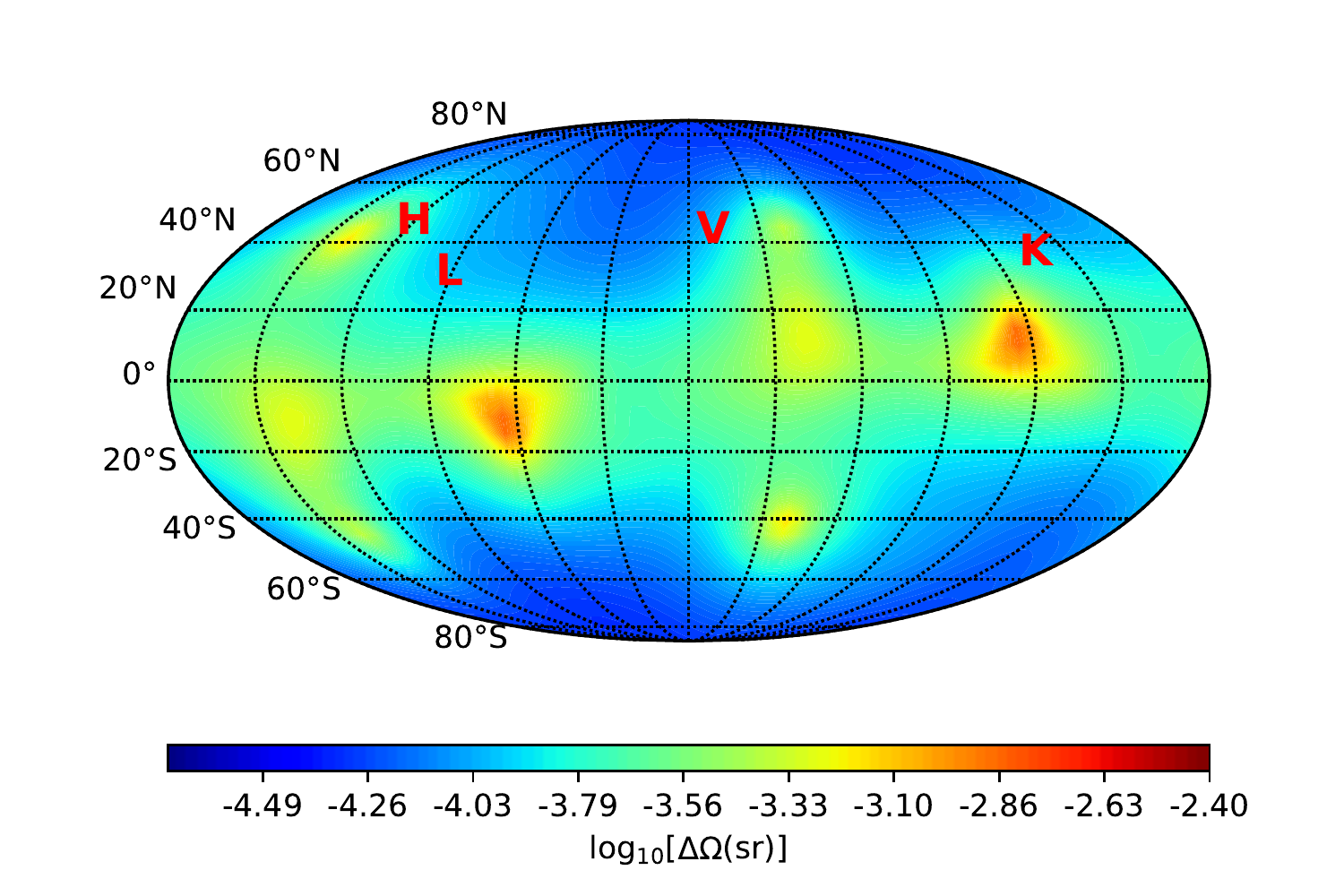}
\label{fig:22-space-0.4-LHVK} &
\includegraphics[width=0.33\textwidth]{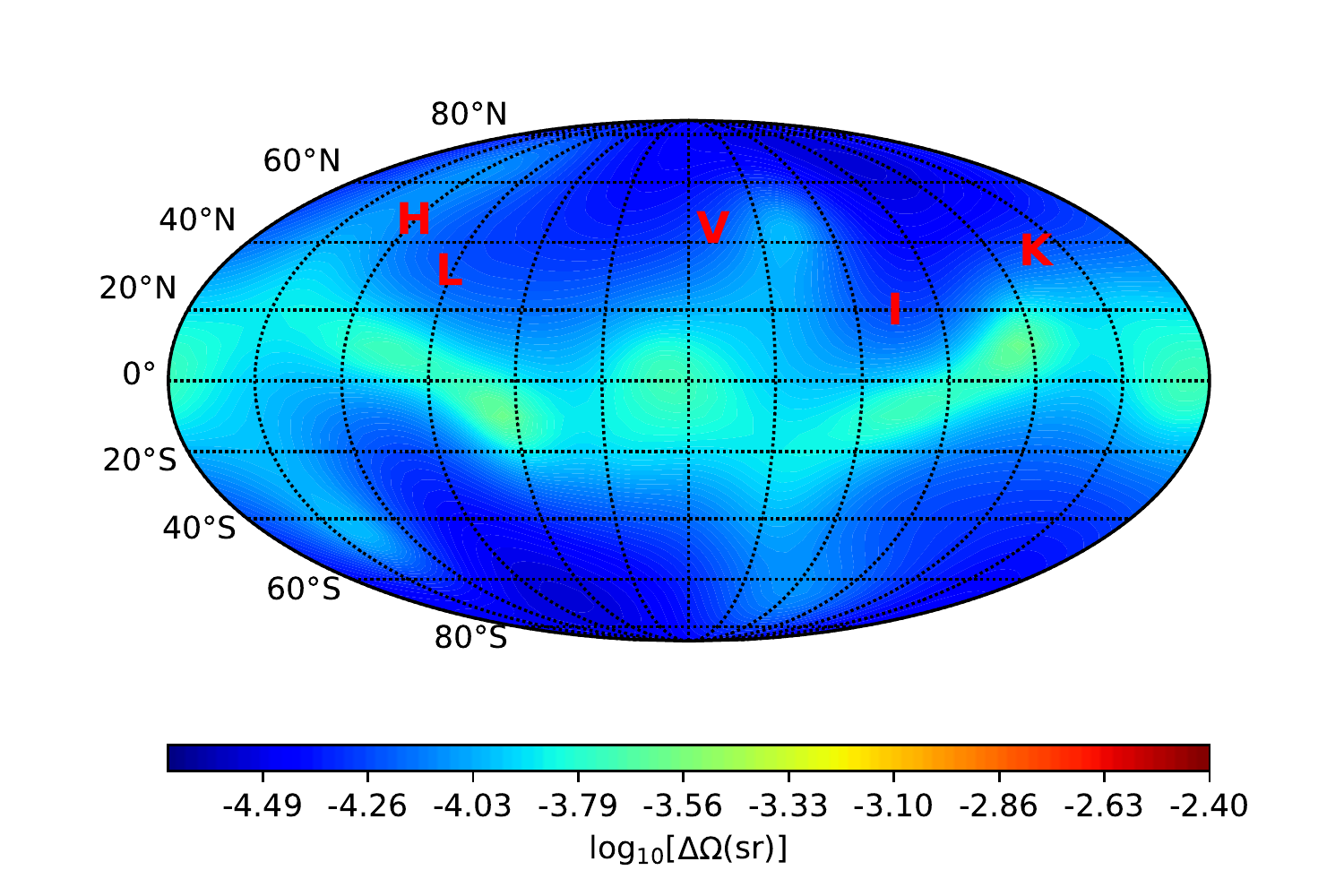}
\label{fig:22-space-0.4-LHVKI}
\end{tabular}
\caption{Estimated error $\Delta\Omega$ of the source localization for the
GW151226-like BBH case.
The panels in the upper and the lower rows correspond to the eccentricities
$e_{0.0}$ and $e_{0.4}$, respectively.
We show the $\Delta\Omega$'s for the LHV, LHVK, and LHVKI cases in the left,
middle, and right columns, respectively.}
\label{fig:22-space}
\end{figure*}

\begin{figure*}
\begin{tabular}{ccc}
\includegraphics[width=0.33\textwidth]{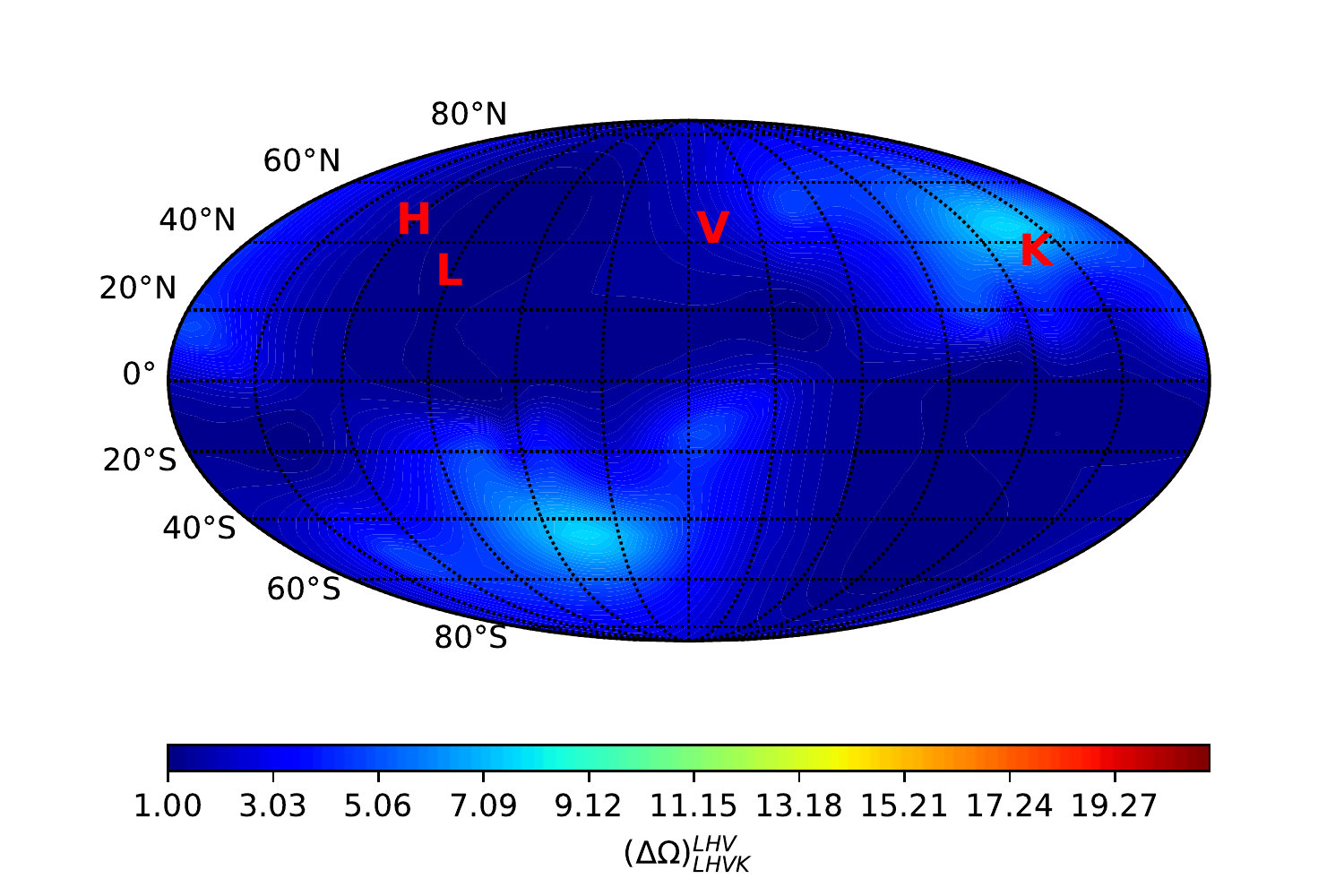}
\label{fig:22-space-0-3-4} &
\includegraphics[width=0.33\textwidth]{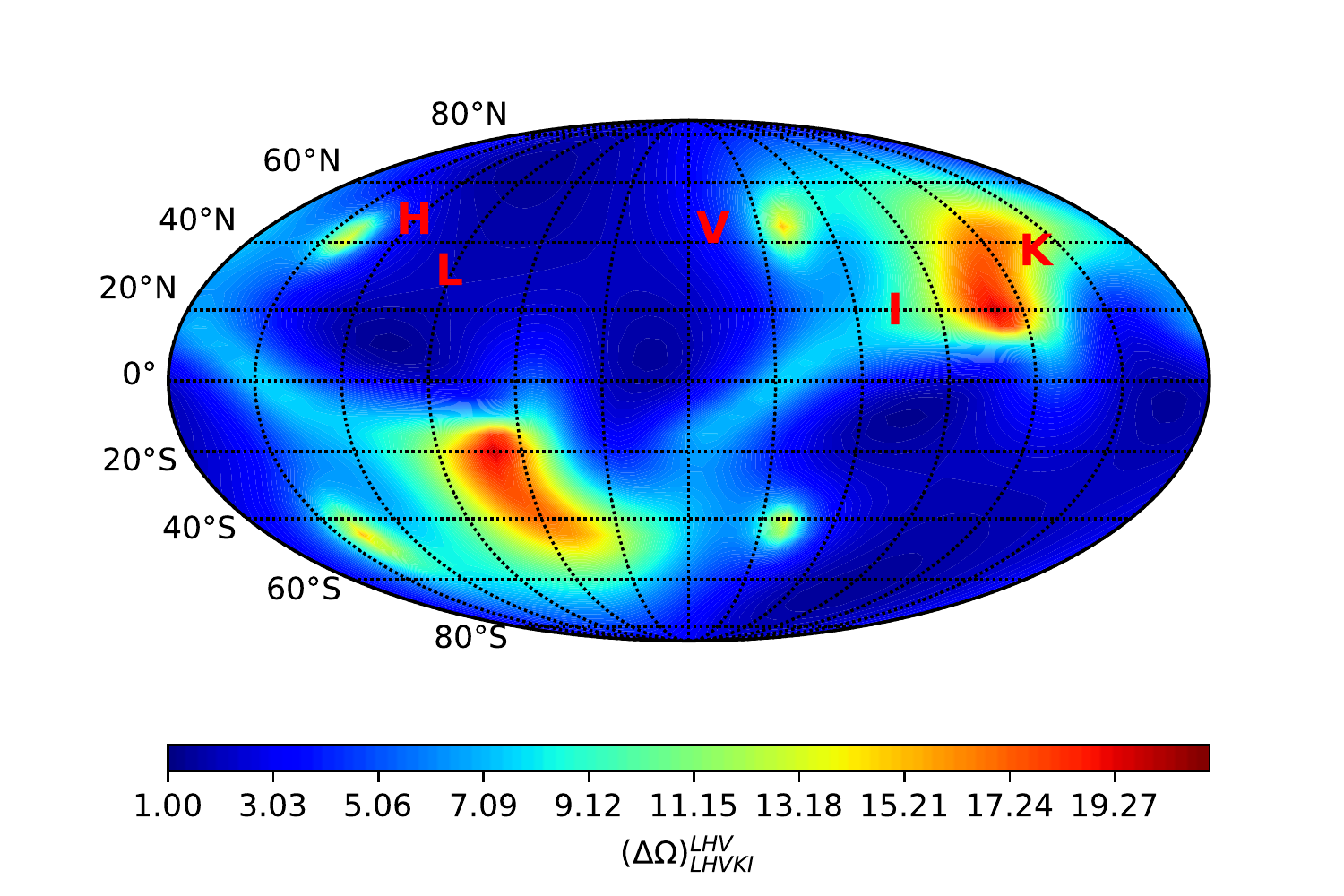}
\label{fig:22-space-0-3-5} &
\includegraphics[width=0.33\textwidth]{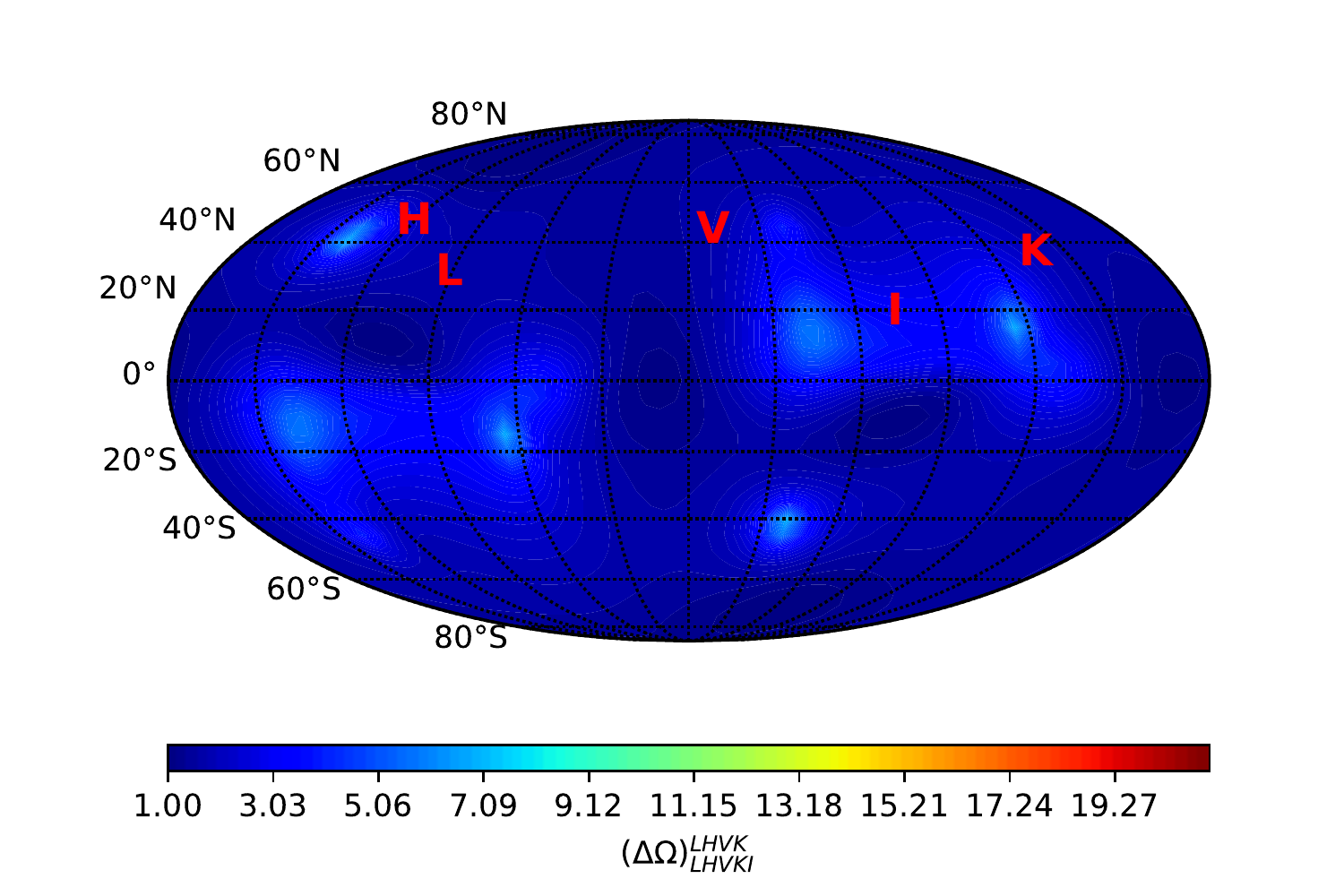}
\label{fig:22-space-0-4-5} \\
\includegraphics[width=0.33\textwidth]{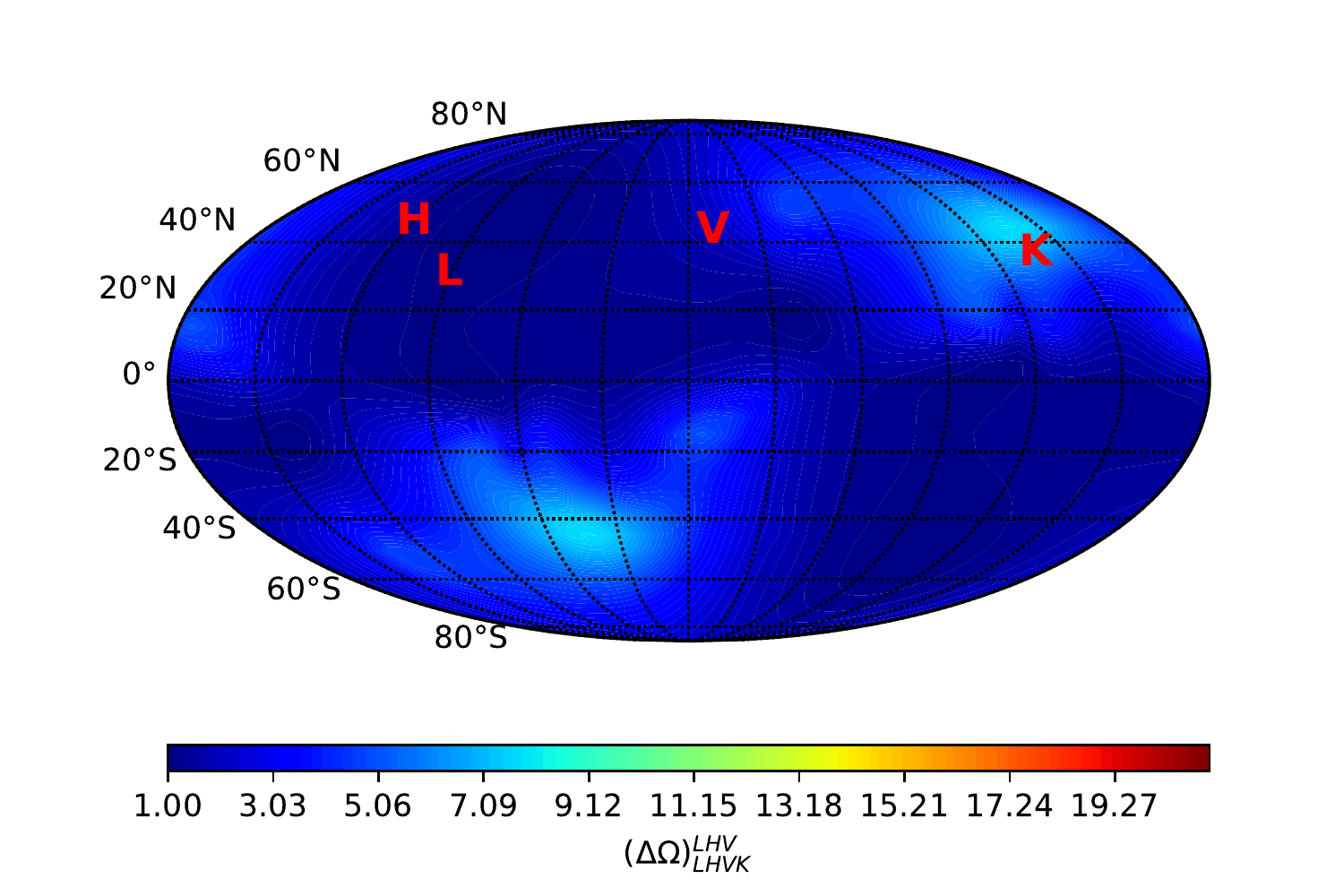}
\label{fig:22-space-0.4-3-4} &
\includegraphics[width=0.33\textwidth]{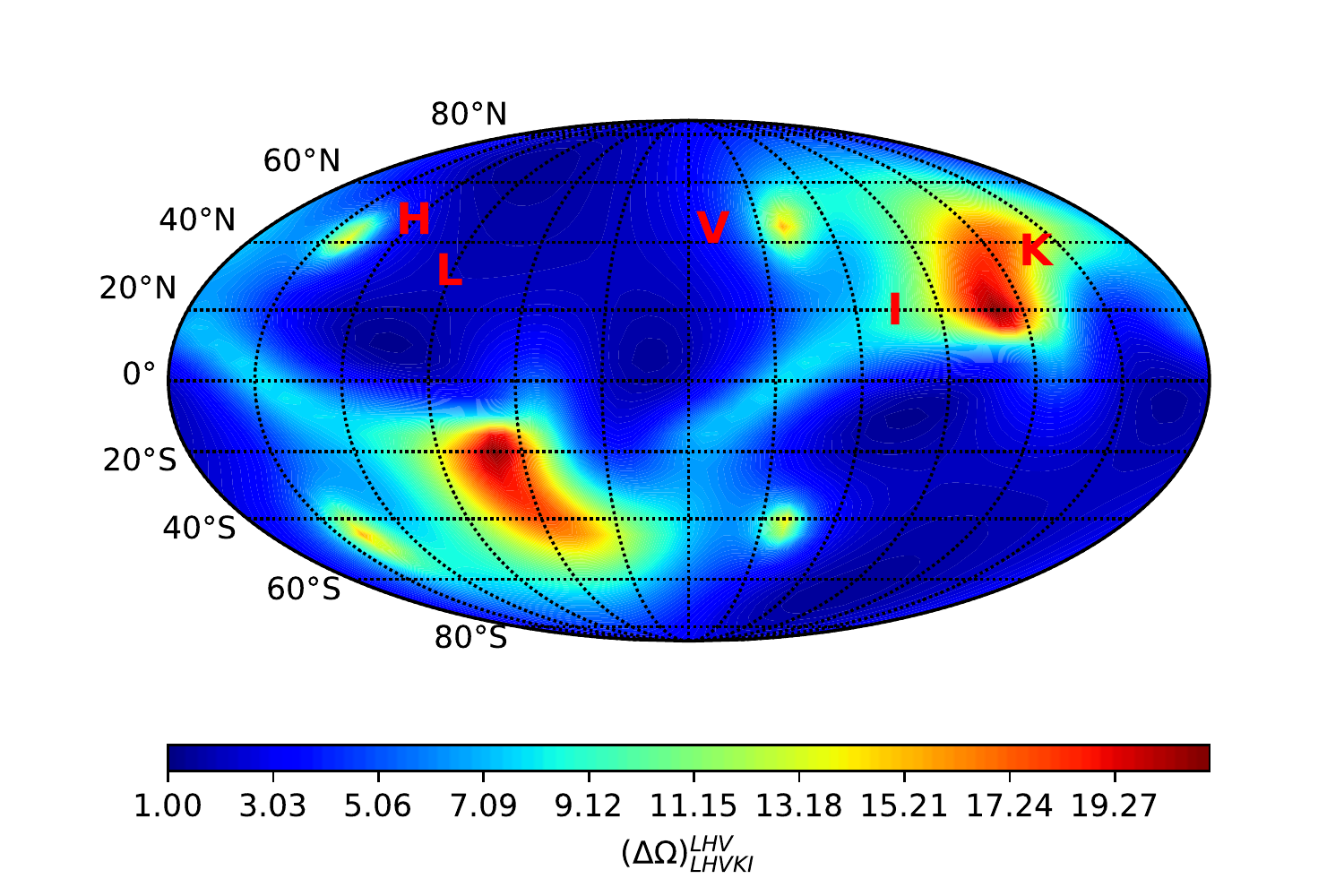}
\label{fig:22-space-0.4-3-5} &
\includegraphics[width=0.33\textwidth]{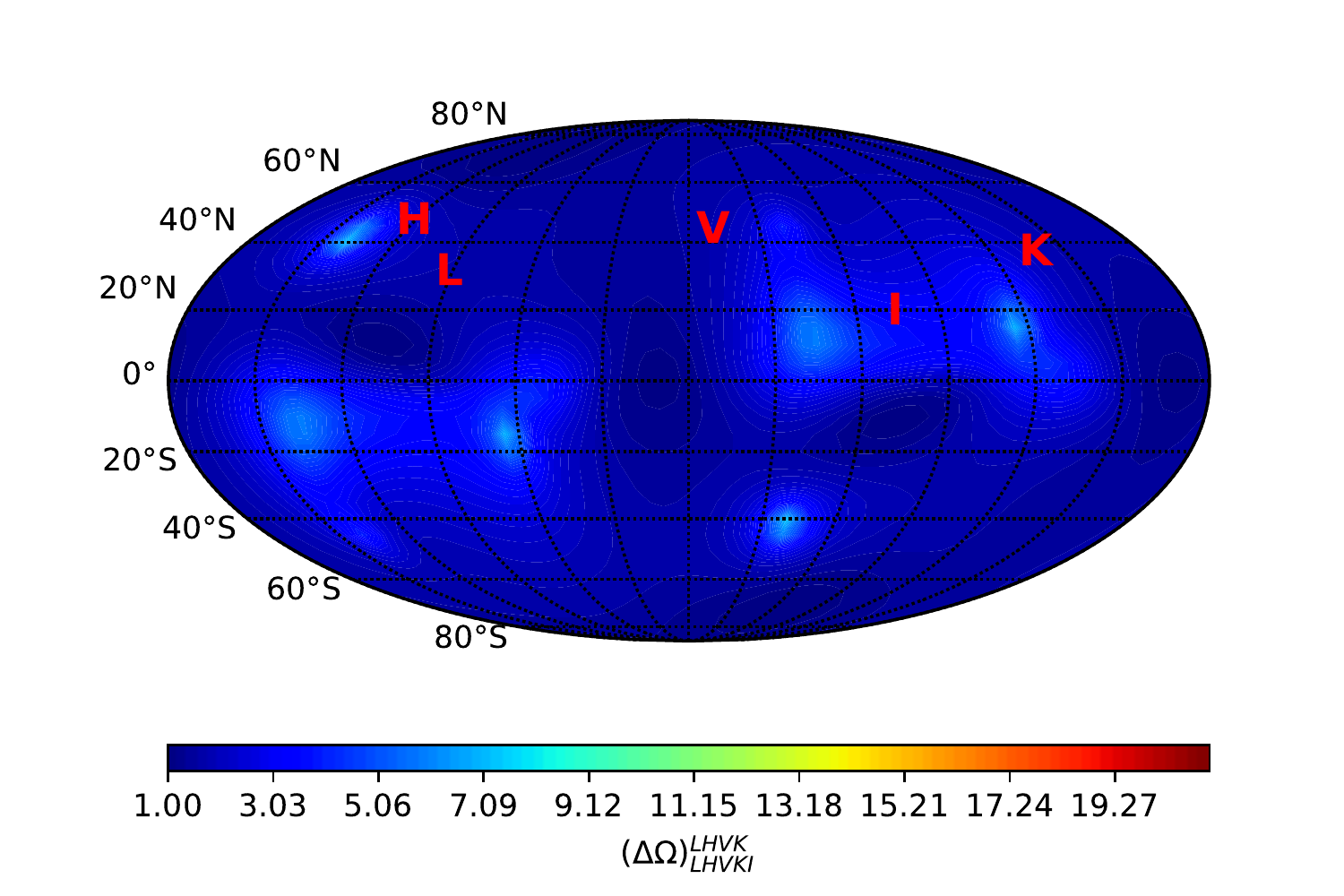}
\label{fig:22-space-0.4-4-5}
\end{tabular}
\caption{Ratios of $\Delta\Omega$ among different networks, defined in
Eq.~(\ref{omegaratio}), for the GW151226-like BBH case.
The panels in the upper and the lower rows correspond to the eccentricities
$e_{0.0}$ and $e_{0.4}$, respectively.
$\Delta\Omega^{\text{LHV}}_{\text{LHVK}}$,
$\Delta\Omega^{\text{LHV}}_{\text{LHVKI}}$, and
$\Delta\Omega^{\text{LHVK}}_{\text{LHVKI}}$ are shown in the left, middle,
and right columns, respectively.}
\label{fig:22-space-ratio-net}
\end{figure*}

In the above result, we have fixed $\psi_e$, $\beta_e$ and $\iota_e$ to be zero.
To investigate the effect of these parameters on the source location
improvement with different eccentricities and networks, we apply the Monte Carlo
method with $10^4$ samples.
We take uniform random values for $\theta_e$ within (0, $\pi$); $\phi_e$,
$\psi_e$, and $\beta_e$ within (0, $2\pi$); and $\iota_e$ within (0, $\pi/2$).
We show the statistical results with the histograms in
Fig.~\ref{fig:100-histograms}.
For the distributions, we can see that the peaks all move leftward with the
initial eccentricity increasing.
By comparing the median value of the statistical data for each eccentricity case,
we can determine the improvement of the accuracy of the source localization. 
Compared with the one in $e_{0.0}$, the source localization accuracy improves
as 1.7 times better with $e_{0.1}$, 2.2 times better with $e_{0.2}$, 2.7 times
better with $e_{0.3}$, and 3.1 times better with $e_{0.4}$ for the LHV case.
For the LHVK case, the source localization accuracy improves as 1.5 times
better with $e_{0.1}$, 1.9 times better with $e_{0.2}$, 2.4 times better with
$e_{0.3}$ and 2.8 times better with $e_{0.4}$.
And it improves as 1.4 times better with $e_{0.1}$, 1.7 times better with
$e_{0.2}$, 2.1 times better with $e_{0.3}$ and 2.5 times better with $e_{0.4}$
for the LHVKI case. 

\subsection{GW151226-like BBH case}
In this subsection, we consider a GW151226-like BBH with a total mass
$22 M_\odot$.
We fix the parameters $D_{Le}=410$Mpc, $\eta=1/4$,
$\mathcal{M}=M\eta^{3/5}=28.3 M_\odot$,
and $t_{ce}=\phi_c=\iota_e=\beta_e=\psi_e=0$, while we vary $\theta_e$, $\phi_e$ and
$e_0$ to investigate the resulting accuracy of the source localization.
Compared with the setting of the big BBH in the previous subsection,
only the value of the chirp mass $\mathcal{M}$ is changed.
Similar to the result in Fig.~\ref{fig:100-ellipse}, we study the
error ellipses for different $\theta_e$ and $\phi_e$.
We show the results of the 5$\sigma$ error region ellipses for the
GW151226-like BBH case in Fig.~\ref{fig:22-ellipse}.
Compared with the ellipses in Fig.~\ref{fig:100-ellipse}, the accuracy of the
source localization in the GW151226-like BBH case is better than in the big BBH
case.
This is consistent with our previous result \cite{PhysRevD.96.084046}.
We can also see that, with more detectors, the accuracy of the  source
localization becomes better.
And we find that the accuracy of the source localization is raised by
increasing the initial eccentricity, but the improvement by the eccentricity
could be negligible, in contrast to the big BBH case.

\begin{figure}
\begin{tabular}{c}
\includegraphics[width=0.49\textwidth]{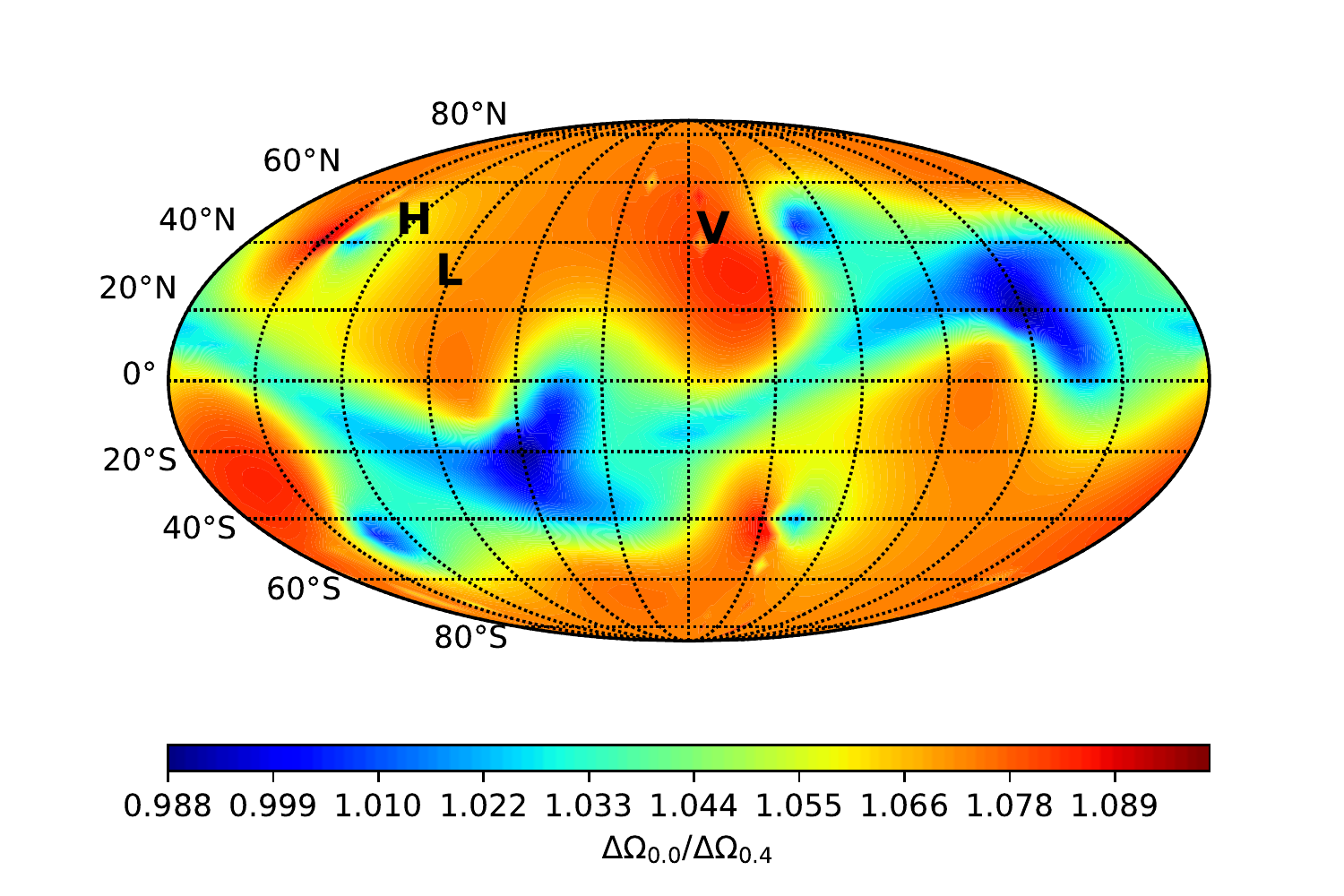}
\label{fig:22-space-ratio-LHV} \\
\includegraphics[width=0.49\textwidth]{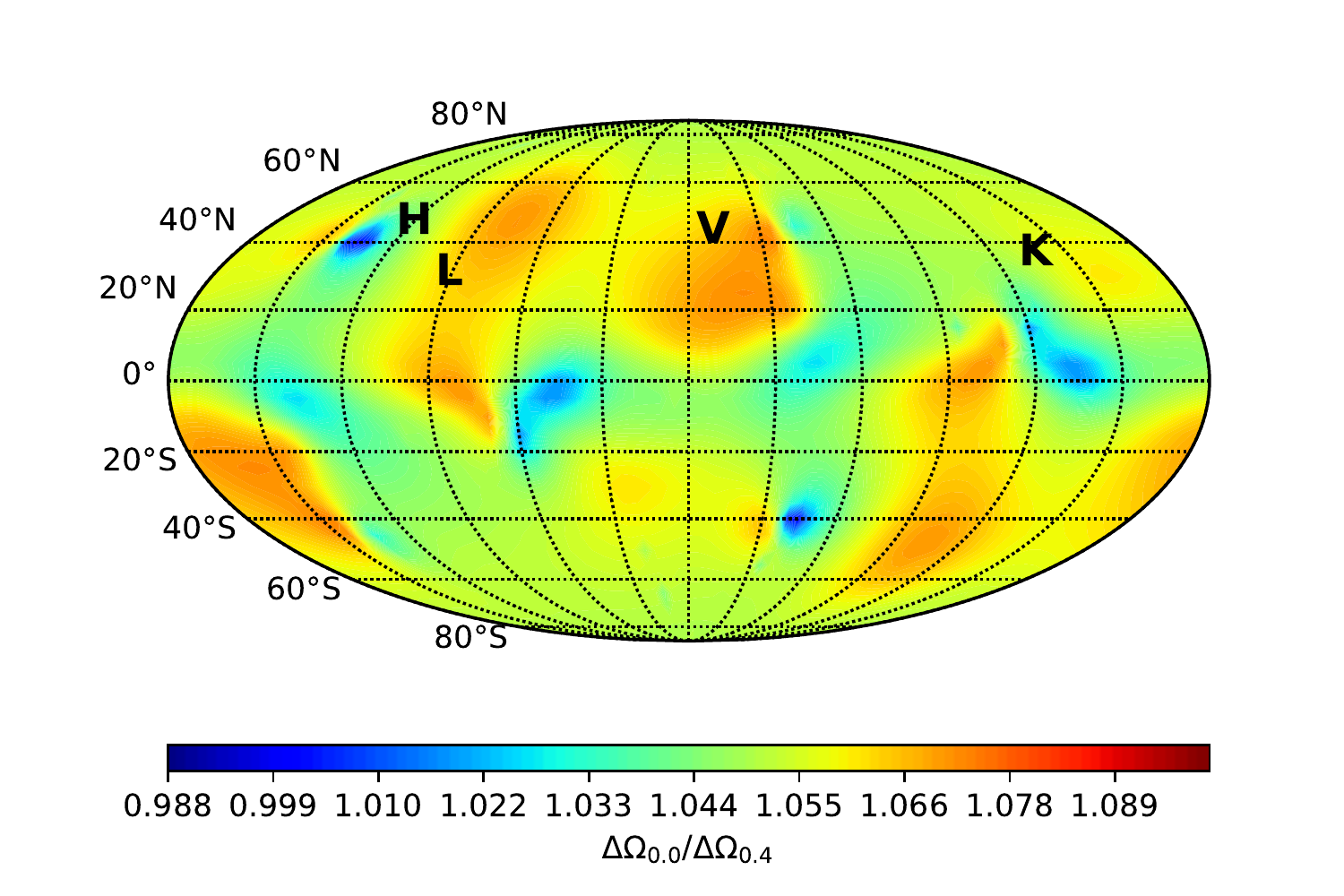}
\label{fig:22-space-ratio-LHVK} \\
\includegraphics[width=0.49\textwidth]{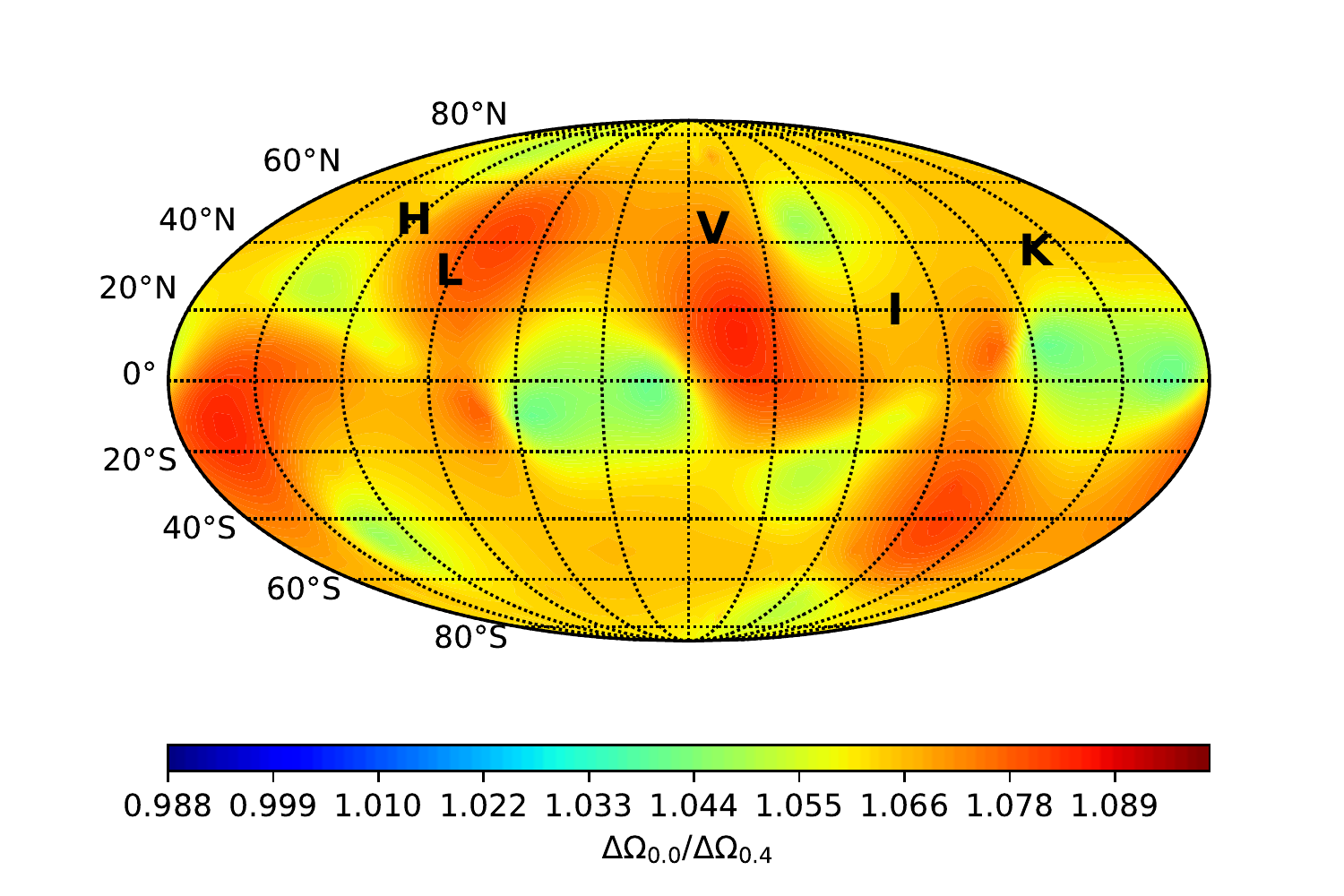}
\label{fig:22-space-ratio-LHVKI}
\end{tabular}
\caption{$\Delta\Omega^{0.0}_{0.4}$ for the GW151226-like BBH case.
The plots in the upper, middle, and lower panels correspond to the LHV, LHVK,
and LHVKI cases, respectively.}
\label{fig:22-space-ratio-e0}
\end{figure}

\begin{figure}
\begin{tabular}{c}
\includegraphics[width=0.49\textwidth]{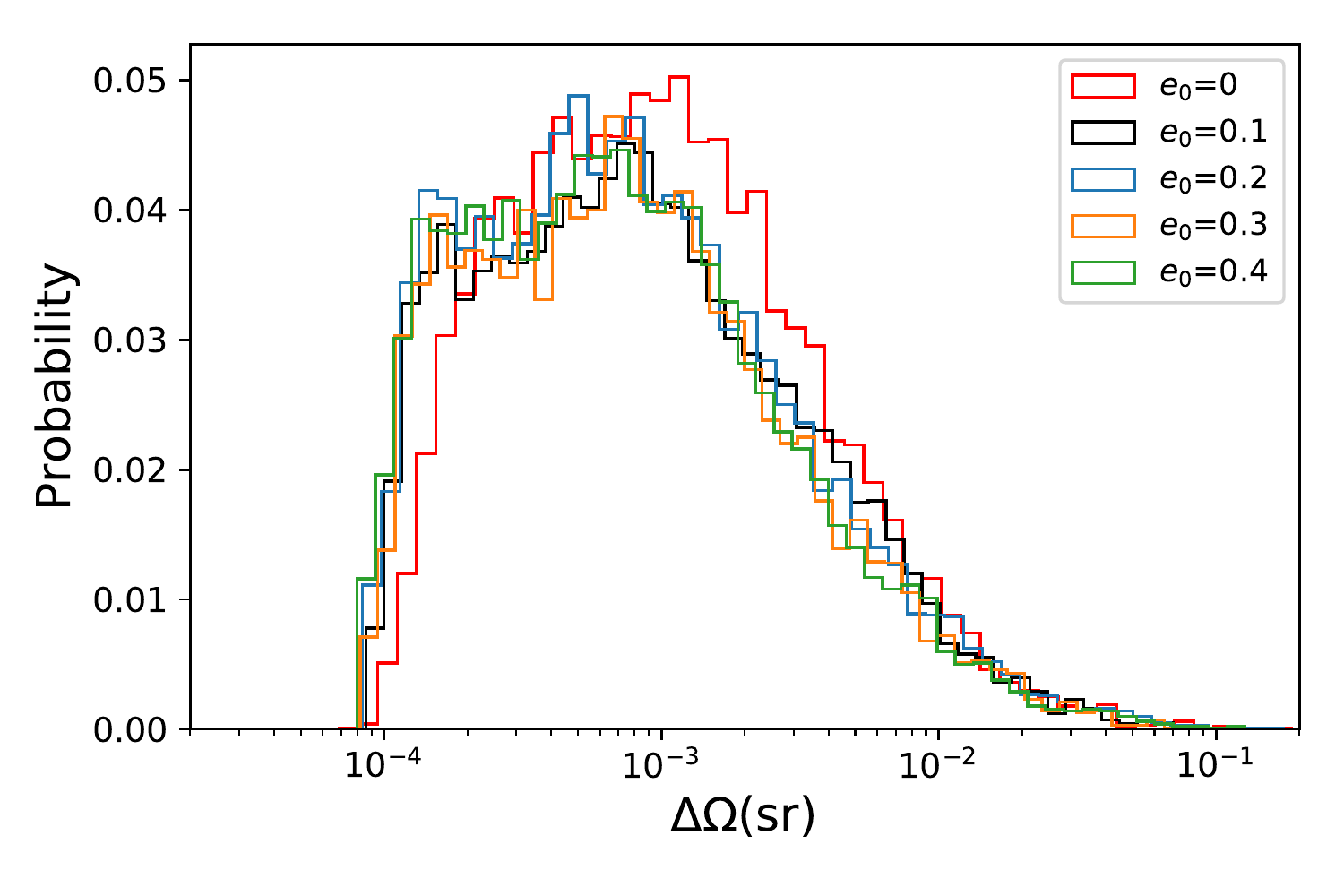} \\
\includegraphics[width=0.49\textwidth]{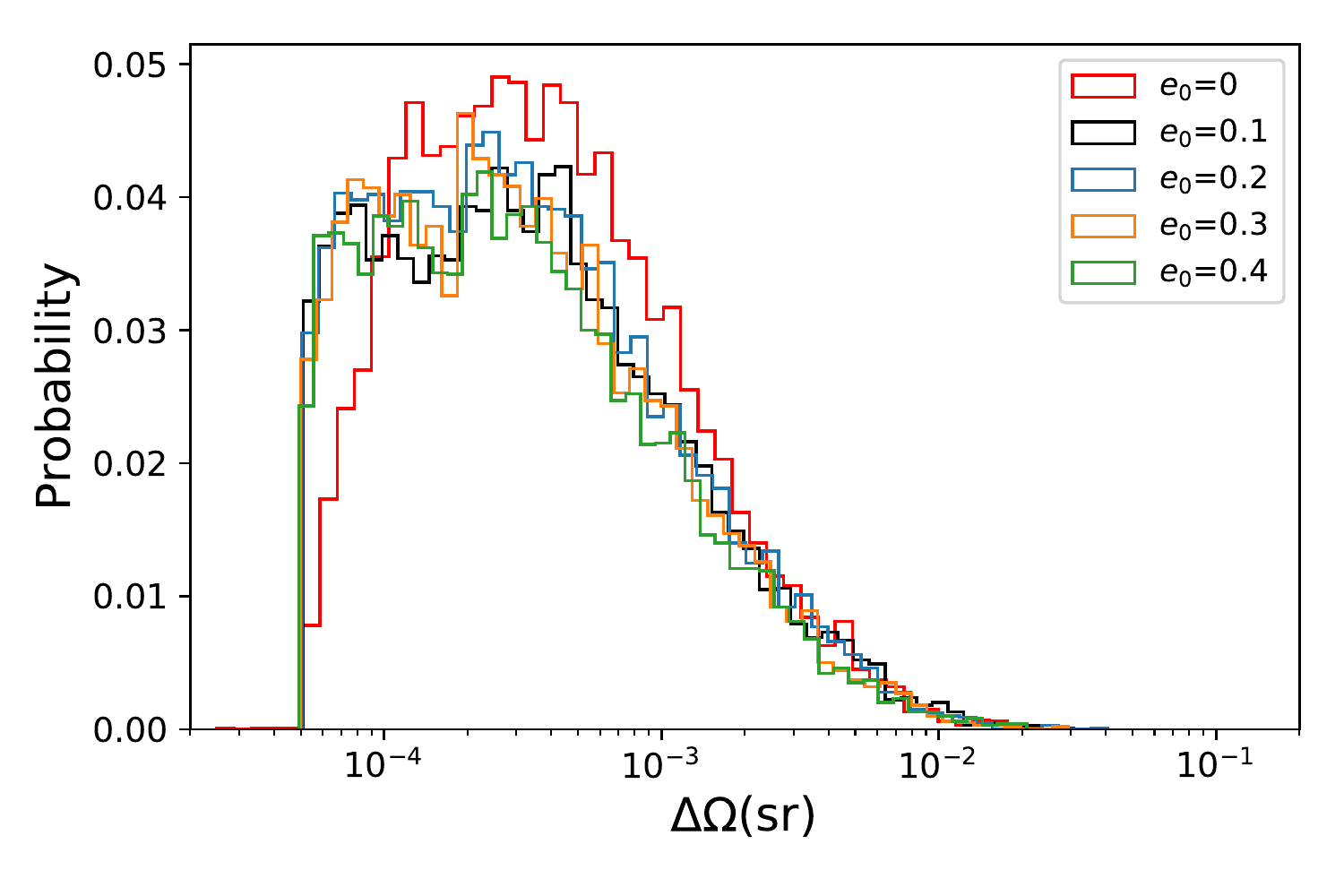} \\
\includegraphics[width=0.49\textwidth]{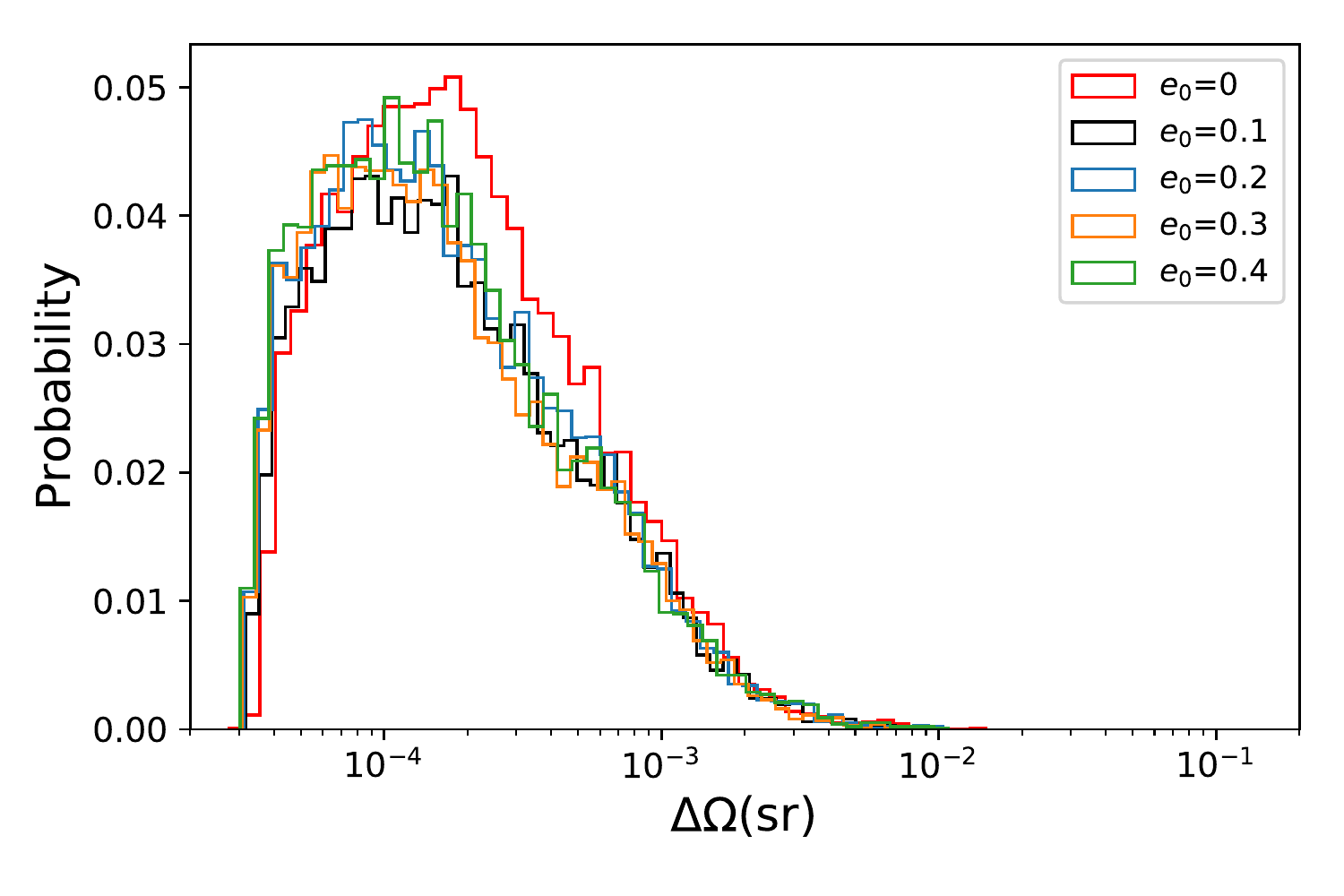}
\end{tabular}
\caption{Histograms of $\Delta\Omega$ for varying angle parameters $\iota_e$,
$\beta_e$, $\psi_e$, $\theta_e$ and $\phi_e$ with $10^4$ Monte Carlo samples,
for the GW151226-like BBH case.
The plots in the upper, middle, and lower panels correspond to the LHV, LHVK,
and LHVKI cases, respectively.}
\label{fig:22-histograms}
\end{figure}

\begin{figure*}
\begin{tabular}{cc}
\subfloat[]{\includegraphics[width=0.5\textwidth]{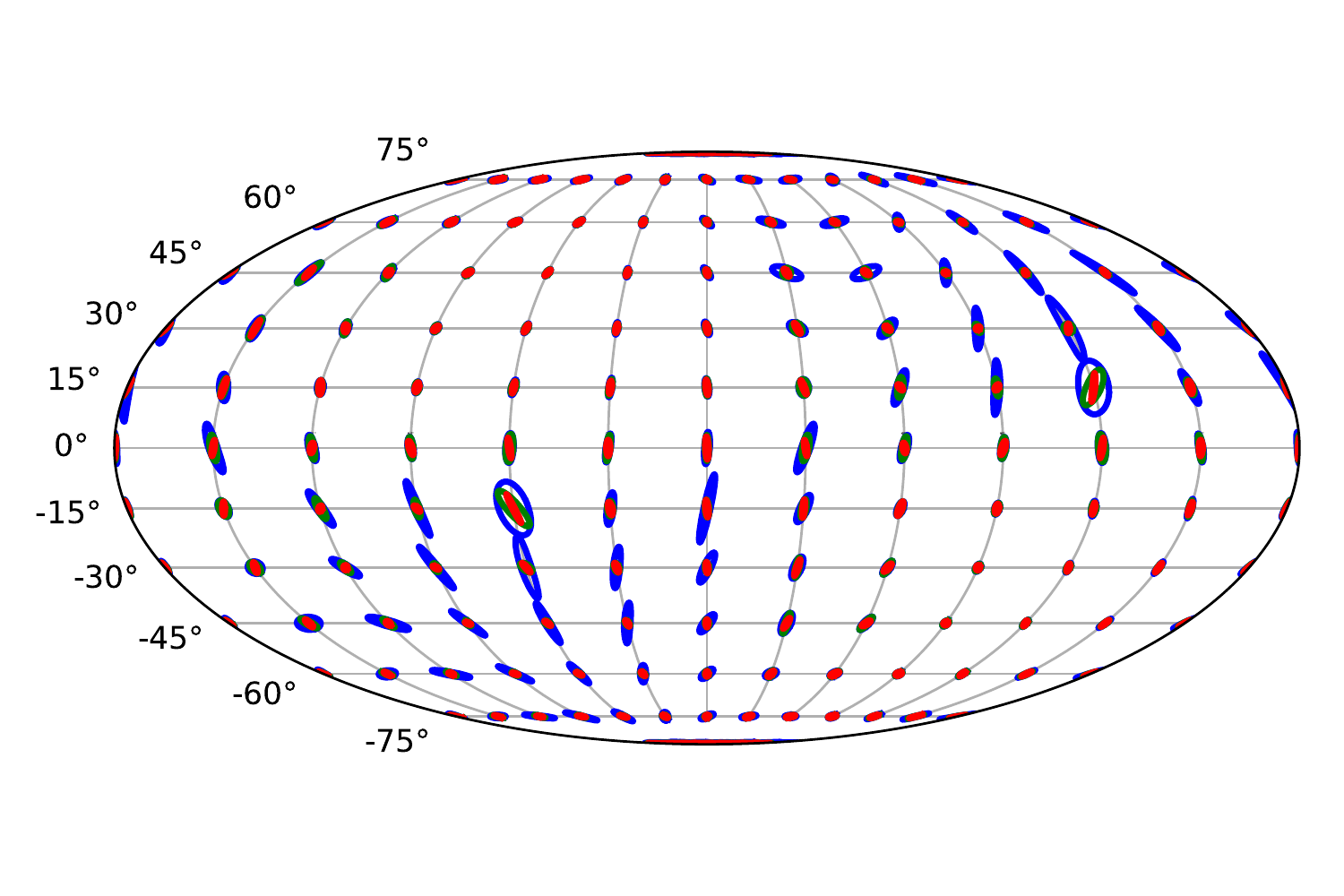}} &
\subfloat[]{\includegraphics[width=0.5\textwidth]{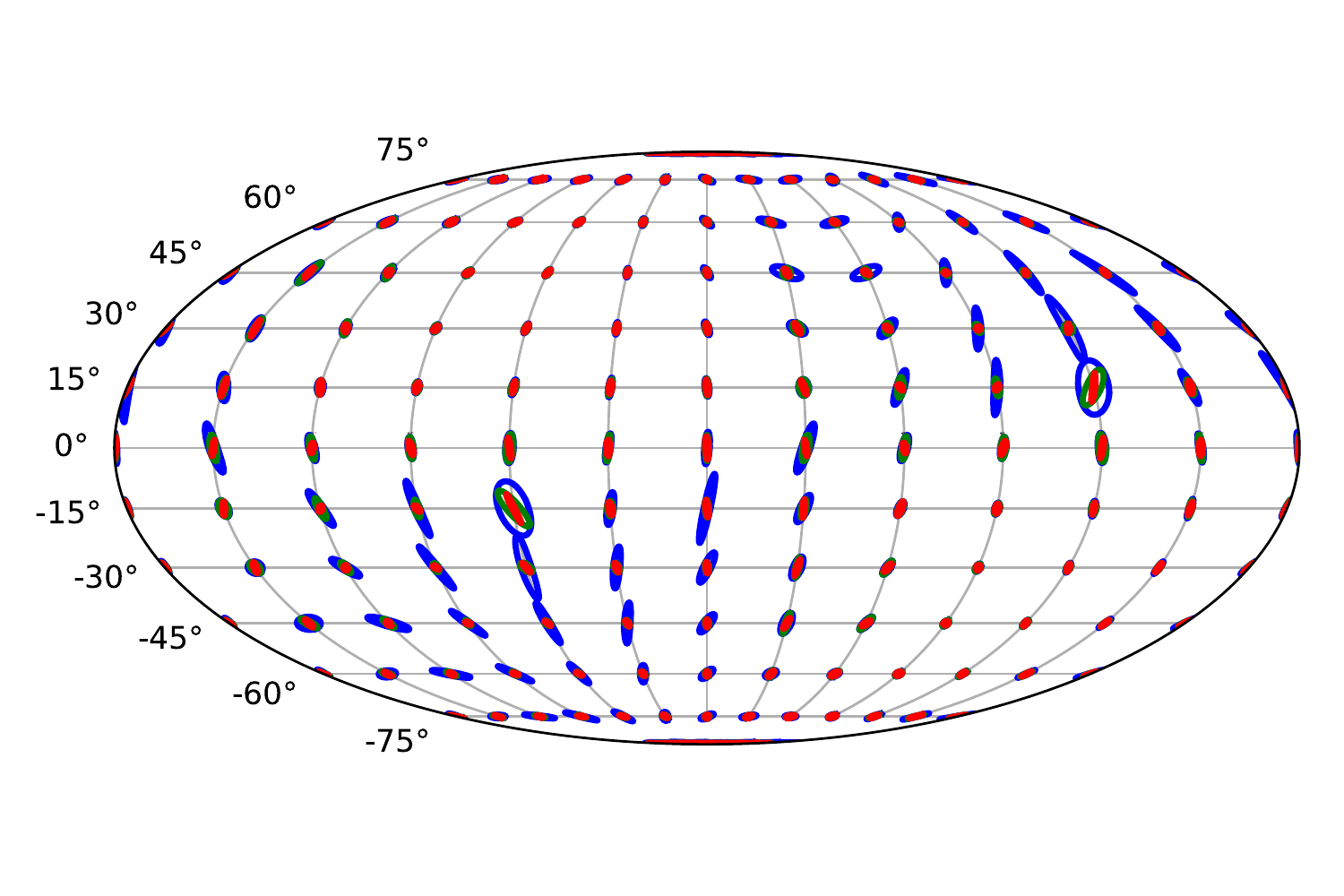}}
\end{tabular}
\caption{Error ellipses of the source localization in the GW170817-like BNS case,
with a total mass $2.74 M_\odot$, thus a chirp mass $\mathcal{M}=1.188 M_\odot$,
and the luminosity distance $D_{Le}=40$Mpc, for (a) $e_{0.0}$ and (b) $e_{0.4}$.
The blue, green, and red ellipses correspond to the LHV, LHVK, and LHVKI cases,
respectively.
The major and minor axes of the ellipses are both doubled, otherwise they are
too small to be recognized.}
\label{fig:BNS-ellipse}
\end{figure*}

\begin{table}[htbp]
\caption{The best/worst accuracy of source localization and the corresponding
sky location for the GW151226-like BBH case.}
\resizebox{0.49\textwidth}{\height}{
\linespread{1.3}\selectfont
\begin{tabular}{c|c|c|c} 
\toprule[0.5pt]
Network&$e_0$&($\theta_e$, $\phi_e$)&$\Delta\Omega$\\ \hline
\multirow{2}{*}{LHV}&0.0&(2.53, 2.18)/(1.31, 2.01)
&$8.56\times 10^{-5}$/$3.64\times 10^{-3}$\\ 
&0.4&(0.61, 5.32)/(1.31, 2.01)&$7.97\times 10^{-5}$/$3.63\times 10^{-3}$
\\ \hline
\multirow{2}{*}{LHVK}&0.0&(2.97, 4.89)/(1.75, 5.15)
&$5.17\times 10^{-5}$/$1.46\times 10^{-3}$\\ 
&0.4&(2.97, 4.89)/(1.75, 5.15)&$4.92\times 10^{-5}$/$1.39\times 10^{-3}$
\\ \hline
\multirow{2}{*}{LHVKI}&0.0&(2.71, 4.97)/(1.75, 5.15)
&$3.21\times 10^{-5}$/$2.77\times 10^{-4}$\\ 
&0.4&(0.44, 1.83)/(1.40, 2.01)&$3.01\times 10^{-5}$/$2.61\times 10^{-4}$\\ 
\bottomrule[0.5pt]
\end{tabular}}     
\label{tab:22BBH_Omega}
\end{table}

Similarly to in Fig.~\ref{fig:100-space}, we plot the distribution of
$\Delta\Omega$ for the GW151226-like BBH case in Fig.~\ref{fig:22-space} for
$e_{0.0}$ (the upper panels) and $e_{0.4}$ (the lower panels).
Compared with Fig.~\ref{fig:100-space}, we can see that the overall distribution
behavior is similar to the one in the big BBH case.
However, with the overall smaller $\Delta\Omega$ in the plots,
the accuracy of the source localization for the GW151226-like BBH
case is better than that for the big BBH case.
This is because there is more gravitational wave signal falling within the
most sensitive frequency band of the detectors in the GW151226-like BBH case
than in the big BBH case.
We can see this clearly in Fig.~\ref{fig:ASD_Flso}.
We postpone the discussion about the related issues until
Sec.~\ref{sec:discussion}.
Table \ref{tab:22BBH_Omega} shows the best and worst $\Delta\Omega$'s and
the corresponding sky locations for the GW151226-like BBH case.

In Fig.~\ref{fig:22-space-ratio-net}, we compare $\Delta\Omega$ among the
three networks by plotting the ratio of $\Delta\Omega$ among them for each
($\theta_e$, $\phi_e$).
The distribution with respect to $(\theta_e,\phi_e)$ is similar to the one in
Fig.~\ref{fig:100-space-ratio-net}.
The maximal value of the ratio is larger than the one in the big BBH case.
However, the minimal value of the ratio turns out to be smaller than the one
in the big BBH case.
We will find the trend clearer in the GW170817-like BNS case in the next 
subsection.
We show the best and worst improvement factors among the networks 
in the third row of Table \ref{tab:improve_det} for this case.

In Fig.~\ref{fig:22-space-ratio-e0}, we show the improvement factor
$\displaystyle\frac{\Delta\Omega_{0.0}}{\Delta\Omega_{0.4}}$ for each
($\theta_e$, $\phi_e$).
The distribution is also similar to the one in
Fig.~\ref{fig:100-space-ratio-e0}, but the range of the improvement factor is
quite a bit smaller than in the big BBH case.
The best and worst improvement factors between the two eccentricities
for the three networks are given in the third row of Table.~\ref{tab:improve_e}
for this case.
We can see that the improvement factors are at most only 1.10 in the best case.
And for the LHV network, the worst improvement factor is less than 1;
this means that $\Delta\Omega$ increases at some regions when the initial
eccentricity $e_0$ increases from 0.0 to 0.4.

Figure \ref{fig:22-histograms} shows the statistics of $\Delta\Omega$ by using
Monte Carlo samplings.
The profiles of the plots are similar to those in
Fig~\ref{fig:100-histograms}, 
other than that $\Delta\Omega$ in this case is smaller than in the big BBH case.
Based on a comparison between the median value in each eccentricity case with the one in $e_{0.0}$, the accuracy of the source localization improves 1.21
times better with $e_{0.1}$, 1.31 times better with $e_{0.2}$, 1.32 times better
with $e_{0.3}$, and 1.4 times better with $e_{0.4}$ for the LHV case.
For the LHVK case, the source localization accuracy improves 1.16 times better
with $e_{0.1}$, 1.21 times better with $e_{0.2}$, 1.27 times better with
$e_{0.3}$, and 1.29 times better with $e_{0.4}$.
It improves 1.13 times better with $e_{0.1}$, 1.19 better times with $e_{0.2}$,
1.22 times better with $e_{0.3}$, and 1.25 times better with $e_{0.4}$ for the
LHVKI case.
From the above result, we can see that the improvement from the eccentricity
increasing in the GW151226-like BBH case is smaller than in the big BBH case,
no matter which detector network we use. 

\begin{figure*}[htbp]
\begin{tabular}{ccc}
\includegraphics[width=0.33\textwidth]{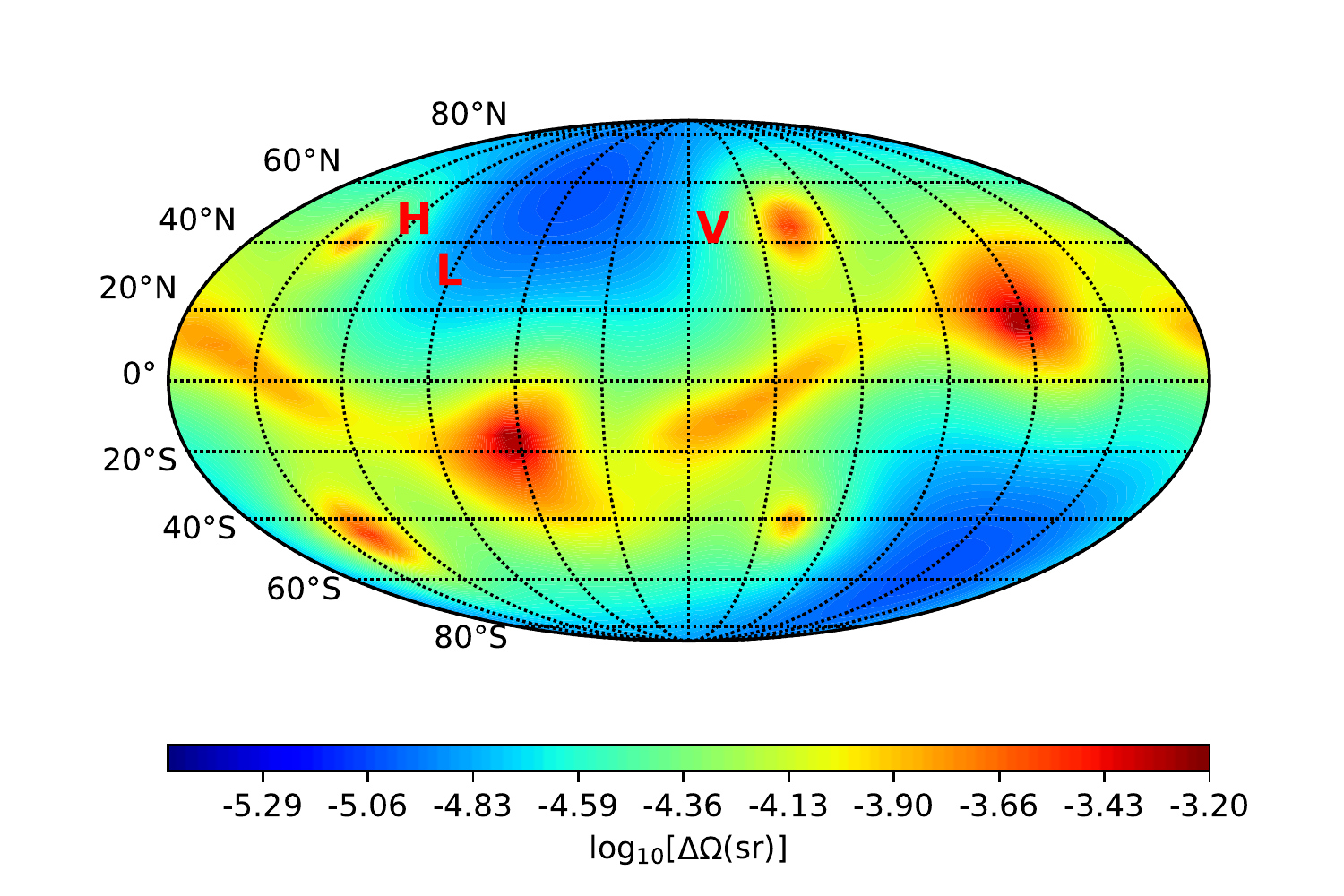}
\label{fig:2p74-space-0-LHV} &
\includegraphics[width=0.33\textwidth]{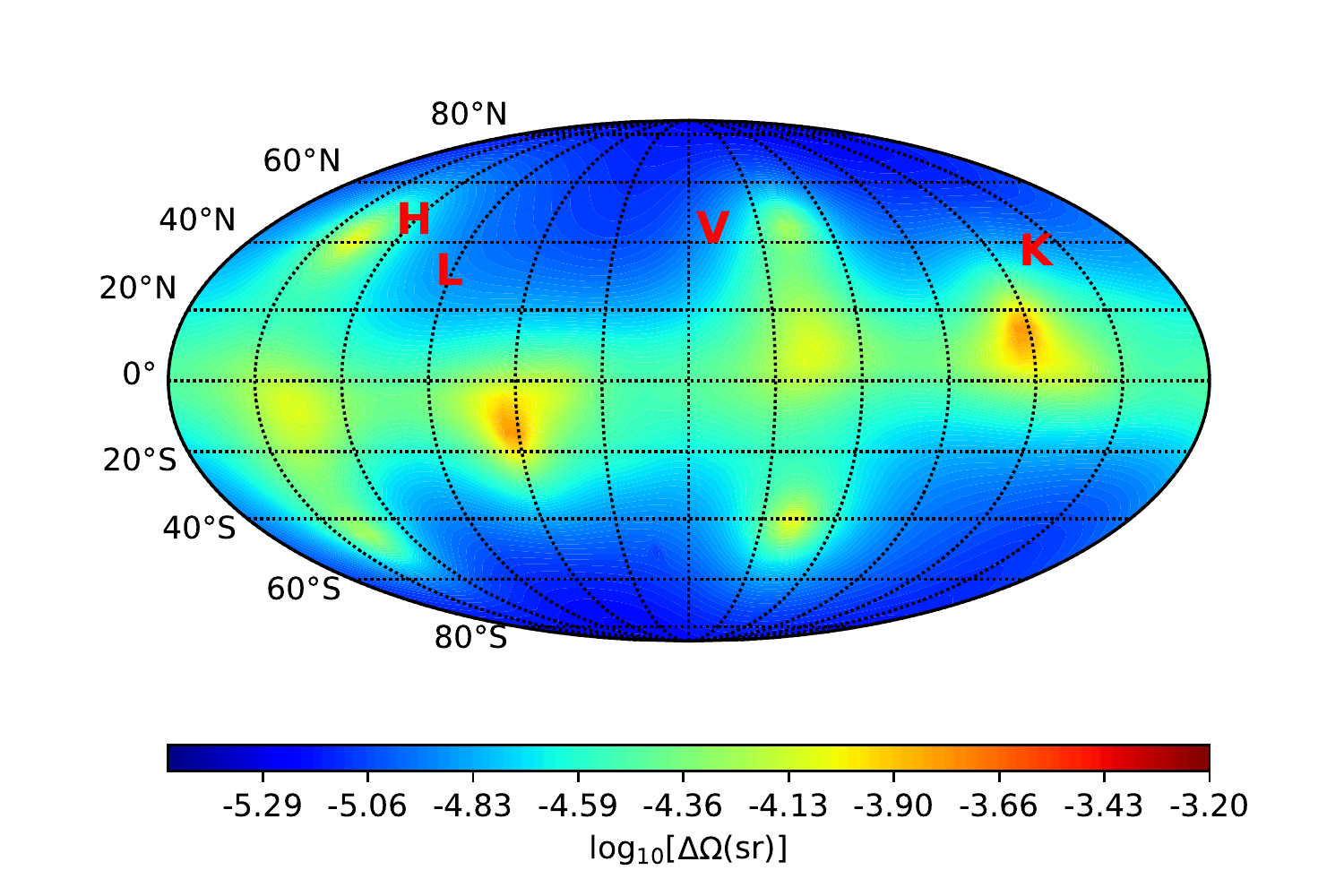}
\label{fig:2p74-space-0-LHVK} &
\includegraphics[width=0.33\textwidth]{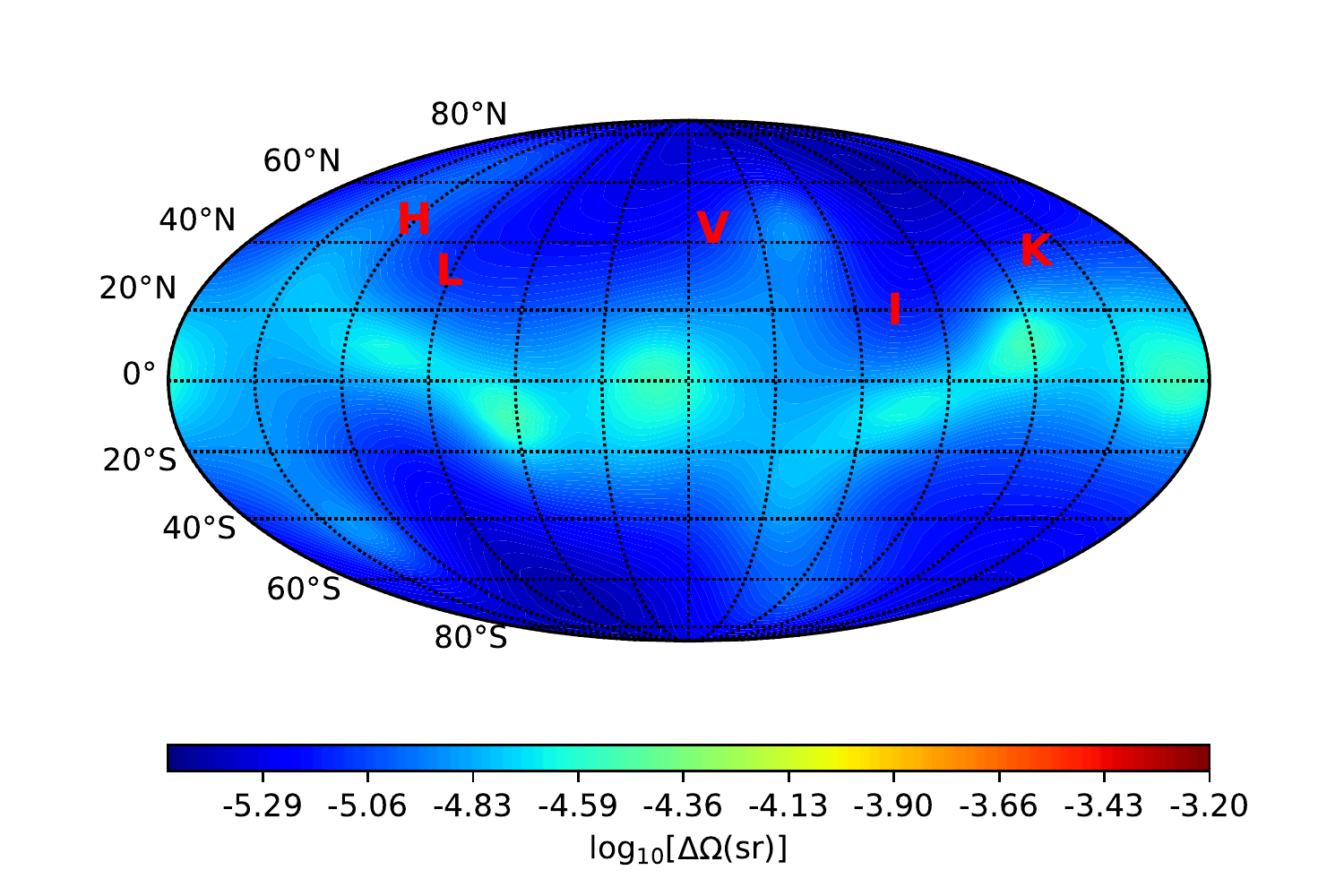}
\label{fig:2p74-space-0-LHVKI} \\
\includegraphics[width=0.33\textwidth]{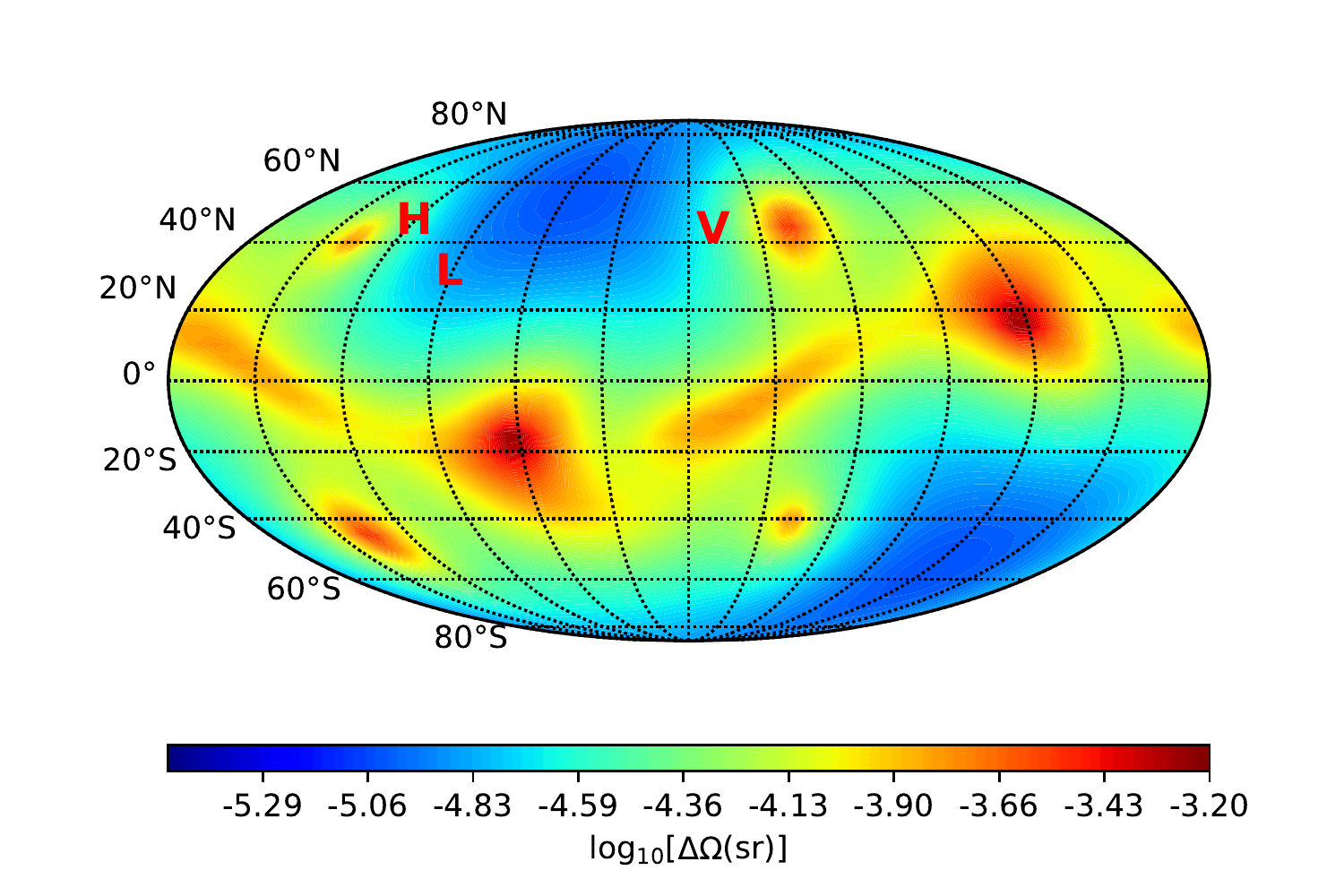}
\label{fig:2p74-space-0.4-LHV} &
\includegraphics[width=0.33\textwidth]{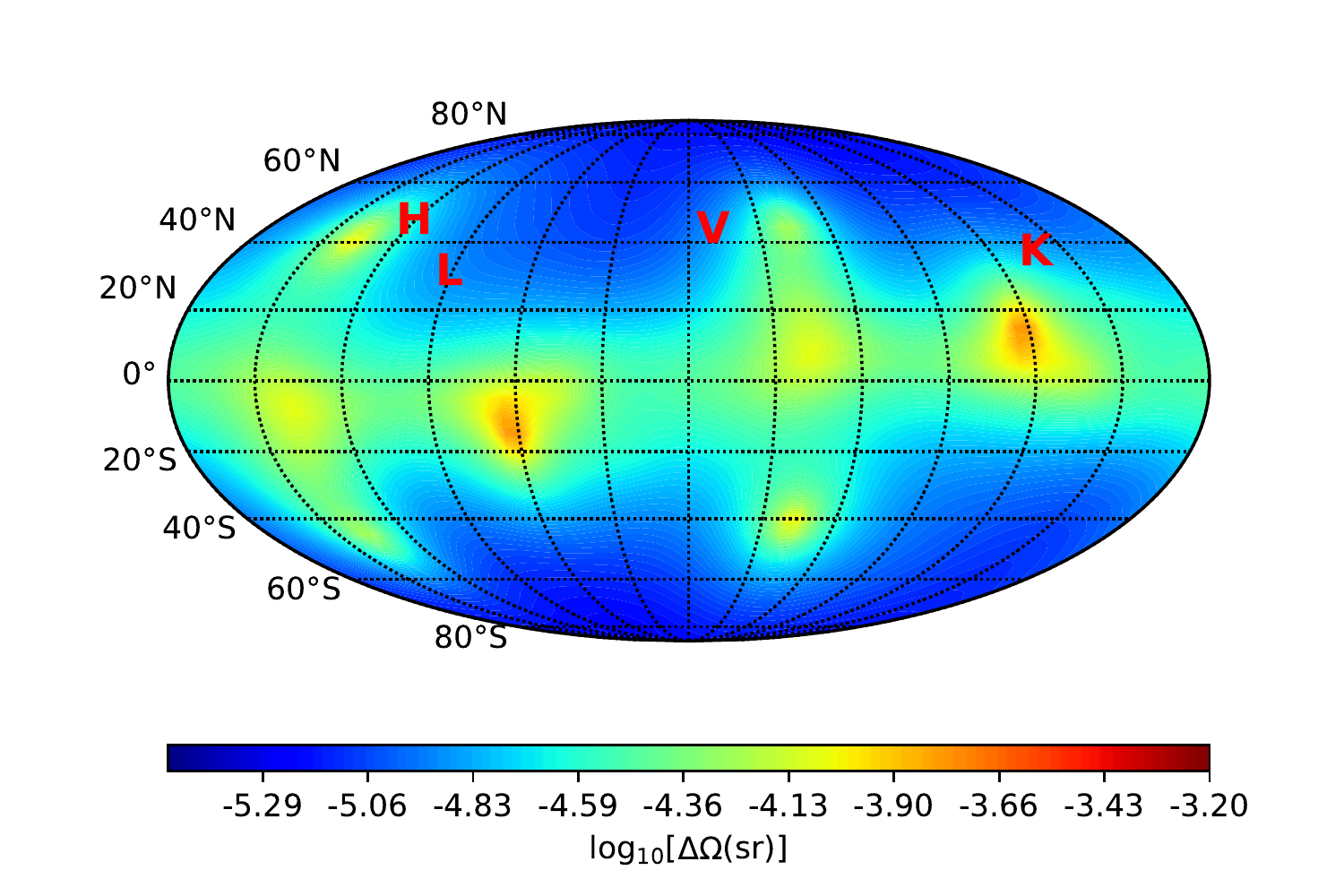}
\label{fig:2p74-space-0.4-LHVK} &
\includegraphics[width=0.33\textwidth]{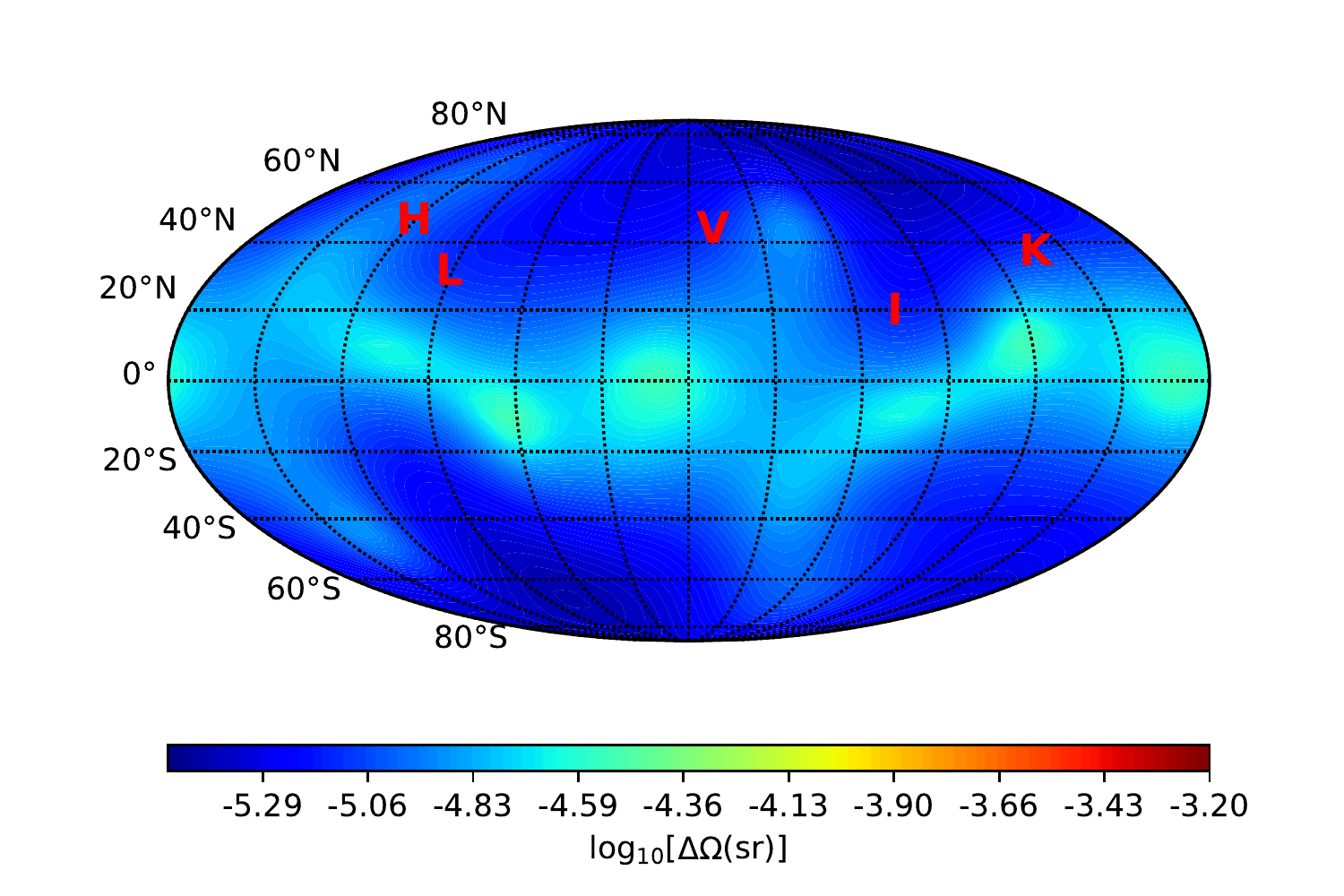}
\label{fig:2p74-space-0.4-LHVKI}
\end{tabular}
\caption{Estimated error $\Delta\Omega$ of the source localization for the
GW170817-like BNS case.
The panels in the upper and the lower rows correspond to the eccentricities
$e_{0.0}$ and $e_{0.4}$, respectively.
We show the $\Delta\Omega$'s for the LHV, LHVK, and LHVKI cases in the left,
middle, and right columns, respectively.}
\label{fig:2p74-space}
\end{figure*}

\begin{figure*}[htbp]
\begin{tabular}{ccc}
\includegraphics[width=0.33\textwidth]{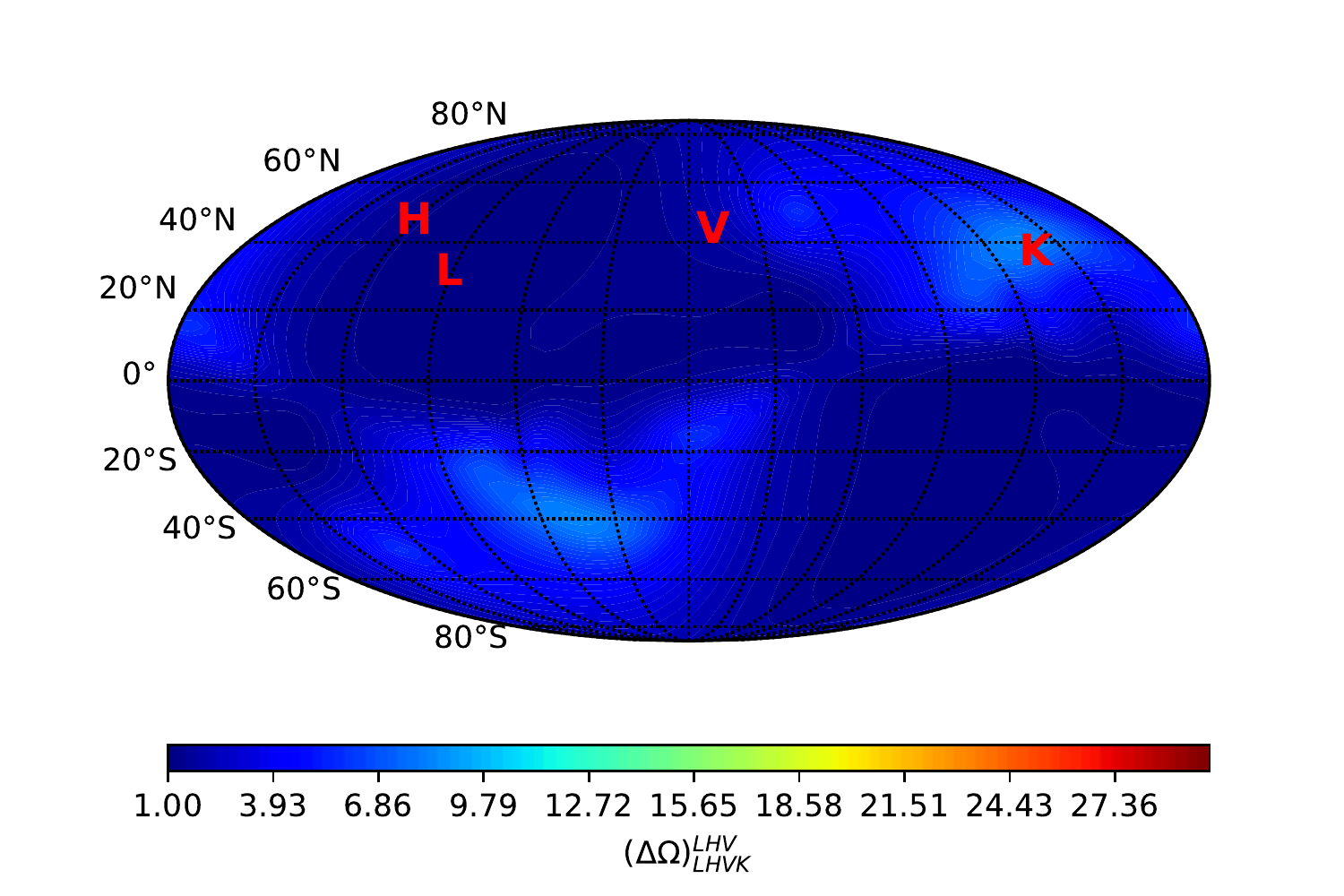}
\label{fig:2p74-space-0-3-4} &
\includegraphics[width=0.33\textwidth]{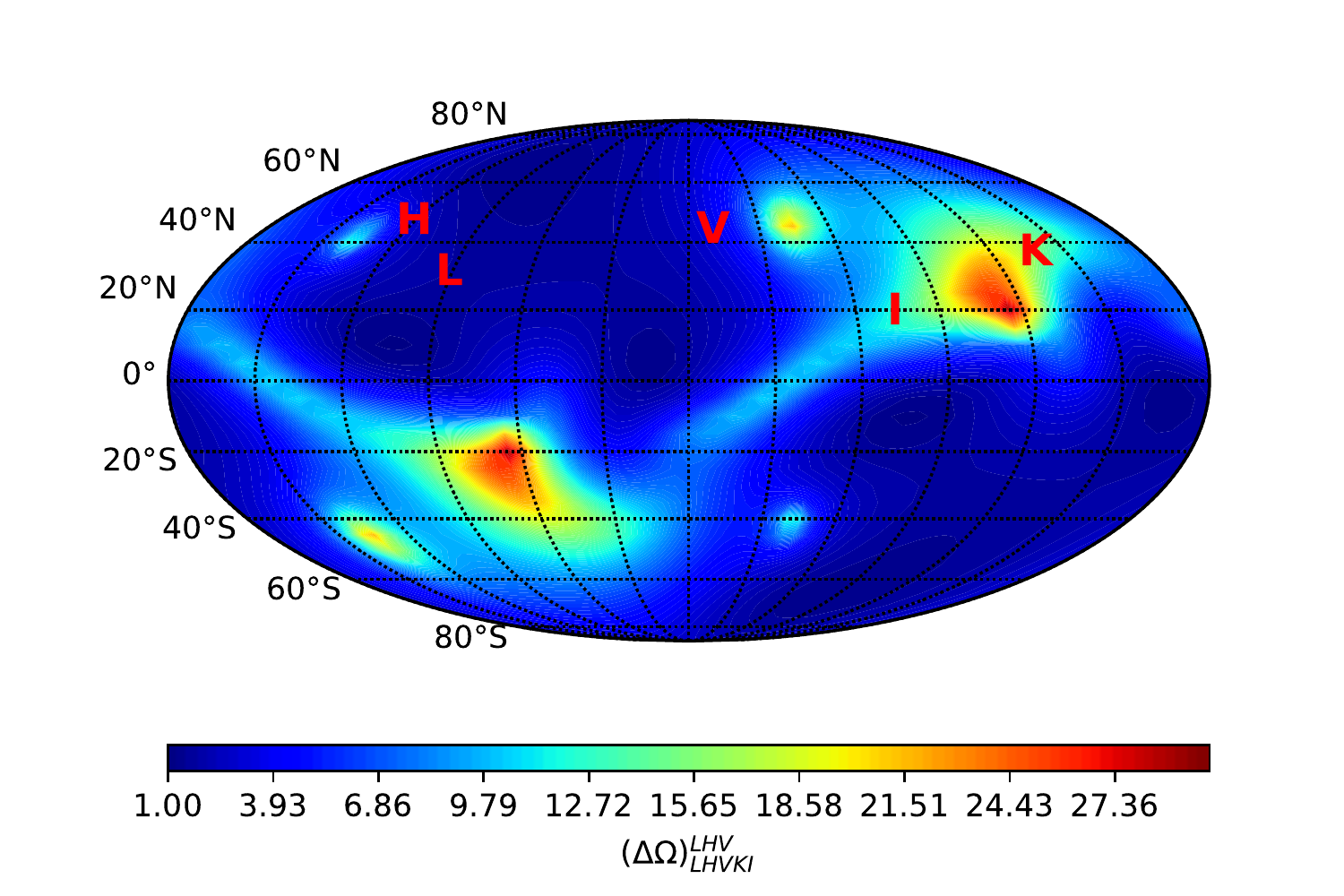}
\label{fig:2p74-space-0-3-5} &
\includegraphics[width=0.33\textwidth]{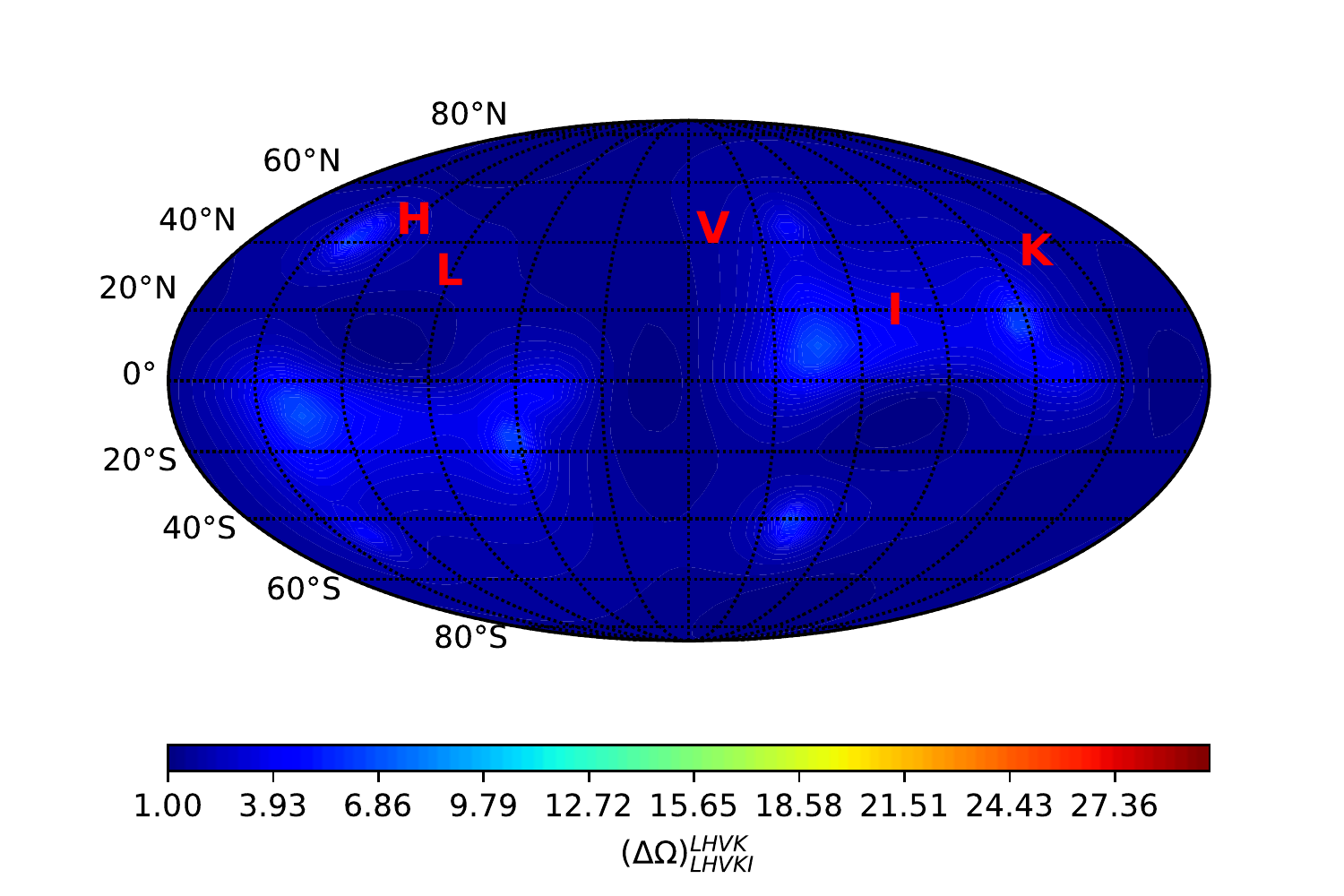}
\label{fig:2p74-space-0-4-5} \\
\includegraphics[width=0.33\textwidth]{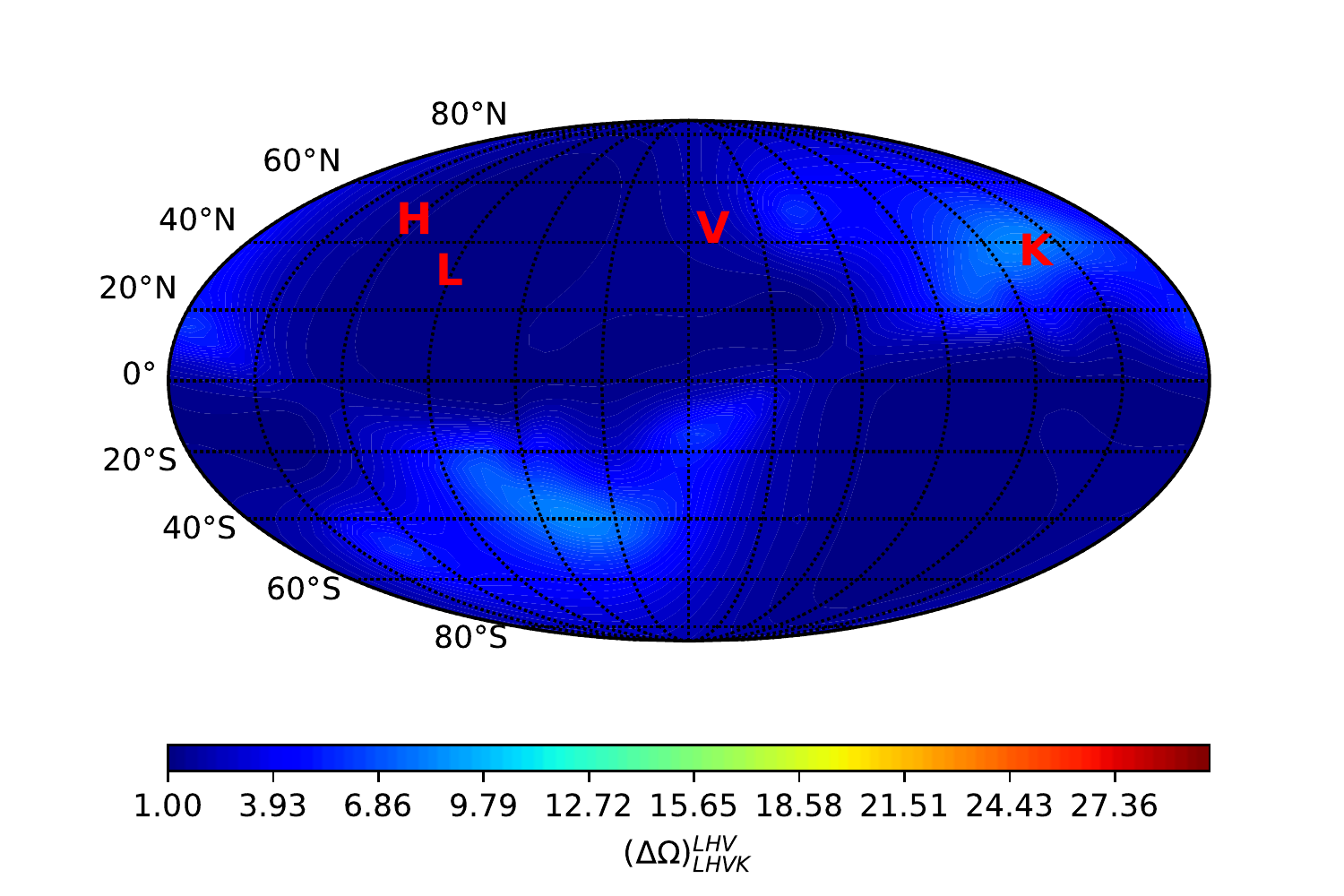}
\label{fig:2p74-space-0.4-3-4} &
\includegraphics[width=0.33\textwidth]{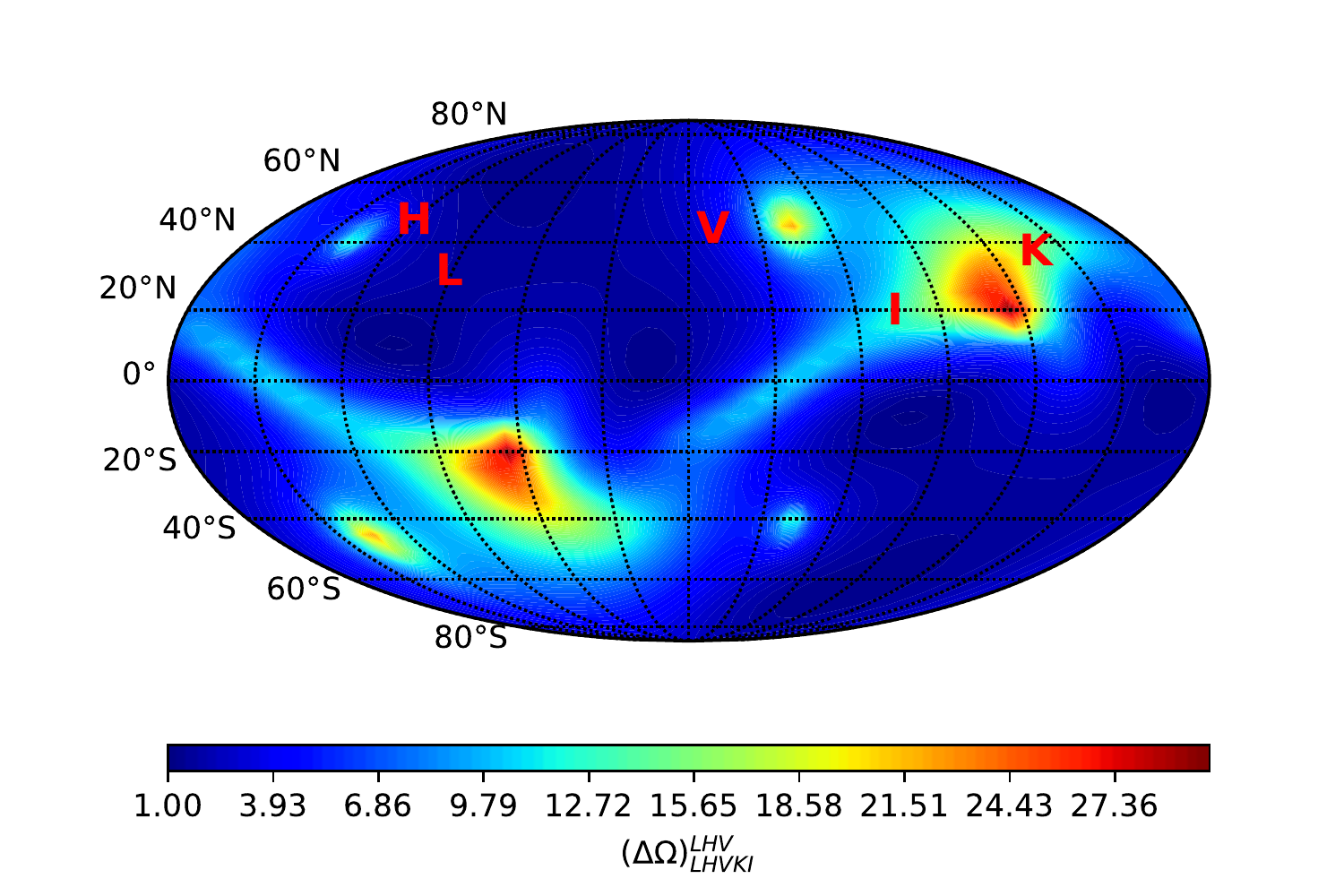}
\label{fig:2p74-space-0.4-3-5} &
\includegraphics[width=0.33\textwidth]{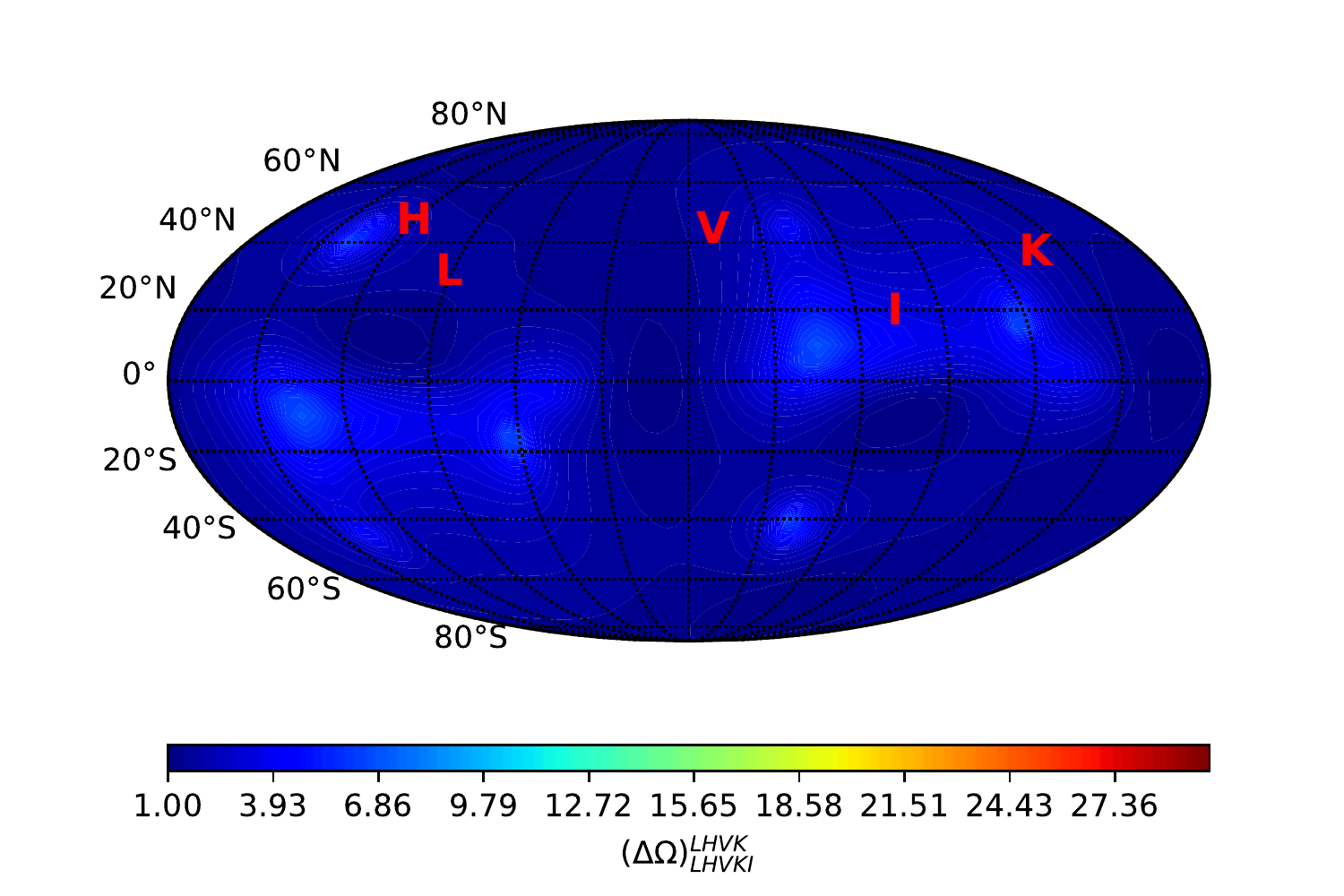}
\label{fig:2p74-space-0.4-4-5}
\end{tabular}
\caption{ Ratios of $\Delta\Omega$ among different networks for the
GW170817-like BNS case.
The panels in the upper and the lower rows correspond to the eccentricities
$e_{0.0}$ and $e_{0.4}$, respectively.
$\Delta\Omega^{\text{LHV}}_{\text{LHVK}}$,
$\Delta\Omega^{\text{LHV}}_{\text{LHVKI}}$, and
$\Delta\Omega^{\text{LHVK}}_{\text{LHVKI}}$ are shown in the left, middle,
and right columns, respectively.}
\label{fig:2p74-space-ratio-net}
\end{figure*}

\begin{figure}
\begin{tabular}{c}
\includegraphics[width=0.49\textwidth]{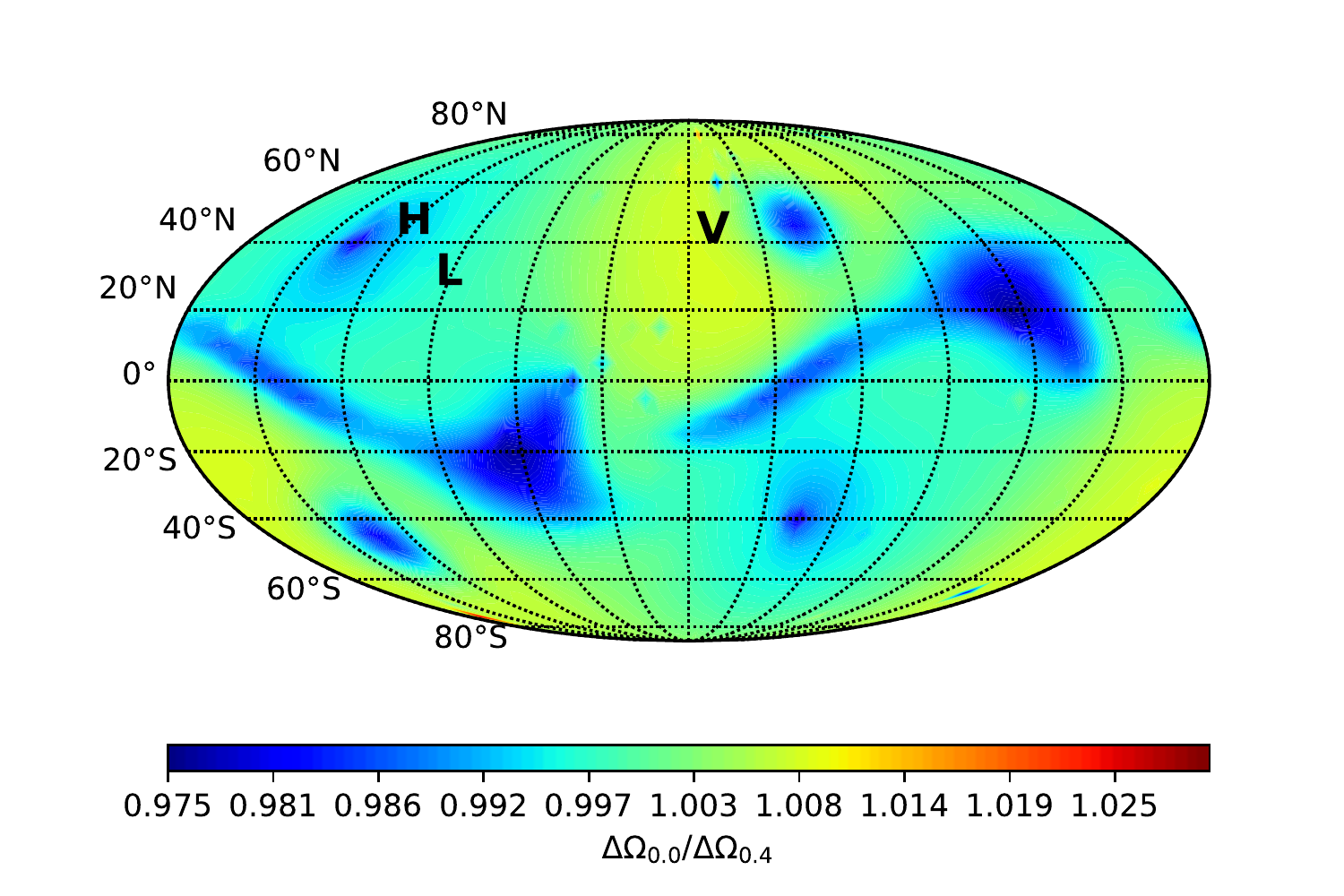}
\label{fig:2p74-space-ratio-LHV} \\
\includegraphics[width=0.49\textwidth]{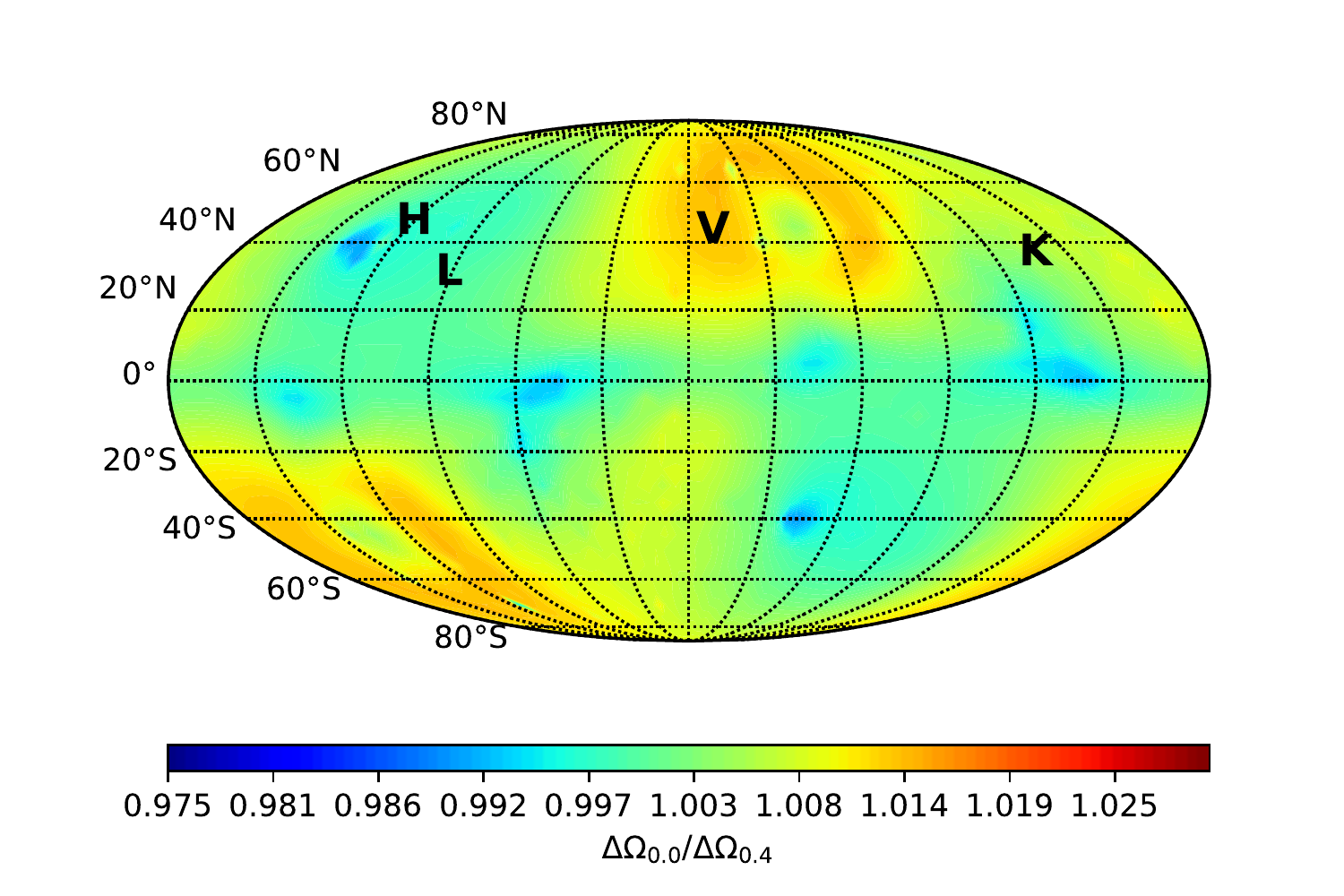}
\label{fig:2p74-space-ratio-LHVK} \\
\includegraphics[width=0.49\textwidth]{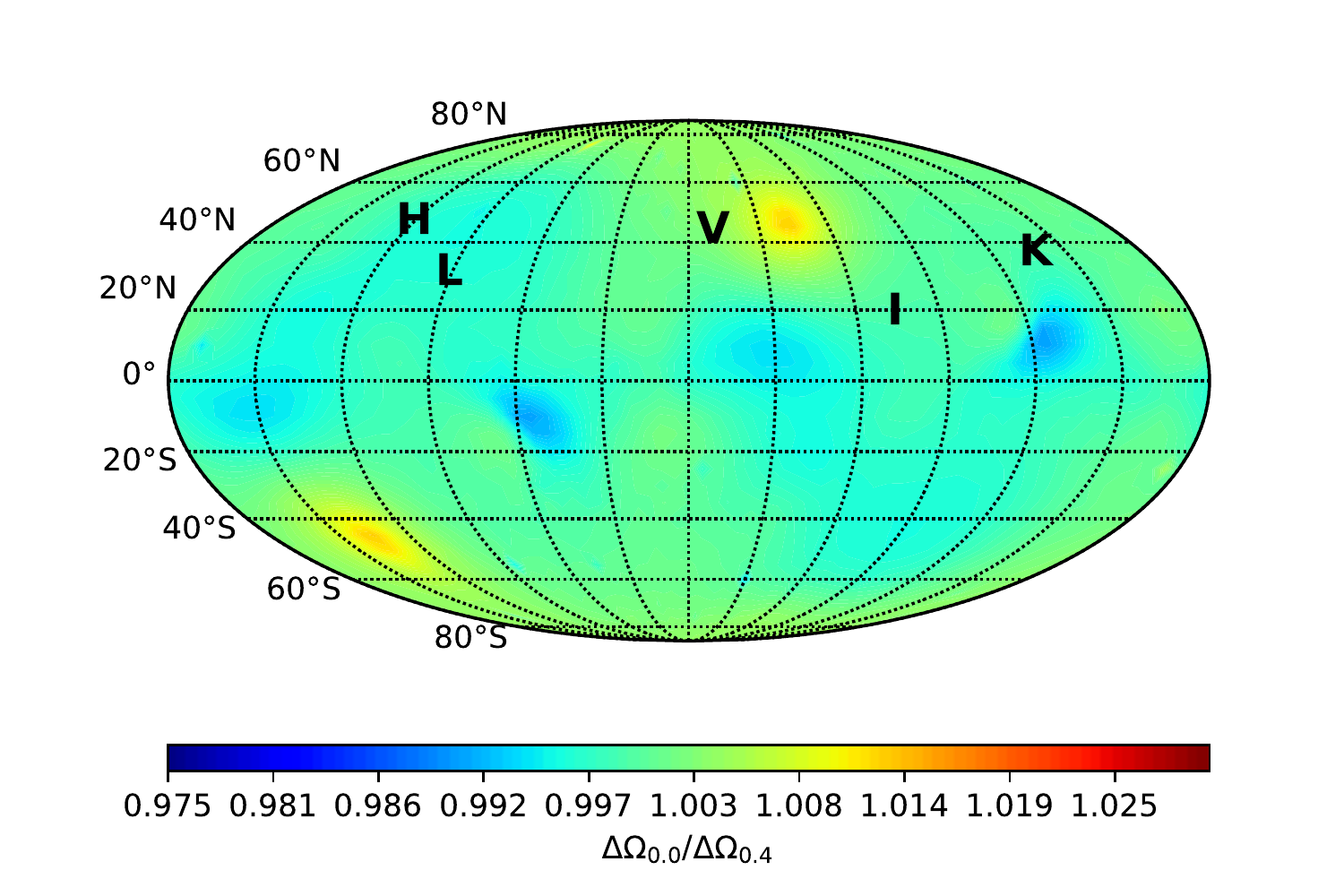}
\label{fig:2p74-space-ratio-LHVKI}
\end{tabular}
\caption{$\Delta\Omega^{0.0}_{0.4}$ for the GW170817-like BNS case.
The plots in the upper, middle, and lower panels correspond to the LHV, LHVK,
and LHVKI cases, respectively.}
\label{fig:2p74-space-ratio-e0}
\end{figure}

\begin{figure}
\begin{tabular}{c}
\includegraphics[width=0.49\textwidth]{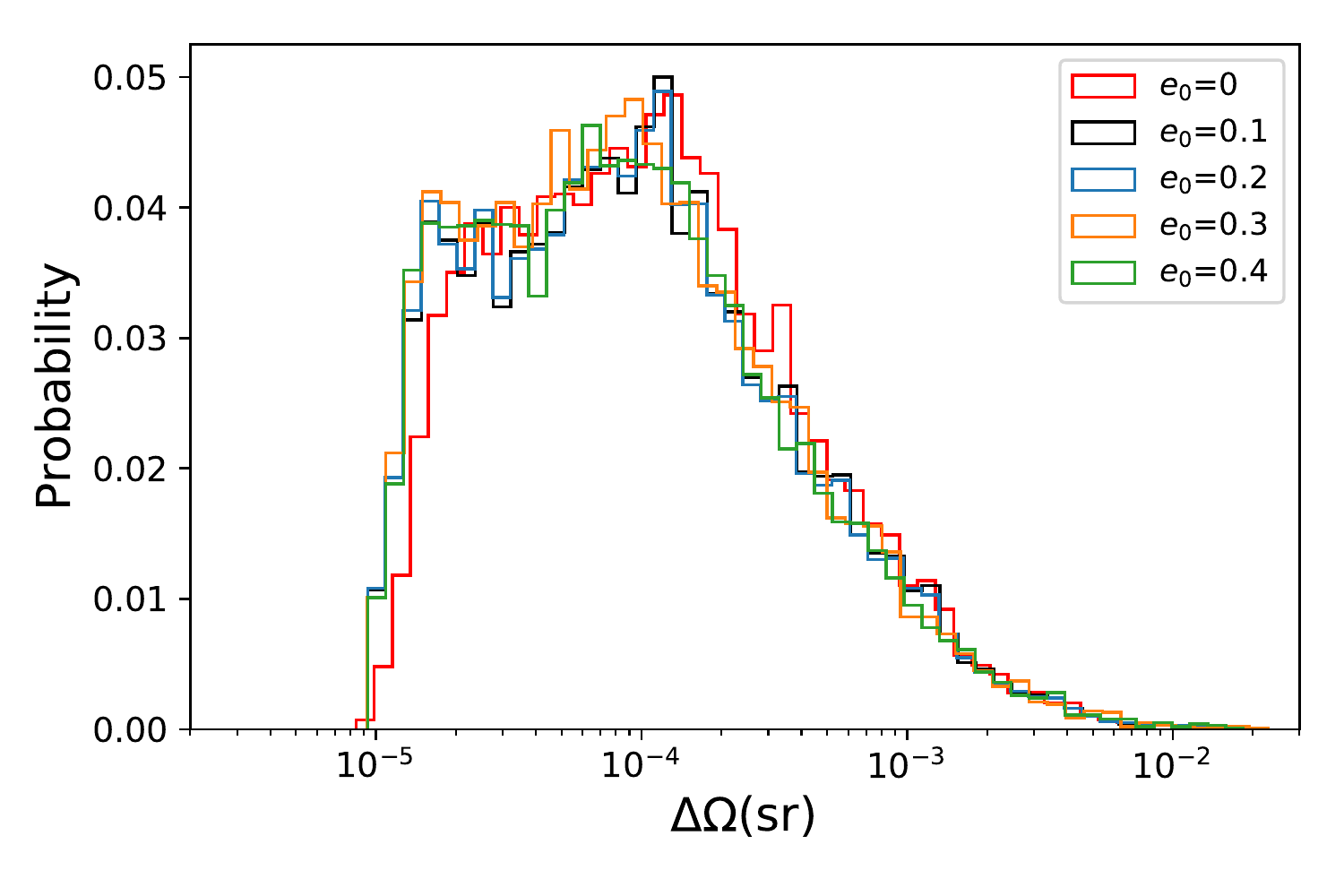} \\
\includegraphics[width=0.49\textwidth]{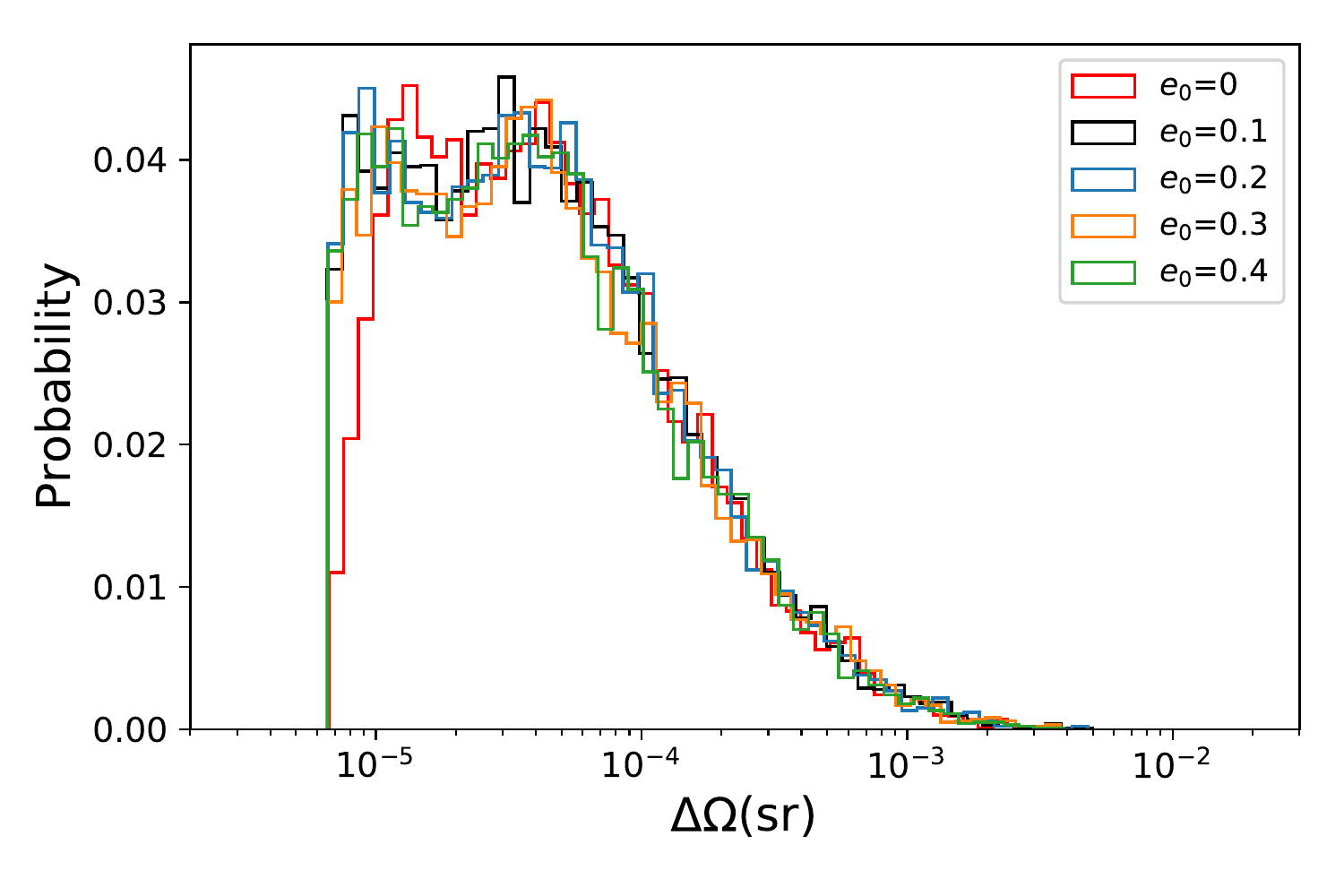} \\
\includegraphics[width=0.49\textwidth]{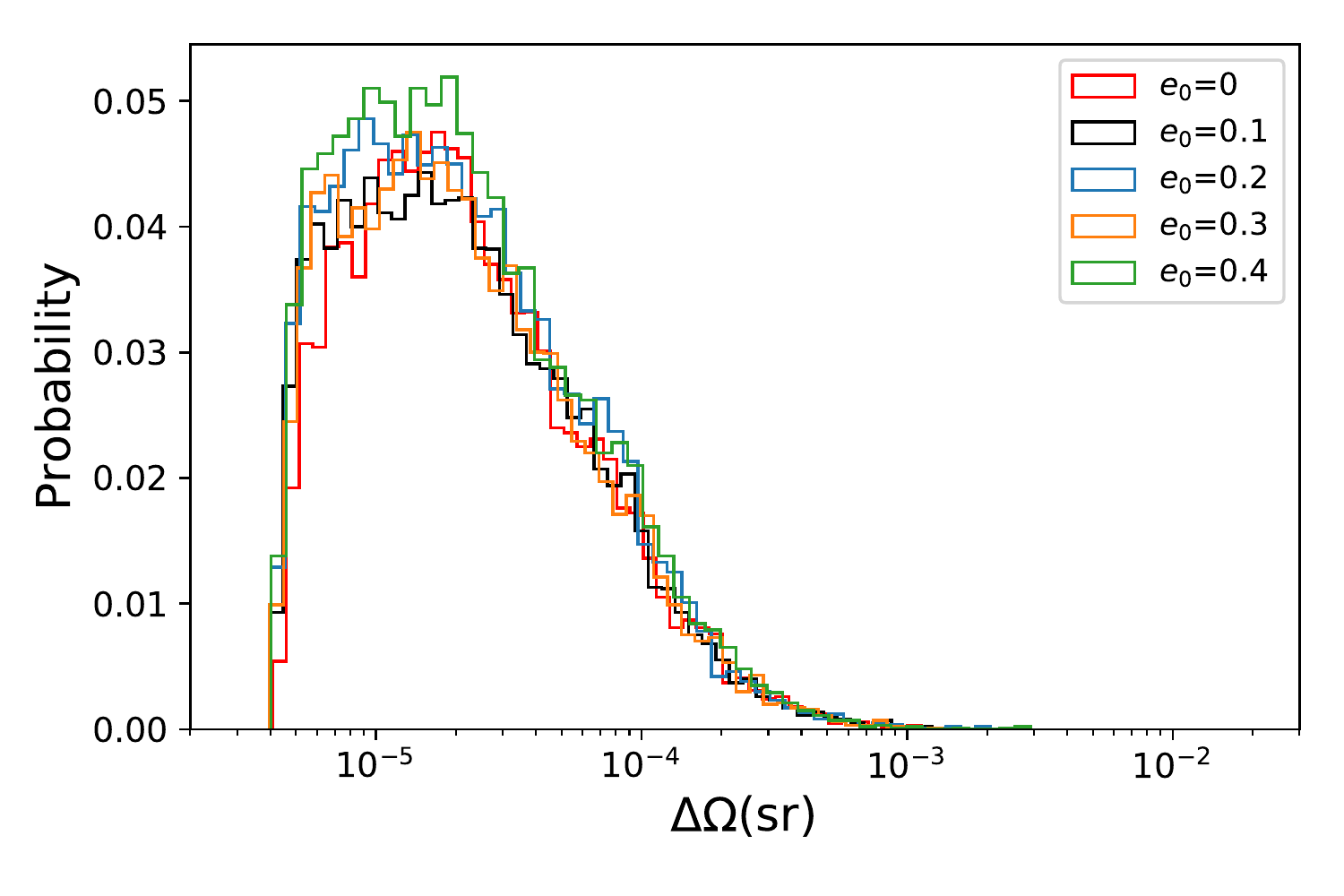}
\end{tabular}
\caption{Histograms of $\Delta\Omega$ for varying angle parameters $\iota_e$,
$\beta_e$, $\psi_e$, $\theta_e$ and $\phi_e$ with $10^4$ Monte Carlo samples,
for the GW170817-like BNS case.
The plots in the upper, middle, and lower panels correspond to the LHV, LHVK,
and LHVKI cases, respectively.}
\label{fig:BNS-histograms}
\end{figure}

\begin{table}
\caption{The best/worst accuracy of source localization and the corresponding
sky location for the GW170817-like BNS case.}
\resizebox{0.49\textwidth}{\height}{
\linespread{1.3}\selectfont
\begin{tabular}{c|c|c|c} 
\toprule[0.5pt]
Network&$e_0$&($\theta_e$, $\phi_e$)&$\Delta\Omega$\\ \hline
\multirow{2}{*}{LHV}&0.0&(2.53, 2.18)/(1.31, 2.01)
&$9.26\times 10^{-6}$/$5.34\times 10^{-4}$ \\ 
&0.4&(2.53, 2.18)/(1.83, 5.15)&$9.24\times 10^{-6}$/$5.46\times 10^{-4}$
\\ \hline
\multirow{2}{*}{LHVK}&0.0&(2.97, 5.06)/(1.83, 5.24)
&$6.29\times 10^{-6}$/$1.75\times 10^{-4}$ \\ 
&0.4&(2.97, 5.06)/(1.83, 5.24)&$6.22\times 10^{-6}$/$1.76\times 10^{-4}$
\\ \hline
\multirow{2}{*}{LHVKI}&0.0&(0.35, 1.83)/(1.75, 5.15)
&$3.90\times 10^{-6}$/$3.15\times 10^{-5}$\\ 
&0.4&(0.35, 1.83)/(1.75, 5.15)&$3.89\times 10^{-6}$/$3.16\times 10^{-5}$\\ 
\bottomrule[0.5pt]
\end{tabular}}     
\label{tab:BNS_Omega}
\end{table}

\begin{figure}[htbp]
\begin{tabular}{c}
\includegraphics[width=0.49\textwidth]{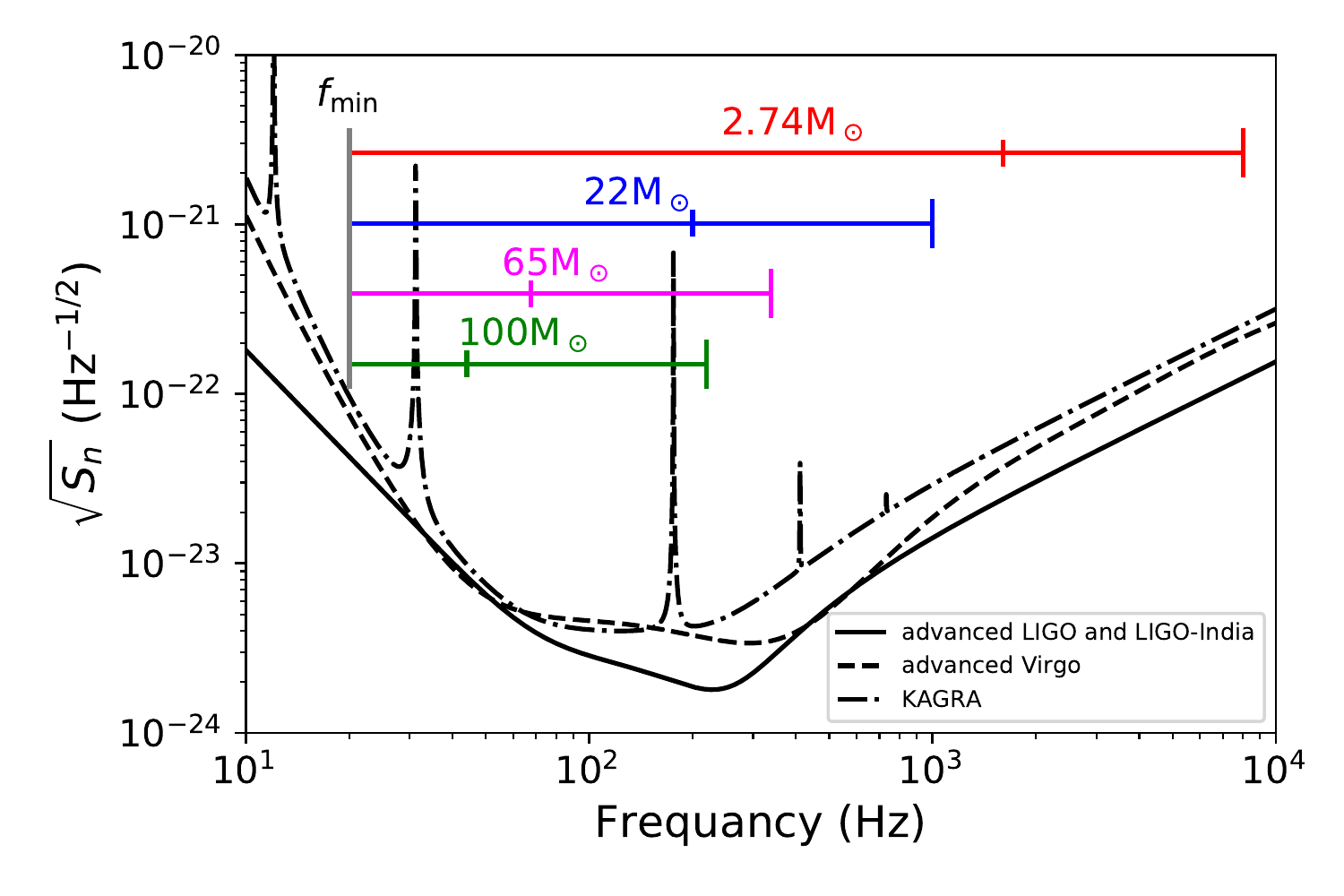}
\end{tabular}
\caption{Domains of the frequency consideration for different binary compact
object systems.
The total masses of the systems are $2.74 M_\odot$, $22 M_\odot$, $65 M_\odot$,
and $100 M_\odot$.
The black lines are the sensitivity curves for the detectors.
The colorful horizontal lines are the frequency domain considered in these
cases.
The mark on each horizontal line indicates the frequency of the lowest mode
($\ell=2$) gravitational wave, $2 F_{\rm LSO}$, during the last stable orbit
for each case.}
\label{fig:ASD_Flso}
\end{figure}

\subsection{GW170817-like BNS case}
In this subsection, we consider a binary neutron star system.
We choose a GW1170817-like BNS with a total mass $2.74 M_\odot$, thus a chirp
mass $\mathcal{M}=1.188 M_\odot$, and a luminosity distance $D_{Le}=40$Mpc.
We fix the parameters $\eta=(\mathcal{M}/M)^{5/3}=0.2484$,
and $t_{ce}=\phi_c=\iota_e=\beta_e=\psi_e=0$, while varying $\theta_e$, $\phi_e$,
and $e_0$ to investigate the resulting accuracy of the source localization.
We plot the 5$\sigma$ error region ellipses in Fig.~\ref{fig:BNS-ellipse}.
However, we double the magnitude of the major and minor axes of the ellipses;
otherwise, they are too small to be recognized (due to the shorter luminosity distance).
Similar to the two BBH cases aforementioned, we can also see that the more
detectors one uses the better the accuracy of the source localization obtained is.
The accuracy of the source localization is in general raised by increasing the
initial eccentricity.
However, similar to the GW151226-like BBH case, the improvement from 
increasing the initial eccentricity is quite negligible.
We can see this phenomenon more clearly in the following context.

We plot the distribution of $\Delta\Omega$ for the GW170817-like BNS in
Fig.~\ref{fig:2p74-space} for $e_{0.0}$ and $e_{0.4}$.
The overall distribution behavior is similar to the previous two BBH cases
for both the initial eccentricities.
And the accuracy of the source localization is better than that for both
the BBH cases, mainly due to a smaller luminosity distance.
We list the best and worst $\Delta\Omega$'s and the corresponding sky
localizations for this case in Table ~\ref{tab:BNS_Omega}.

The ratios of $\Delta\Omega$ among the three networks with respect to
($\theta_e$, $\phi_e$) in Fig.~\ref{fig:2p74-space-ratio-net} are also similar
to the ones in Fig.~\ref{fig:100-space-ratio-net} and in
Fig.~\ref{fig:22-space-ratio-net}.
From the results in the last row of Table \ref{tab:improve_det},
one can find that the accuracy of source localization is still improved
significantly by adding more detectors into the network, for this case.

In Fig.~\ref{fig:2p74-space-ratio-e0}, we show the improvement factor
$\displaystyle\frac{\Delta\Omega_{0.0}}{\Delta\Omega_{0.4}}$ for each
($\theta_e$, $\phi_e$).
From the results in the last row of Table \ref{tab:improve_e},
one can see that the improvement factors are less than 1.05 in the best case.
Moreover, the improvement factors in the worst cases are less than 1 for
the all three networks.
This says that higher eccentricity does not necessarily give more accuracy
on the source localization in the case of a compact binary with a small total
mass.
We will elaborate on this point in more detail in the next subsection.

Figure \ref{fig:BNS-histograms} shows the statistics of $\Delta\Omega$ by using
Monte Carlo samplings.
The profiles of the plots are similar to those in
 Figs.~\ref{fig:100-histograms} and \ref{fig:22-histograms}.
We find that the improvement on the localization accuracy by the eccentricity
is negligible.
Nevertheless, we can see from the figure that the localization accuracy is still
improved significantly by adding more detectors into the network.

\subsection{Discussion} \label{sec:discussion}
We found that the accuracy of the source localization is improved significantly
by raising the initial eccentricity for the binaries with a larger total mass,
but not much for the binaries with a smaller total mass.
For the GW151226-like BBH and the GW170817-like BBH cases,
the localization accuracy is even worse at some locations with a larger
initial eccentricity.
We suspect that the weakened improvement of the localization accuracy might be
related to the SNR with respect to the frequency domain involved in each case.
As in the earlier studies, the lowest frequency of a gravitational wave is
$2 F_{\rm LSO}$, and a binary system may excite higher-frequency modes with
nonvanishing eccentricity.
In the EPC model, the highest mode considered is up to $\ell=10$; 
thus, the frequency of the gravitational wave can reach $10 F_{\rm LSO}$.
If the initial eccentricity is larger,
the contribution from the higher harmonic modes to the waveform becomes larger.
In contrast, the contribution from the lower modes to the waveform becomes
smaller if compared with the gravitational waveform for the one with zero
eccentricity.
We have shown the frequency band ($f_{\rm min}$, $10 F_{\rm LSO}$)
of the binary systems with different total masses in Fig.~\ref{fig:ASD_Flso}.
In the figure,
the frequency range ($f_{\rm min}$, $2 F_{\rm LSO}$) is the frequency band
for the circular ($\ell=2$) waveform,
and the frequency range ($2 F_{\rm LSO}$, $10 F_{\rm LSO}$) could be reached
with nonzero eccentricity.

For the big BBH case with the total mass $100 M_\odot$,
the major frequency range considered falls on the most sensitive area of
the detectors' frequency band.
This allows one to extract the most that the detectors can offer about 
the information from the higher modes of the gravitational wave,
besides the $\ell=2$ mode, due to the nonzero eccentricity.
For the GW151226-like BBH case with the total mass $22 M_\odot$ and
the GW170817-like BBH case with the total mass $2.74 M_\odot$,
their major frequency ranges fall on the quite insensitive domain of the
detectors' frequency band.
In such cases, the loud high-frequency noise of the detectors ruins the
information from the higher modes of the gravitational wave.
Even worse, the $\ell=2$ mode wave from such systems with nonzero eccentricity
could be weakened, compared with the zero-eccentricity wave,
because the total energy has also to be partitioned to the higher modes. 
Therefore, the SNR, and thus the localization accuracy, is improved strongly
with nonzero eccentricity in the big BBH case;
meanwhile, the improvement of the localization accuracy becomes negligible,
even weakened, for the lower-total-mass BBH and BNS cases.

\begin{figure}[htbp]
\begin{tabular}{c}
\includegraphics[width=0.49\textwidth]{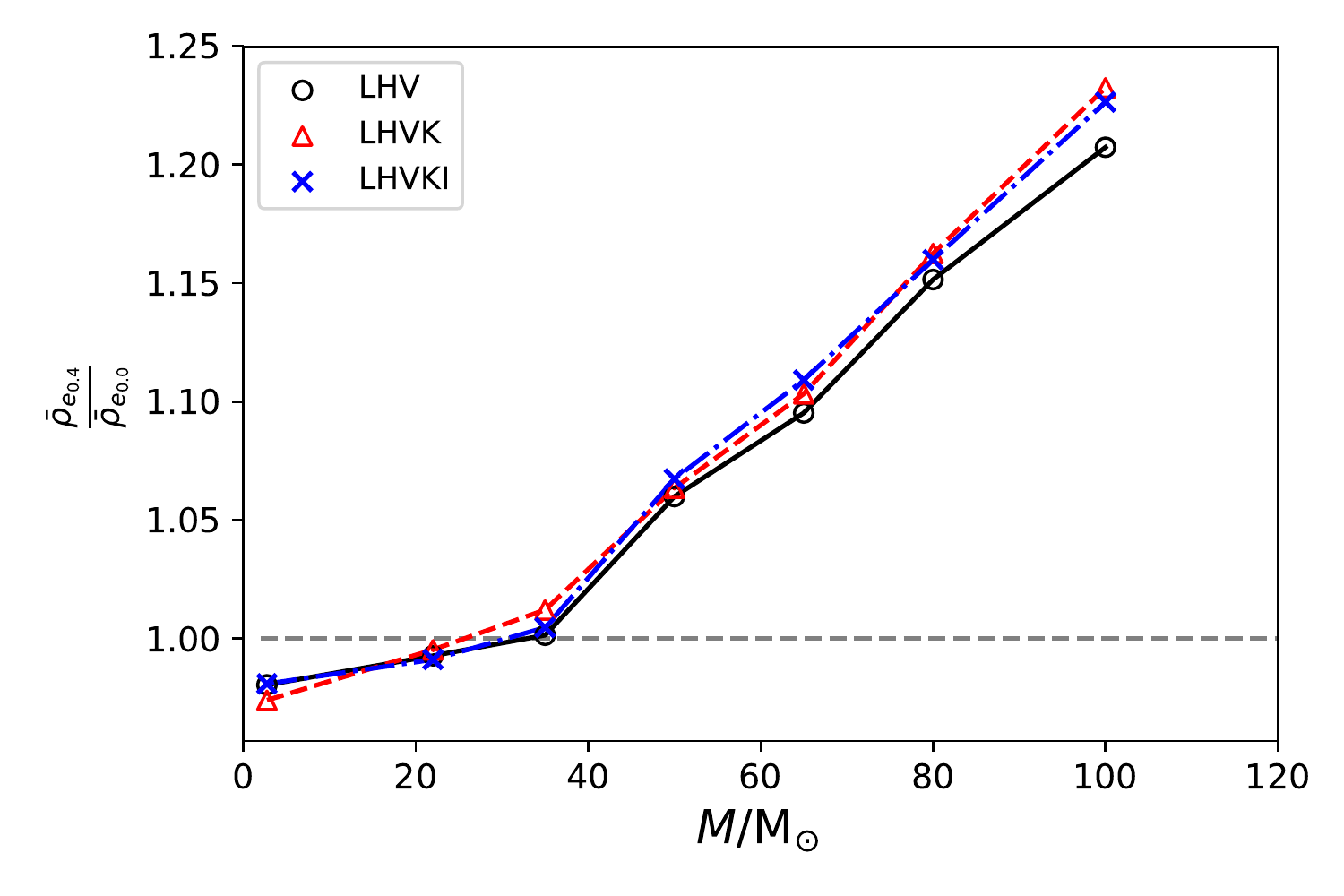}
\end{tabular}
\caption{The relative difference of the averaged SNR
$\displaystyle\frac{\bar{\rho}_{e_{0.4}}}{\bar{\rho}_{e_{0.0}}}$ for the binary
compact objects systems with different total masses.}
\label{fig:rho-vs-M}
\end{figure}

To check the dependence of the SNR improvement on the eccentricity
for the binary systems with different total masses,
we evaluate the averaged SNR for $e_{0.0}$ and $e_{0.4}$ with
the Monte Carlo method using $10^4$ samplings,
then obtain their ratio
$\displaystyle\frac{\bar{\rho}_{e_{0.4}}}{\bar{\rho}_{e_{0.0}}}$.
We have considered the binary systems with total mass $2.74 M_\odot$,
$22 M_\odot$, $35 M_\odot$, $50 M_\odot$, $65 M_\odot$, $80 M_\odot$,
and $100 M_\odot$.
The results for the LHV, LHVK and LHVKI networks are plotted in
Fig.~\ref{fig:rho-vs-M}.
Overall, we can see that the improvement of the SNR from the eccentricity
depends on the total mass of the binary system for all networks.
Especially, it shows that the ratio of the average SNR is less than 1 for
the $2.74 M_\odot$ and the $22 M_\odot$ cases.
This leads to the weakened accuracy of the source localization at some
orientations for the small total mass binary systems.

\begin{table*}[htbp]
\caption{The comparison of the sky localization error $\Delta\Omega$'s of
neutron star binaries between this work and those in \cite{Nissanke_2011},
and their ratios.}
\linespread{1.3}\selectfont
\renewcommand\arraystretch{1.7}
\begin{threeparttable}
\begin{tabular}{c|c|c|c|c}
\toprule[0.5pt]
Case & Distance (Mpc) & LHV & LHVK & LHVKI(LHVKA)\tnote{a}\\
\hline
This work & 40 & $2.62\times 10^{-4}$sr & $9.23\times 10^{-5}$sr  &
$3.92\times 10^{-5}$sr \\ \hline
Fig.~1 in Ref.~\cite{Nissanke_2011} & 180 &
\tabincell[0.7]{c}{$3.66\times 10^{-3}$sr \\ (12 $\text{deg}^2$)\tnote{b}} &
\tabincell[0.7]{c}{$2.13\times 10^{-3}$sr \\ (7 $\text{deg}^2$)} &
\tabincell[0.7]{c}{$3.96\times 10^{-4}$sr \\ (1.3 $\text{deg}^2$)} \\ \hline
Fig.~2 in Ref.~\cite{Nissanke_2011} & 567 & 
\tabincell[0.7]{c}{$8.38\times 10^{-2}$sr \\ (275 $\text{deg}^2$)} &
\tabincell[0.7]{c}{$4.14\times 10^{-2}$sr \\ (136 $\text{deg}^2$)} &
\tabincell[0.7]{c}{$5.79\times 10^{-3}$sr \\ (19 $\text{deg}^2$)}
\\ \hline\hline
Case ratio &(Distance ratio$)^2$&
\multicolumn{3}{c}{Sky localization error ratio}\\ \hline
$\displaystyle\left(\frac{\text{Fig.~1 in Ref.~\cite{Nissanke_2011}}}
{\text{This work}}\right)$ & 20 & 14 & 23 & 10 \\ \hline
$\displaystyle\left(\frac{\text{Fig.~2 in Ref.~\cite{Nissanke_2011}}}
{\text{This work}}\right)$ & 201 & 320 & 448 & 148 \\ \hline
\end{tabular}
\begin{tablenotes}
\item[a] This work uses LHVKI while LHVKA is used in \cite{Nissanke_2011}.
\item[b] 1$\text{deg}^2=3.046174\times 10^{-4}$sr.
\end{tablenotes}
\end{threeparttable}
\label{tab:comMCMC}
\end{table*}

The nonzero initial eccentricity can be used to improve the accuracy of the
source localization.
But there is a certain binary mass range in which the corrections due to the
EPC model are expected to be futile.
This phenomenon could happen to the other more accurate waveform models.
However, this is not those waveform models' fault.
The problem really comes from the sensitivity of the GW detectors.
This is because most corrections from the eccentric-binary waveform models
show up in the higher-frequency bandwidth.
Below a certain mass range, the higher-frequency bandwidth coincides with
the lower sensitivity part of those detectors;
thus, the effort from the corrections is contaminated with the high-frequency
noise from the detectors, as shown in Fig.~\ref{fig:ASD_Flso}.
Nevertheless, this situation could be improved and thus the corrections can
be revived if the (high-frequency bandwidth) sensitivity of the GW detectors
is enhanced in the (near) future.

Here, we would like to compare our result with the previous investigations in
Ref.~\cite{Gond_n_2018_APJ_855_34, Gond_n_2018_APJ_860_5, Gond_n_2019_APJ_871_178}
about this field which form a series studying the effect of eccentric
compact binary on GW detection. However,
due to the different waveform model used and different initial eccentricities
considered on both sides, we can only make a qualitative comparison between
them:
\begin{itemize}
\item In this work, we mostly focus on the localization improvement with the
EPC model within a moderate initial eccentricity range, i.e., $0\le e_0\le0.4$,
via the networks with three, four, and five GW detectors.
In contrast, a general estimation for parameters, especially the initial
eccentricity $e_{\rm 10 Hz}$
\footnote{$e_{\rm 10 Hz}$ is close to the initial eccentricity $e_0$
in this work which is considered at the frequency $f_{\rm min}=20$Hz.}
at 10 Hz when entering aLIGO’s band, the eccentricity $e_{\rm LSO}$ at LSO,
and the pericenter distance $\rho_{\rm p 0}$, is performed with the network
of four GW detectors under the consideration of almost the whole eccentricity range,
especially the higher initial eccentricity $e\ge 0.9$, in
Refs.~\cite{Gond_n_2018_APJ_855_34, Gond_n_2018_APJ_860_5, Gond_n_2019_APJ_871_178}.
This makes these two works to be quite complementary in covering various
aspects for studying GW from eccentric binaries.
\item In general, we both see that more accurate parameter estimation of an
eccentric binary can be obtained with higher initial eccentricity,
although the factor of accuracy improvement varies for different parameters.
Therefore, the chirp mass measurement precision can improve by a factor of
$20$ for eccentric neutron star binaries with the initial eccentricity 
$e=0.9$ in Ref.~\cite{Gond_n_2019_APJ_871_178} while the improvements of their
localizations are shown to be negligible and could be even less somewhere,
with $e_0=0.4$ in this work, as mentioned before.
As in
Refs.~\cite{Gond_n_2018_APJ_855_34, Gond_n_2018_APJ_860_5, Gond_n_2019_APJ_871_178},
the estimate of slow parameters,
which appear in the slowly varying amplitude of GW signal,
is more accurate with more eccentric waveform.
This observation is consistent with the result in this work.
\end{itemize}

Finally, we compare our result with those in Ref.~\cite{Nissanke_2011}.
The work in Ref.~\cite{Nissanke_2011} studies sky localization for both individual
systems and populations of BNSs with the MCMC techniques using different
networks of advanced GW detectors.
In their results of normalized cumulative distributions of sky-error area,
the sky localization can be improved by increasing the number of detectors
in a network, which is consistent with our results.

We continue the comparison between these two by checking their sky localization
errors $\Delta\Omega$'s and their ratios in Table~\ref{tab:comMCMC}.
The $\Delta\Omega$'s shown in the second row of Table~\ref{tab:comMCMC}
are the average values of the integrals from the result of
Fig.~\ref{fig:BNS-histograms} for the listed five different eccentricities.
The third and fourth rows in Table~\ref{tab:comMCMC} give the results shown 
in Figs.~1 and 2 of Ref.~\cite{Nissanke_2011}.
Note that the data in the last column come from different networks. 
That is, we use the five-GW-detector network LHVKI,
while the authors in Ref.~\cite{Nissanke_2011} form another
five-GW-detector network by using LHVK + LIGO-Australia (LHVKA)
\cite{Barriga_2010,Munch_AIGO_2011}.
As we all know, the LIGO-Australia project no longer exists.

For a fair comparison of $\Delta\Omega$,
we need to compensate for the effect of the distance $D$.
As we already know, $\Delta\Omega\propto D^2$.
Therefore the bottom half of Table~\ref{tab:comMCMC} is dedicated to this 
purpose.
The bottom half shows the ratios of the $D^2$ and of $\Delta\Omega$ between the 
result in the two figures of Ref.~\cite{Nissanke_2011} and the one in this work.
The logic is that if the error ratio is larger than the distance square ratio,
our sky localization error is smaller than the one in Ref.~\cite{Nissanke_2011},
and vice versa.
We can see that the numbers in the each row for the last two rows of the table
are close, by considering the order of magnitude, except for the ones in the
last column for which the detection networks are different.
This indicates that the results in this work are quantitatively consistent with
(and could be slightly better than) those in Ref.~\cite{Nissanke_2011}.
In addition, the smaller ratio values in the last column assure us that
LIGO-Australia is a better location than LIGO-India for forming a global
network for detecting GW.

\section{conclusion}
\label{sec:conclusion}
Source localization is always an important issue for gravitational wave
astronomy.
This topic has been widely studied on quasicircular binary compact objects.
But for accurate localization the nonzero eccentricity of the orbit of a compact binary cannot be 
overlooked.
Currently, the two LIGO detectors and the VIRGO detector are operating
for the O3 run.
KAGRA in Japan will also join the O3 run soon at the end of this year.
In the near future the LIGO-India detector will be constructed.
So it will be quite interesting to see how these new detectors enhance
the accuracy of the source localization by forming a larger detector network.

In this work, we have studied the effects of the networks formed by
the gravitational detectors, nonzero eccentricity,
and the total mass of a compact binary, on the accuracy of localization, 
using the matched filtering technique and the Fisher information matrix method.
Recalling the results in Ref.~\cite{PhysRevD.96.084046},
we have found that the accuracy of the source localization can be improved by 
nonzero eccentricity with the three-detector LHV network.
And the improvement also depends on the total mass of the observed binary.
We extended our study in this work to the four-detector LHVK network
by adding KAGRA into the LHV network
and the five-detector LHVKI network
by adding LIGO-Indian into the LHVK network,
with the enhanced postcircular waveform model.
We find that the accuracy of the localization is improved considerably
with more detectors in a network, as expected.
Also, we find that the accuracy is improved significantly by increasing the
eccentricity for the large total mass,
roughly estimated as $M\ge 40 M_\odot$ from Fig.~\ref{fig:rho-vs-M},
binaries with all three networks.
For the small total mass, roughly $M<40 M_\odot$, binaries,
this effect is negligible.
For the smaller total mass, roughly $M<5 M_\odot$, binaries,
the accuracy could be even worse at some orientations with
increasing eccentricity.

According to the discussion in Sec.~\ref{sec:discussion},
this phenomenon mainly comes from how well the frequency of the higher
harmonic modes induced by the increased eccentricity coincides with the
sensitive bandwidth of the detectors.
One can read this quite clearly from Fig.~\ref{fig:ASD_Flso}.
For the big BBH case with the total mass $100 M_\odot$,
the improvement factor is about 2 in general, when the eccentricity grows
from zero to $0.4$.
For the GW151216-like BBH case with total mass $22 M_\odot$ and 
the GW170817-like BNS case with total mass $2.74 M_\odot$,
the improvement factor is less than $1.1$,
and it could be less than 1 at some orientations.
From our analysis we can expect that this limitation could be largely relieved 
once the sensitivity of the gravitational wave detectors on the high-frequency
bandwidth and/or the overall frequency bandwidth are improved
in the future.
And as the GW detectors improve their sensitivities,
the result in this work can serve as a comparison point
for more accurate models.

\section*{Acknowledgments}
We are very grateful to the anonymous referee for his useful comments that
improved the quality of this paper.
This work was supported by the Ministry of Science and Technology under
Grant No. MOST 106-2112-M-006-011.
Z.~Cao was supported by the NSFC (Grants No.~11690023 and No.~11622546) and ``the Fundamental Research Funds for the Central Universities" and the ``Interdiscipline Research Funds of Beijing Normal University."
We are grateful to the National Center for High-Performance Computing and
Institute of Astronomy and Astrophysics, Academia Sinica for providing
the computing resource.




\end{document}